\documentclass[11pt]{article}
\pdfoutput=1
\usepackage{jheppub,amsmath,amssymb,mathabx}
\usepackage{enumerate}
\usepackage{bm}
\usepackage{bbm}
\usepackage[usenames,dvipsnames]{xcolor}
\usepackage{dsfont}
\usepackage{graphics}
\usepackage{tikz,todonotes}
\usepackage{amssymb}
\usepackage{amsmath}
\usepackage[utf8]{inputenc}
\usetikzlibrary{arrows,positioning,decorations.markings,decorations.pathmorphing,calc}

\definecolor{hyperref}{RGB}{026,028,087}

\def\gsim{ \lower .75ex \hbox{$\sim$} \llap{\raise .27ex \hbox{$>$}} }
\def\lsim{ \lower .75ex \hbox{$\sim$} \llap{\raise .27ex \hbox{$<$}} }
\def\be{\begin{equation}}
\def\ee{\end{equation}}
\def\bea{\begin{eqnarray}}
\def\eea{\end{eqnarray}}
\newcommand{\nn}{\nonumber}

\newcommand{\Sc}{\mathcal{S}}

\newcommand{\Uc}{\mathcal{U}}

\newcommand{\Tc}{\mathcal{T}}
\newcommand{\Hc}{\mathcal{H}}

\def \bal#1\eal  {\begin{align} #1 \end{align}}
\newcommand{\ud} {\mathrm{d}}

\newcommand{\bfz} {{\bf z}}

\newcommand{\bfp} {{\bf p}}

\newcommand{\li}{{\lambda}}

\newcommand{\ba}{\begin{array}}
\newcommand{\ea}{\end{array}}

\newcommand{\commentout}[1]{}

\newcommand{\comment}[1]{}

\newcommand{\bs}{\begin{split}}

\def\ba{\begin{eqnarray}}
\def\ea{\end{eqnarray}}

\def\nn{\nonumber}

\def\d{\mathrm{d}}

\def\mn{_{\mu \nu}}

\def\({\left(}
\def\){\right)}
\def\ie{{\it i.e. }}

\newcommand*{\mathcolor}{}
\def\mathcolor#1#{\mathcoloraux{#1}}
\newcommand*{\mathcoloraux}[3]{%
  \protect\leavevmode
  \begingroup
    \color#1{#2}#3%
  \endgroup
}
\newlength{\stheight}
\newcommand\textst[1][fu-grey]{
	\ifmmode\setlength{\stheight}{+1.0ex}
	\else\setlength{\stheight}{+0.5ex}
	\fi
	\bgroup\markoverwith{\textcolor{#1}{\rule[\the\stheight]{2pt}{1.0pt}}}\ULon
}
\newcommand{\textins}[2][fu-grey]{
	\ifmmode\mathcolor{#1}{#2}
	\else\textcolor{#1}{#2}\@\,
	\fi
}

\def\({\left(}
\def\){\right)}

\def\L{{\cal L}}

\begin{document}

\title{UV complete me: \\
Positivity Bounds for Particles with Spin}

\author[a,b]{Claudia de Rham}
\author[a]{Scott Melville}
\author[a,b]{Andrew J. Tolley}
\author[a]{Shuang-Yong Zhou}
\affiliation[a]{Theoretical Physics, Blackett Laboratory, Imperial College, London, SW7 2AZ, U.K.}
\affiliation[b]{CERCA, Department of Physics, Case Western Reserve University, 10900 Euclid Ave, Cleveland, OH 44106, USA}

\emailAdd{c.de-rham@imperial.ac.uk}
\emailAdd{s.melville16@imperial.ac.uk}
\emailAdd{a.tolley@imperial.ac.uk}
\emailAdd{shuangyong.zhou@imperial.ac.uk}

\abstract{
For a low energy effective theory to admit a standard local, unitary, analytic and Lorentz-invariant UV completion, its scattering amplitudes must satisfy certain inequalities. While these bounds are known in the forward limit for real polarizations, any extension beyond this for particles with nonzero spin is subtle due to their non-trivial crossing relations. Using the transversity formalism (i.e. spin projections orthogonal to the scattering plane), in which the crossing relations become diagonal, these inequalities can be derived for 2-to-2 scattering between any pair of massive particles, for a complete set of polarizations at and away from the forward scattering limit. This provides a set of powerful criteria which can be used to restrict the parameter space of any effective field theory, often considerably more so than its forward limit subset alone.
}


\maketitle


\section{Introduction}

Low energy effective field theories (EFTs) are ubiquitous in modern physics, used to describe everything from fundamental particle interactions to phenomenological models of the Universe. The universal mechanism that underlies this is the decoupling of high energy physics from low energies whenever there is a clear hierarchy of scales. This decoupling is built into any local quantum field theory as a consequence of the uncertainty principle.
In the modern perspective, non-renormalizable theories such as General Relativity, far from being inconsistent with the tenets of quantum mechanics, should be viewed as EFTs which can be consistently quantized at low energies to any desired order of accuracy up to, at a given order, a finite number of undetermined matching coefficients. \\

Although decoupling guarantees that we do not need to know the explicit UV completion of a given low energy EFT to make predictions, not all information about the high energy physics is lost. A given EFT will by definition break down in predictivity at some energy scale, the cutoff. The standard approach to UV completion introduces new degrees of freedom at energies at and above the cutoff in such a way that the $ S$-matrix for the theory remains Lorentz invariant, analytic (causality), polynomially/exponentially bounded (locality) and unitary (predictive). If no such UV completion is possible, then the original EFT is describing something which is inherently incompatible with a local quantum field theory (QFT) description. In recent years it has been recognized that not all EFTs admit a local Lorentz invariant UV completion \cite{Vafa:2005ui,Adams:2006sv}. This can be demonstrated by showing that the requirement that the $ S$-matrix is analytic imposes nontrivial constraints, {\it positivity bounds}, on the scattering amplitudes of the low energy effective theory, which in turn place constraints on the form of the EFT Lagrangian which could not have been determined by low energy symmetries and unitary alone. For spin-0 particle scattering the explicit constraints that apply to the 2-to-2 scattering amplitude are straightforward to derive in the forward scattering limit \cite{Adams:2006sv} and have recently been generalized by the authors to an infinite number of bounds that apply away from the forward scattering limit \cite{deRham:2017avq}. These bounds have for instance been applied to Galileon EFTs in \cite{deRham:2017imi}.  \\

More recently, a similar program is being brought to bear on gravity, \ie the scattering of spin-2 particles. Remaining agnostic about the explicit form of a UV complete description of gravity, it is possible to derive constraints on gravitational EFTs by nevertheless demanding that such a UV completion should exist \cite{Bellazzini:2015cra, Cheung:2016wjt}. For example, this has been shown to severely constrain the parameter space of massive gravity \cite{Cheung:2016yqr}, and pseudo-linear massive gravity \cite{Bonifacio:2016wcb}. However, the majority of bounds in the literature to date rely on one crucial assumption: a trivial crossing relation between the processes
\begin{equation}
(\text{$s$-channel}) \quad  A + B \to C + D \;\;\;\; \text{and} \;\;\;\; (\text{$u$-channel}) \quad A + \bar D   \to  C + \bar B .
\label{eqn:su}
\end{equation}
For spins $S > 0$, this requirement forces one to consider only \emph{real polarizations} in the \emph{forward limit}. 
A discussion of the bounds for spinning particles in the forward scattering limit, which applies to more general polarizations, addressing the nontrivial issues with analyticity and statistics for fermions has been given recently in \cite{Bellazzini:2016xrt}. Fermion-boson scattering includes additional branch cuts, which significantly simplify in the forward scattering limit, and \cite{Bellazzini:2016xrt} demonstrates that forward limit positivity bounds can be meaningfully extended to elastic scattering of particles of any spin. In addition to these extra branch cuts,  another problem with extending these results away from forward scattering, is if the crossing relation is not sufficiently simple, it is not possible to guarantee positivity of the discontinuity of the scattering amplitude along the left hand branch cut which is a crucial ingredient\footnote{When crossing is simple, the left hand discontinuity is related to the discontinuity on the right hand cut which is guaranteed positive by unitarity.} in the proof of the positivity bounds. For example, the real polarizations (used in the $t \to 0$ limit) have a non-trivial optical theorem when $t \neq 0$ \cite{Jacob:1959at}. The more common \emph{helicity formalism}, while having clear unitarity properties, transforms in a complicated way under crossing \cite{Trueman:1964zzb}.  \\

Historically, a number of approaches have been taken to deal with this problem. One approach is to expand a general spin scattering amplitude in terms of scalar invariant amplitudes which have simple crossing properties \cite{Hepp:1964,Williams:1963zz}. In practice however this approach is cumbersome for general spins. Closely related to the helicity amplitudes are the $M$-amplitudes which transform covariantly as tensor-spinors and so have more straightforward crossing properties \cite{PhysRev.160.1251,Barut:1963zzb,Scadron:1969rw,PhysRev.130.436} (for a general discussion on the relation of these approaches see \cite{barut1967theory}). Although simplifying the crossing relations helps, it is also necessary to deal with quantities which have positive discontinuities. A further approach that addresses this is to consider linear combinations of helicity amplitudes which respect positivity along both the left and right hand cuts \cite{Mahoux:1969um}. \\

In this work, we demonstrate how the \emph{transversity formalism} \cite{Kotanski:1965zz}, in which the crossing relation is (semi)-diagonal, resolves these issues most straightforwardly. The transversity formalism is simply a change of polarization basis in which the spin of the particle is projected onto the normal to the scattering plane (hence the name transversity). The simplicity of the crossing relation in the transversity basis will allow us to infer a dispersion relation which is positive on both the left and right hand branch cuts, for any choice of transversities, and in so doing derive for scattering of particles of general spin direct analogues of the scalar positivity bounds derived in \cite{deRham:2017avq}. \\

In section~\ref{sec:formalisms} we briefly review the key properties of the helicity and transversity formalisms, and discuss the more complicated analyticity structure of spin scattering, and how to remove kinematic singularities.
Then in section~\ref{sec:identical} we prove the positivity properties of the dispersion relation, which in turn allows us to prove general positivity bounds for particles of arbitrary spin. For the sake of clarity, in section~\ref{sec:identical} we will give the expressions for identical particles of mass $m$ and spin $S$. In section~\ref{sec:different}, we discuss how this bound is altered in the case of different particle masses and spins. A summary and closing remarks are given in section~\ref{sec:conc}. Much of the formalism is given in the appendices. After reviewing the connection between analyticity and causality in appendix~\ref{app:AnalyticCausal}, we give a novel derivation of the crucial crossing formula for general spin scattering in appendix~\ref{app:multispinor}, and illustrate this in explicit examples in appendix~\ref{app:examples} which can be used to check the claimed analyticity properties. We connect this derivation with the historical approach in appendix~\ref{app:crossing} and finally we collect various technical properties of the amplitudes (appendix~\ref{app:symm}) and Wigner matrices (appendix~\ref{app:dJ}) for convenience.

\section{From Helicity to Transversity}
\label{sec:formalisms}

In this section, we first review the commonly used helicity formalism for calculating the scattering amplitude. Our principal concern is the scattering of massive particles although statements about the analyticity of tree level scattering amplitudes will also apply in the massless limit.
The helicity formalism is however not convenient to establish positivity bounds (except for the case of the forward scattering limit or the pure scalar interactions), as the crossing relations for nonzero spin particles are highly nontrivial. Fortunately, the crossing relations can be diagonalized by using the so-called transversity formalism, which we will introduce via a rotation from the helicity formalism. See Fig.~\ref{fig:h2t} for a pictorial view of both approaches.  \\

For concreteness, and to focus on the main points, we first look at the simple case where all the four particle masses are equal and the spin of particle $C$ $(S_3)$ and $D$ $(S_4)$ equal to that of particle A $(S_1)$ and B $(S_2)$ respectively:
\be
m_1 = m_2 = m_3 = m_4 =m, \quad S_3=S_1,~~S_4=S_2 \, .
\ee
We will then discuss extensions to more general cases where the masses can differ in Section~\ref{sec:different}.
\\

As we are dealing with 2-to-2 scattering amplitudes, we will make great use of the Mandelstam variables:
\ba
\label{eq:Mandelstam}
s = - ( k_1 + k_2 )^2 , \;\;\;\; t = - ( k_1 -  k_3 )^2  , \;\;\;\; u = - ( k_1 - k_4)^2 = 4m^2 - s-t\, .
\ea
For later convenience, it will also be useful to introduce the associated variables:
\ba
\label{eq:CalS}
{\cal S}= s(s-4m^2) \quad {\rm and }\quad {\cal U}= u(u-4m^2)\,,
\ea
which are both positive in the physical $s$ and $u$-channel regions.

\subsection{Helicity Formalism}

When computing 2-to-2 scattering amplitudes, it is standard to consider plane wave 2-particle incoming and outgoing states. However, unitarity is better expressed using spherical waves as we can exploit angular momentum conservation. In what follows we shall review the standard spherical and plane wave states before relating them by means of the partial wave expansion. For definiteness, we consider the scattering plane to be the $xz$ plane, and the $y$ direction to be orthogonal to the scattering plane. We further fix coordinates so that the incoming particles move along the $z$-axis without loss of generality. \\

\paragraph{Spherical wave states:}
Irreducible representations of the $SO(3)$
rotational symmetry provide the basis of the `spherical wave' states
\begin{equation}
J^2 | j m \rangle = j (j+1) | j m \rangle , \;\;\;\; J_z | j m \rangle = m | j m \rangle   \,,
\end{equation}
where of course, here $m$ is the spin projection along the $z$ direction, rather than the particle mass $m$. Any three-dimensional rotation can be characterized by three Euler angles $(\alpha, \beta, \gamma)$, and implemented on a state via the operator $R (\alpha, \beta, \gamma) = e^{-i \alpha J_z} e^{-i \beta J_y} e^{-i \gamma J_z} $, where $J_x$, $J_y$ and $J_z$ are the angular momentum operators. The action of $R (\alpha, \beta, \gamma)$ on the spherical wave states can be expressed in terms of the Wigner $D$ matrices \cite{wigner_algebra_1931}
\begin{equation}
 R ( \alpha, \beta, \gamma) | j m \rangle = \sum_{m'=-j}^j D^j_{m' m} (\alpha , \beta, \gamma ) | j m' \rangle  \, ,
\end{equation}
where
\begin{align}
D^j_{m' m} (\alpha, \beta, \gamma )
&= e^{-i \alpha m'} d^j_{m' m} (\beta) e^{-i \gamma m}   ,
{\rm ~~with~~}
d^j_{m' m} (\beta) = \langle j m' | e^{-i \beta J_y} | j m \rangle \label{eqn:dJ} \, .
\end{align}
Explicit expressions for the small $d$ matrix are given in Appendix~\ref{app:dJ}.  \\

\paragraph{Plane wave states:}
On the other hand, one particle `plane wave' states are eigenstates of momentum, with well-defined angular momentum in the rest frame
\begin{align}
J^2 | \mathbf{p}=0, S , \lambda \rangle = S (S+1) | \mathbf{p}=0, S , \lambda \rangle , \;\;\;\; J_z | \mathbf{p} =0 , S, \lambda \rangle = \lambda | \mathbf{p}=0 ,S , \lambda \rangle  \,,
\end{align}
where $S$ is the spin of the particle. These transform into each other under boosts and rotations. For example, a nonzero momentum state is constructed from the rest frame as
\begin{align}
| \mathbf{p}, S, \lambda \rangle =  R ( \phi, \theta, 0 ) L( p ) | 0, S, \lambda \rangle \;\;\;\; \text{for} \;\;\;\; \mathbf{p} = (p\sin \theta \cos \phi, p\sin\theta \sin \phi, p \cos \theta )  \, ,
\end{align}
where $L(p)$ is the boost along the $z$ direction to momentum $p\hat{\bfz}$. Note that a finite momentum state no longer has well-defined angular momentum, \emph{except} along the momentum axis
\begin{equation}
 \frac{ \mathbf{J} \cdot \mathbf{p}}{|\bfp|}  \;  | \mathbf{p} , S, \lambda \rangle = \lambda | \mathbf{p} , S , \lambda \rangle .
\end{equation}
Physically, this is because the orbital angular momentum $\mathbf{L} = \mathbf{r} \times \mathbf{p}$ is zero along this axis. $\lambda$ is called {\it helicity}, and is a good quantum number in all reference frames \cite{Jacob:1959at}. \\

Two particle states are constructed simply as the tensor product of one particle states
$ | \mathbf{p}_1 \mathbf{p}_2  \lambda_1 \lambda_2 \rangle   = | \mathbf{p}_1 , S_1 , \lambda_1 \rangle \otimes | \mathbf{p}_2 , S_2, \lambda_2 \rangle $ .
The particle spins are some fixed, known quantities, so one usually omits them in the kets, and one can also factor out the center of mass motion and write \cite{Jacob:1959at, Richman:1984gh}
\begin{equation}
 | \mathbf{p}_1 \mathbf{p}_2  \lambda_1 \lambda_2 \rangle = 2 \pi  \sqrt{ \frac{4 \sqrt{s}}{p} } |p \theta \phi \lambda_1 \lambda_2 \rangle | P \rangle \, ,
\end{equation}
where $P = ( 2 \sqrt{m^2+p^2} ,0,0,0)$ is the center of mass 4-momentum of the two particles, $p$ is the 3-momentum value of particle A (or B), and the normalizations are $\langle p \theta' \phi' \lambda_1' \lambda_2' |p \theta \phi \lambda_1 \lambda_2 \rangle  = \delta(\cos \theta'-\cos \theta) \delta(\phi'-\phi)\delta_{\lambda_1'\lambda_1}\delta_{\lambda_2'\lambda_2}$ and  $\langle P | P' \rangle = (2\pi)^4 \delta^4 (P' - P)$.
$\theta$ and $\phi$ are the angles of particle A in the center of mass system, and $P$ contains the information of the total momentum and just goes along for the ride.  \\

\paragraph{Partial wave expansion:}
Now, we want to relate the plane wave 2-particle states $| p \theta \phi \lambda_1 \lambda_2 \rangle$, which we use when calculating scattering amplitudes, to the spherical wave 2-particle states $| p J M \lambda_1 \lambda_2 \rangle$, which provide a convenient expression of unitarity. To do this, note that when the two particles collide along the $z$-axis, we have $J_z | p 0 0 \lambda_1 \lambda_2 \rangle =  (\lambda_1 - \lambda_2 ) | p 0 0 \lambda_1 \lambda_2 \rangle $, which implies
\be
 | p 0 0  \lambda_1 \lambda_2 \rangle = \sum_J c_J \; |p J \lambda  \lambda_1 \lambda_2 \rangle , ~~{\rm with~~} \lambda = \lambda_1 - \lambda_2    .
\ee
One may normalize the spherical wave states as $\langle p J'M'\lambda_1'\lambda_2'| p JM\lambda_1\lambda_2\rangle = \delta_{J'J}\delta_{M'M}\delta_{\lambda_1' \lambda_1}\delta_{\lambda_2'\lambda_2}$, and this then fixes $c_J$ up to a phase $ | c_J |^2 = (2 J+1)/4 \pi$. We will choose this phase to be zero. Now we can use a rotation to go to any desired collision axis
\be
 | p  \theta  \phi   \lambda_1 \lambda_2 \rangle = R (\phi, \theta, 0 ) | p  0  0  \lambda_1 \lambda_2 \rangle = \sum_{J,M} \sqrt{ \frac{2J+1}{4 \pi} } D^J_{M \lambda} (\phi, \theta,  0 ) | p  J  M   \lambda_1 \lambda_2  \rangle   \, .
 \ee

Consider scattering between initial state $ |i \rangle = |p_i  0  0  \lambda_1 \lambda_2 \rangle   $ and final state $| f \rangle =  | p_f  \theta  \phi  \lambda_3 \lambda_4 \rangle  $ that conserves the total energy momentum. Splitting the $S$ matrix into $\hat S =1 + i \hat T$, and remembering that we can factor out an overall momentum conserving delta function, we define the helicity amplitude $ \Hc_{\lambda_1 \lambda_2 \lambda_3 \lambda_4} ( s, \theta)$ via
\be
\langle f |\hat T | i \rangle =  (2\pi)^4  \delta^4(P_f-P_i)  \Hc_{\lambda_1 \lambda_2 \lambda_3 \lambda_4} ( s, \theta) \, ,
\ee where
\begin{align}
\Hc_{\lambda_1 \lambda_2 \lambda_3 \lambda_4} ( s, \theta) = 16 \pi^2 \sqrt{ \frac{s}{p_i p_f} } \langle  p_f \theta \phi \lambda_3 \lambda_4 | \hat T | p_i0 0 \lambda_1 \lambda_2 \rangle \, ,
\end{align}
and where $s$ is the center of mass energy square defined in \eqref{eq:Mandelstam}. \\

Then inserting the complete spherical wave basis, we have
\ba
\Hc_{\lambda_1 \lambda_2 \lambda_3 \lambda_4} ( s, \theta) &=&  16 \pi^2 \sqrt{\frac{s}{p_i p_f}} \sum_{JM}  \langle p_f \theta \phi \lambda_3 \lambda_4 | p_f J M \lambda_3 \lambda_4 \rangle T_{\lambda_1 \lambda_2 \lambda_3 \lambda_4}^J \langle p_i J M \lambda_1 \lambda_2 |p_i 0 0 \lambda_1 \lambda_2 \rangle \nn  \\
 &=&  4 \pi \sqrt{\frac{s}{p_i p_f}}  \sum_J (2J+1)\, e^{i \lambda\phi} d_{\lambda \mu}^J ( \theta) T^J_{\lambda_1 \lambda_2 \lambda_3 \lambda_4}(s)   \,,
 \label{eq:sphericalWave}
\ea
where
\be
\lambda = \lambda_1 - \lambda_2 {,~~~~~~~}\mu = \lambda_3 - \lambda_4
\ee
and $T^J_{\lambda_1 \lambda_2 \lambda_3 \lambda_4}(s)$ is the partial wave helicity amplitude
\be
 T^J_{\lambda_1 \lambda_2 \lambda_3 \lambda_4}(s) = \langle p_f J M \lambda_3 \lambda_4 | \hat T | p_i J M \lambda_1 \lambda_2 \rangle \,.
\ee
Since the angle $\phi$ is unimportant as the system is symmetric with respect to rotations about the collision axis, we may set $\phi=0$. Physically, $T^J_{\lambda_1 \lambda_2 \lambda_3 \lambda_4}(s)$ is the scattering amplitude between two particles states of definite total angular momentum and definite individual helicities. \\

\paragraph{Kinematical singularities:}
In going from the $s$ channel scattering angle $\theta$ to the Mandelstam variables\footnote{The amplitude is analytic in Mandelstam variables \cite{Mahoux:1969um}, up to known kinematic branch cuts \cite{Hara:1964zza}.} defined in \eqref{eq:Mandelstam}, we have
\begin{equation}
\cos \theta = 1 + \frac{2t}{s-4m^2} , \;\;\;\; \sin \theta = \frac{ 2 \sqrt{t u}  }{ s-4m^2 }   .
\label{eqn:kinematical}
\end{equation}
We see that additional singularities at $s=4m^2$ may be introduced. In the physical region, $t \to 0$ whenever $s \to 4m^2$, and so these residues vanish--- \ie these poles are unphysical. These are known as `kinematical singularities' \cite{Trueman:1964zzb}, and can be systematically removed \cite{Wang:1966zza,cohen-tannoudji_kinematical_1968}. We will return to this in more detail when we construct our transversity amplitudes. \\

\paragraph{Unitarity:}
Since angular momentum is conserved in the scattering process, the S-matrix can be block-diagonalized to different partial waves labelled by $J$. The partial wave unitarity condition then gives $i (\hat T^{J\dagger} - \hat T^J) = \hat T^{J\dagger} \hat T^J$, where $\hat T^J$ is the $J$ component of the partial wave expansion of the transition matrix $\hat T$. For the helicity amplitudes, this implies
\begin{align}
\label{pwuc}
{\text{Abs}}_s \, T^J_{\lambda_1 \lambda_2 \lambda_3 \lambda_4} &= \frac 12   \sum_N \, \langle p_f J M \lambda_3 \lambda_4 | \hat T^{J \dagger} | N \rangle \langle N | \hat T^J | p_i J M \lambda_1 \lambda_2 \rangle  ,
\end{align}
where we have omitted the momentum labels in the bras and kets, as we will do in the following, the sum runs over all intermediate states and we have defined the $s$-channel ``absorptive'' part of an arbitrary function $f$ as\footnote{Strictly speaking, the partial wave unitarity only gives $(T^J_{\lambda_1 \lambda_2 \lambda_3 \lambda_4}(s+i\epsilon)-T^J_{\lambda_3 \lambda_4 \lambda_1 \lambda_2}(s+i\epsilon)^*)/2i$ as the left hand side of Eq.~(\ref{pwuc}). However, since the $S$ matrix (and $T$ matrix) is Hermitian analytic, we have $[T^J_{\lambda_3 \lambda_4 \lambda_1 \lambda_2}(s+i\epsilon)]^*=T^J_{\lambda_1 \lambda_2\lambda_3 \lambda_4 }(s-i\epsilon)$. }
\bal
{\text{Abs}}_s \, f(s)  = \frac{1}{2i}\text{Disc} \, f(s)   =  \frac{1}{2i}\lim_{\epsilon \to 0} \left[ f(s+i\epsilon)  - f(s-i\epsilon) \right] , \quad \quad \text{for } s\ge 4m^2  .
\eal
If the scattering process is time reversal invariant, the $S$ matrix (and $T$ matrix) is real analytic, so we further have $T^J_{\lambda_1 \lambda_2 \lambda_3 \lambda_4}=T^J_{ \lambda_3 \lambda_4\lambda_1 \lambda_2}$. Then the absorptive part is just the imaginary part
\be
{\text{Abs}}_s \, T^J_{\lambda_1 \lambda_2 \lambda_3 \lambda_4} = \text{Im} \, T^J_{\lambda_1 \lambda_2 \lambda_3 \lambda_4} ~~~~~~~\text{if time reversal invariant}  .
\ee
From Eq.~(\ref{pwuc}), we see that ${\text{Abs}}_s \, T^J_{\lambda_1 \lambda_2 \lambda_3 \lambda_4}$ has the form $\sum_\alpha \alpha_{\lambda_3 \lambda_4}^* \alpha_{\lambda_1 \lambda_2}$. If one regards $\{\lambda_1 \lambda_2\}$ and $\{\lambda_3 \lambda_4\}$ as two indices, then it is clear that
\be
\label{uniT}
{\text{Abs}}_s\, T^J_{\lambda_1 \lambda_2 \lambda_3 \lambda_4} {\text{ is a positive definite Hermitian matrix}}.
 \ee

\paragraph{Crossing:}
The amplitude associated with the $s$-channel ($A + B \to C + D$) is denoted $\Hc^s_{\lambda_1 \lambda_2 \lambda_3 \lambda_4}$, and the corresponding $u$-channel amplitude is $\Hc^u_{\lambda_1 \lambda_4 \lambda_3 \lambda_2}$, describing the process $A + \bar D \to C + \bar B$, where $\bar B$ and $\bar D$ denote the antiparticles of $B$ and $D$. Then under crossing particles $B \leftrightarrow D$ (when $S_1=S_3$ and $S_2=S_4$ and the particles have equal mass $m$), we have  \cite{Trueman:1964zzb,kotanski_transversity_1970,Hara:1970gc,Hara:1971kj}
\ba
\Hc_{\lambda_1 \lambda_2 \lambda_3 \lambda_4}^s ( s, t , u)
&=& (-1)^{2S_2}  \sum_{\lambda_i'}  e^{i \pi(\lambda_1'-\lambda_3')}   d^{S_1}_{\lambda_1' \lambda_1} ( \chi_u )  d^{S_2}_{\lambda_2' \lambda_2} ( -\pi + \chi_u )   \nn  \\
&&~~~~~~~\cdot  d^{S_1}_{\lambda_3' \lambda_3} (- \chi_u )  d^{S_2}_{\lambda_4' \lambda_4} ( \pi -  \chi_u ) \Hc^u_{\lambda_1' \lambda_4' \lambda_3' \lambda_2'} (u, t, s )     ,
\label{eqn:helicity_crossing}
\ea
where the angle $\chi_u$ is given by
\be
\label{chianglesame}
 \cos \chi_u = \frac{-su}{ \sqrt{\Sc \Uc } } \,, \quad \sin \chi_u =  \frac{- 2m \sqrt{stu}}{ \sqrt{ \Sc \Uc }  } \,, \quad {\rm or }\quad e^{\pm i \chi_u} = \frac{-su \mp 2 i m \sqrt{s t u}}{\sqrt{\Sc \Uc}} \,.
\ee
This result, as well as its inelastic generalization, are derived in Appendices \ref{app:multispinor} and \ref{app:crossing} and are checked through examples in \ref{app:examples}. In the forward scattering limit $t=0$, we have $\chi_u = 0$ and hence,
\ba
 \Hc_{\lambda_1 \lambda_2 \lambda_3 \lambda_4}^s  ( s, 0, u ) =    \Hc^u_{\lambda_1 -\lambda_4 \lambda_3 -\lambda_2} (u, 0, s ) \, ,
\ea
which is consistent with the result of \cite{Bellazzini:2016xrt}, where the helicities flip sign because the momenta effectively reverse. To avoid excessive notation, in what follows we will drop the $(s,t,u)$-channel sub/superscript notation when unimportant.

\subsection{Transversity Formalism}
\label{sec:trfm}

\begin{figure}
\centering
\includegraphics[width=0.85\textwidth]{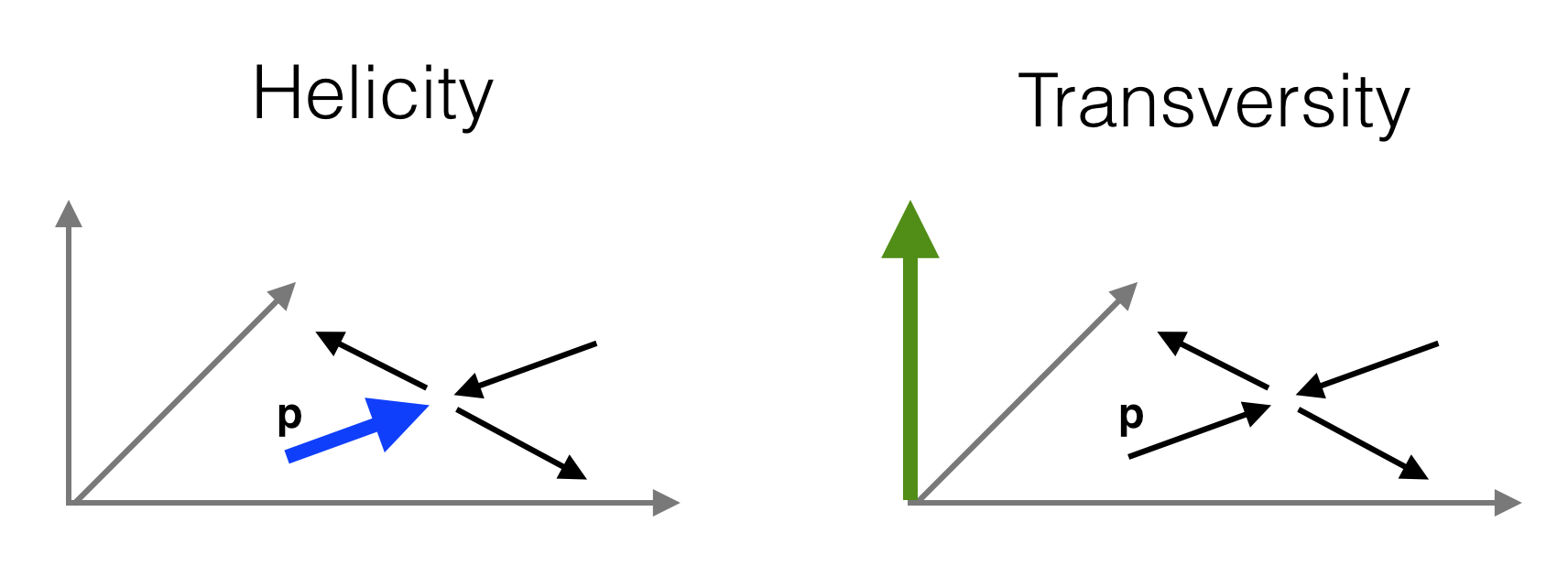}
\caption{The difference between the helicity and transversity formalism. The horizontal plane ($xz$ plane) is the particle interaction plane. In the helicity formalism particle spins are projected onto the direction of motion, while in the transversity formalism particle spins are projected in the vertical direction, which is transverse to the interaction plane. \label{fig:h2t} }
\end{figure}

Since $\Hc_{\lambda_1 \lambda_2 \lambda_3 \lambda_4} (s,t,u)$ contains a branch cut on the real axis of the complex $s$ plane between $s=4m^2$ and $\infty$, the crossing symmetry implies that there is a second branch cut in the real axis between $s=-t$ and $-\infty$. However, this second branch cut has no obvious positivity properties in the helicity formalism, due to the complicated crossing mixing of different helicity amplitudes as can be  seen from Eq.~\eqref{eqn:helicity_crossing} (unless $\chi_u=0$, corresponding to the forward scattering limit $t=0$, or unless all particles have zero spin). To go beyond the forward scattering limit for non-zero spins, we first need to simplify  the crossing relation by going to the transversity basis, see Fig.~\ref{fig:h2t}. \\

\paragraph{Transversity Amplitudes:}
We define the transversity eigenstates \cite{Kotanski:1965zz,kotanski_transversity_1970} as a particular combination of the helicity eigenstates
\be
  | \mathbf{p} , S, \tau \rangle  \equiv \sum_\lambda u^S_{ \lambda \tau } \, | \mathbf{p} , S, \lambda \rangle  \, ,
\ee
where 
the unitary matrix $u^S_{ \lambda \tau}$ is simply the Wigner $D^S$ matrix
associated with the rotation $R=e^{-i \pi/2 J_z}e^{-i \pi/2 J_y}e^{i \pi/2 J_z}$,
\be
 u^S_{ \lambda \tau} = D^S_{ \lambda \tau} \left( \frac{\pi}{2}, \frac{\pi}{2}, - \frac{\pi}{2} \right)\,.
\ee
This unitary $u^S$ matrix has the virtue of diagonalizing any of the Wigner $d^S$ matrix, independently of their angles. See Appendix \ref{app:dJ} for properties of the $u^S$ matrices. \\

The transversity amplitudes are thus related to the helicity amplitude via
\begin{equation}
\Tc_{\tau_1 \tau_2 \tau_3 \tau_4} = \sum_{\lambda_1 \lambda_2 \lambda_3 \lambda_4} u^{S_1}_{\lambda_1 \tau_1} u^{S_2}_{\lambda_2 \tau_2} u^{S_1 *}_{\tau_3 \lambda_3} u^{S_2 *}_{\tau_4 \lambda_4} \Hc_{\lambda_1 \lambda_2 \lambda_3 \lambda_4}  \, .
\label{eqn:transtohel}
\end{equation}
Physically, this corresponds to scattering particles with definite spin projection orthogonal to the scattering plane, \ie eigenvalues of the operator \cite{cohen-tannoudji_kinematical_1968}
\begin{equation}
\tau =-\frac{1}{m} w_\mu W^\mu(k_i)  \;\;\;\; \text{with} \;\;\;\; w_\mu = \frac{- 2 \epsilon_{\mu \nu \rho \sigma} k_1^\nu k_2^\rho k_3^\sigma  }{\sqrt{stu}} \,,
\label{eqn:transversity_operator}
\end{equation}
where $W^\mu(k_i)$ is the Pauli-Lubanksi (pseudo)vector of particle $i$,  $\epsilon_{\mu \nu \rho \sigma}$ is the Levi-Civita tensor and $k_1$, $k_2$ and $k_3$ are the respective momenta of particles A, B and C.
 In short, while the spin quantization axis of the helicity formalism is chosen to be the momentum direction of an incident particle, in the transversity formalism it is chosen to be transverse to the scattering plane (see Fig.~\ref{fig:h2t}).
 An important point to notice, which will become significant later, is that there is in general a $\sqrt{stu}=0$ branch point in the scattering amplitudes. Furthermore under $\sqrt{stu} \rightarrow -\sqrt{stu}$ the transversity $\tau$ flips sign, something that will be clear in the properties of the transversity amplitudes. We will see below how to deal with this kinematic branch point.
  \\

\paragraph{Crossing:}
As already mentioned, the main motivation for considering the transversity amplitudes is that  the unitary matrices $u^S$ diagonalize the Wigner matrices appearing in the helicity crossing relation \eqref{eqn:helicity_crossing}, so the crossing relations in the transversity formalism are much simpler. Explicitly, for $S_3 = S_1$ and $S_4= S_2$ we have
\be
\Tc^s_{\tau_1 \tau_2 \tau_3 \tau_4} (s, t, u )  = (-1)^{2S_1+2S_2} e^{i \pi \sum_i \tau_i}  e^{-i \chi_u \sum_i \tau_i}\, \Tc^u_{-\tau_1 - \tau_4 -\tau_3 - \tau_2} ( u, t, s) \, .
\label{eqn:transversity_crossing}
\ee
where $\chi_u$ is given in \eqref{chianglesame}. Further considering elastic transversities $\tau_1=\tau_3$, $\tau_2=\tau_4$ this is
\be
\Tc^s_{\tau_1 \tau_2 \tau_1 \tau_2} (s, t, u )  =  e^{-i \chi_u \sum_i \tau_i}\, \Tc^u_{-\tau_1 - \tau_2 -\tau_1 - \tau_2} ( u, t, s) \, .
\label{eqn:crossing_elastic}
\ee
This is considerably simpler than the equivalent expression in the helicity basis. 
If we further take the forward scattering limit $t=0$, we then have $\chi_u=0$ and so
\be
\Tc^s_{\tau_1 \tau_2 \tau_3 \tau_4} (s, 0, u )  =  \Tc^u_{-\tau_1 - \tau_4 -\tau_3 - \tau_2} ( u, 0, s) \, ,
\label{eqn:crossing_forwd}
\ee
where the transversities flip sign because the scattering plane normal is reversed.

\paragraph{Kinematical Singularities:}
Note that at first sight the $e^{i \chi_u}$ term appears to introduce additional poles/branch cuts in the complex $s$ plane. For BB or FF scattering, $\sum_j \tau_j$ is an even number, and the worst singularities are additional kinematic poles at $s, u=4m^2$ and a kinematic branch point at $stu=0$. For BF scattering, $\sum_j \tau_j$ is odd\footnote{
In the special case of parity invariant amplitudes, $\Tc_{\tau_1 \tau_2 \tau_3 \tau_4}$ vanishes unless $ \tau_1+\tau_2-\tau_3-\tau_4$ is an even integer \cite{kotanski_transversity_1970}, but we will not assume parity invariance here.
}, and the crossing factor has an additional kinematic branch point at $su=0$ \cite{martin_rigorous_1971}. We shall see below that these branch points and additional kinematical poles are removed by multiplying the scattering amplitudes by an additional regulating factor.   \\

In general potential poles and branch cuts arise only at\footnote{
These are known as `thresholds' (at $s=4m^2$), and `pseudothresholds' (at $s=0$), and the `boundary of the physical space' (at $\sqrt{stu}=0$) in \cite{cohen-tannoudji_kinematical_1968} and others.
}

\begin{itemize}

\item $s=0$: The helicity amplitudes can be shown to be regular at $s=0$ \cite{cohen-tannoudji_kinematical_1968}, and therefore so are the transversity amplitudes (by Eq.~\eqref{eqn:transtohel}).

\item $s=4m^2$: These are factorizable singularities, which can be removed by multiplying an appropriate prefactor
\ba
\label{eq:pref4m2}
\big(\sqrt{-u}\big)^{\xi}  \left( \sqrt{s-4m^2} \right)^{ |\sum_i\tau_i|} \Tc_{\tau_1 \tau_2 \tau_3 \tau_4} (s, t, u)   \, ,
\ea
where $\xi=1$ if $S_1+S_2$ is equal to a half integer (\ie for BF scattering) and $\xi=0$ otherwise.
Essentially, this is subtracting all of the terms that go as negative powers of  $(s-4m^2)^{1/2}$ from the amplitude. On the physical space, $s=4m^2$ requires $t=0$, where it can be shown that the residue associated with these poles then vanish, and  these terms are therefore not physical poles. However, their presence complicates the dispersion relation and so it is more convenient to subtract them. Recall that these come from replacing $\theta_s$ with $s,t,u$ by \eqref{eqn:kinematical}. To be consistent with the crossing relation\footnote{
The crossing relation \eqref{eqn:transversity_crossing} exchanges $1/(s-4m^2)$ with $1/(s+t)$. As there is originally no pole at $u=4m^2$ ($s=-t$) in the scattering amplitude, the crossing relation must be turning \emph{every} $s=4m^2$ kinematical singularity into one at $u=4m^2$.
}, we see that the maximal order of these poles is $N= |\sum_i\tau_i | \leq 2(S_1+S_2)$ for elastic amplitudes.

In practice it will prove convenient to utilize the prefactor
\ba
\label{eq:pref4m22}
\big(\sqrt{-su}\big)^{\xi} {\cal S}^{S_1+S_2}\,  \Tc_{\tau_1 \tau_2 \tau_3 \tau_4} (s, t, u)  \, ,
\ea
where ${\cal S}= s(s-4m^2)$ as defined in \eqref{eq:CalS}.
For elastic scattering, the prefactor in \eqref{eq:pref4m22} has indeed the same analytic structure as that in \eqref{eq:pref4m2}. We emphasize however that \eqref{eq:pref4m2} has the virtue of being applicable in the more general case of inelastic scattering, and preserves the positivity along both the left and right hand cuts.

\item $\sqrt{stu}=0$: There is a potential branch point at $stu=0$, which can be removed by taking an appropriate combination of the transversity amplitudes \cite{martin_rigorous_1971}.
Since $\sqrt{stu} = {\cal S} \sin \theta_s/\sqrt{4s}$, $\sqrt{stu} \leftrightarrow - \sqrt{stu}$ corresponds to $\theta_s \leftrightarrow - \theta_s$. Consequently any even function of $\theta_s$ will not contain the branch cut.
The two natural combinations are
\begin{equation}
 \Tc_{\tau_1 \tau_2 \tau_3 \tau_4}(\theta) + \Tc_{\tau_1 \tau_2 \tau_3 \tau_4}(-\theta) \, ,
\label{eqn:T_sum}
\end{equation}
and
\begin{equation}
\sqrt{stu} \left( \Tc_{\tau_1 \tau_2 \tau_3 \tau_4}(\theta) -\Tc_{\tau_1 \tau_2 \tau_3 \tau_4}(-\theta)  \right) \, .
\label{eqn:T_diff}
\end{equation}
In general as we go around $\sqrt{stu} = 0$, we have \cite{cohen-tannoudji_kinematical_1968}
\be
 \Tc_{\tau_1 \tau_2 \tau_3  \tau_4 } |_{\sqrt{stu}} = (-1)^{S_1+S_3-S_2-S_4} e^{i \pi \sum_i \tau_i} \Tc_{-\tau_1 -\tau_2  -\tau_3 - \tau_4} |_{-\sqrt{stu}} \,  .
 \ee
\ie
\be
\Tc_{\tau_1 \tau_2 \tau_3 \tau_4}(s,-\theta)=(-1)^{S_1+S_3-S_2-S_4} e^{i \pi \sum_i \tau_i} \Tc_{-\tau_1 -\tau_2 -\tau_3 -\tau_4}(s,\theta) \, .
\ee
For elastic scattering $\Tc_{\tau_1 \tau_2 \tau_1 \tau_2}(-\theta)=\Tc_{-\tau_1 -\tau_2 -\tau_1 -\tau_2}(\theta)$, and so in this case the sum \eqref{eqn:T_sum} and difference \eqref{eqn:T_diff} can also be written as
\begin{equation}
 \Tc_{\tau_1 \tau_2 \tau_1 \tau_2}(s,t,u) + \Tc_{-\tau_1 -\tau_2 -\tau_1 -\tau_2}(s,t,u) \, ,
\label{eqn:T_sum_2}
\end{equation}
or
\begin{equation}
\sqrt{stu} \left( \Tc_{\tau_1 \tau_2 \tau_1 \tau_2}(s,t,u) - \Tc_{-\tau_1 -\tau_2 -\tau_1 -\tau_2}(s,t,u)  \right) \, ,
\label{eqn:T_diff_2}
\end{equation}
and have trivial monodromy and carry no branch cut from $stu=0$.

\end{itemize}
In summary, we shall consider the regularized amplitudes\footnote{The expressions \eqref{eqn:T+} and \eqref{eqn:T-} are the most convenient ones when dealing with elastic scattering. As already emphasize, when dealing with inelastic scattering, the prefactor $\big(\sqrt{-su}\big)^{\xi} {\cal S}^{S_1+S_2}$ should instead be replaced by
$\big(\sqrt{-u}\big)^{\xi}  \left( \sqrt{s-4m^2} \right)^{ |\sum_i\tau_i|}$ as determined in \eqref{eq:pref4m2}.}
\bal
 \Tc^{+}_{\tau_1 \tau_2 \tau_3 \tau_4} (s,\theta) &= \big(\sqrt{-su}\big)^{\xi} {\cal S}^{S_1+S_2} \big(   \Tc_{\tau_1 \tau_2 \tau_3 \tau_4} (s,\theta) + \Tc_{\tau_1 \tau_2 \tau_3 \tau_4} (s,-\theta)    \big)   \, ,
\label{eqn:T+}
\\
 \Tc^{-}_{\tau_1 \tau_2 \tau_3 \tau_4}(s,\theta) &= -i \sqrt{stu}\big(\sqrt{-su}\big)^{\xi} {\cal S}^{S_1+S_2}  \big(   \Tc_{\tau_1 \tau_2 \tau_3 \tau_4} (s,\theta) - \Tc_{\tau_1 \tau_2 \tau_3 \tau_4} (s,-\theta)    \big)   \, ,
\label{eqn:T-}
\eal
where ${\cal S}= s(s-4m^2)$ as defined in \eqref{eq:CalS},
$\xi=1$ if $S_1+S_2$ is half integer and $\xi=0$ otherwise.  These have nicer crossing relations than the helicity amplitudes, (see Eq.~\eqref{eqn:transversity_crossing} or even Eq.~\eqref{eqn:crossing_elastic} in the elastic case)  and are also free of all kinematical singularities (poles and branch points).

\section{Positivity Bounds}
\label{sec:identical}

In this section, we make use of the transversity amplitudes to derive an infinite number of positivity bounds for non-forward scattering amplitudes of arbitrary spins.

\subsection{Unitarity and the Right Hand Cut}
\label{sec:rhc}

To begin with we consider the case of elastic scattering of particles of definite transversity, so that
\be
\tau_3=\tau_1 \quad{\rm and } \quad\tau_4=\tau_2  .
\ee
The partial wave expansion for transversity eigenstates is rather complicated \cite{kotanski_transversity_1970,Peters:2004qw}, in essence because one cannot define a rotationally invariant notion of transversity in a state with only two particles. Instead, we use the helicity partial wave expansion
\ba
\Tc_{\tau_1 \tau_2 \tau_1 \tau_2}(s,\theta) = \sum_{J \lambda_1 \lambda_2 \lambda_3 \lambda_4} u^{S_1}_{\lambda_1 \tau_1} u^{S_2}_{\lambda_2 \tau_2} u^{S_1 *}_{\tau_1 \lambda_3} u^{S_2 *}_{\tau_2 \lambda_4} d^J_{\mu \lambda}(\theta) \bar T^J_{\lambda_1 \lambda_2 \lambda_3 \lambda_4}(s)\,  ,
\label{eq:PWTs}
\ea
where we have set the interaction plane to lie along $\phi=0$ and in analogy with \eqref{eq:sphericalWave}, we have  defined
\be
\bar T^J_{\lambda_1 \lambda_2 \lambda_3 \lambda_4} = 4\pi (2J+1) \sqrt{\frac{s}{p_i p_f}} T^J_{\lambda_1 \lambda_2 \lambda_3 \lambda_4}    \,.
\ee
Using the properties of $\Tc_{\tau_1 \tau_2 \tau_1 \tau_2}(s,\theta) $ under $\theta \rightarrow - \theta$ we find
\ba
\label{sumTT}
 \Tc_{\tau_1 \tau_2 \tau_1 \tau_2}(s,\theta) &+& \Tc_{-\tau_1 -\tau_2 -\tau_1 -\tau_2}(s,\theta) =  \Tc_{\tau_1 \tau_2 \tau_1 \tau_2}(s,\theta) +  \Tc_{\tau_1 \tau_2 \tau_1 \tau_2}(s,-\theta) \nn  \\ &=&\sum_{J \lambda_1 \lambda_2 \lambda_3 \lambda_4} u^{S_1}_{\lambda_1 \tau_1} u^{S_2}_{\lambda_2 \tau_2} u^{S_1*}_{\tau_1 \lambda_3} u^{S_2*}_{\tau_2 \lambda_4}  \left( d^J_{\mu \lambda}(\theta)+ d^J_{\mu \lambda}(-\theta)\right)  \bar T^J_{\lambda_1 \lambda_2 \lambda_3 \lambda_4} \, .
\ea
When considering $\Tc^+_{\tau_1 \tau_2 \tau_3 \tau_4}$ all kinematic singularities are removed by construction, and so the remaining discontinuity along the right hand cut arises from the physical partial wave amplitude  $\bar T^J_{\lambda_1 \lambda_2 \lambda_3 \lambda_4}(s)$. Consequently we can take the absorptive part\footnote{Where by ${\text{Abs}}_s \bar T^J_{\lambda_1 \lambda_2 \lambda_3 \lambda_4}(s)$ we mean $ {\text{Abs}}_s \bar T^J_{\lambda_1 \lambda_2 \lambda_3 \lambda_4}(s)=4\pi (2J+1) \sqrt{\frac{s}{p_i p_f}} {\text{Abs}}_s T^J_{\lambda_1 \lambda_2 \lambda_3 \lambda_4} $, i.e. the discontinuity comes from the physical part.}
\ba
{\text{Abs}}_s \Tc^+_{\tau_1 \tau_2 \tau_1 \tau_2} =
  \big(\sqrt{-su}\big)^{\xi} {\cal S}^{S_1+S_2} \hspace{-10pt} \sum_{J \lambda_1 \lambda_2 \lambda_3 \lambda_4} \!\!\!\!\! u^{S_1}_{\lambda_1 \tau_1} u^{S_2}_{\lambda_2 \tau_2} u^{S_1*}_{\tau_1 \lambda_3} u^{S_2*}_{\tau_2 \lambda_4}  \left( d^J_{\mu \lambda}(\theta)+ d^J_{\mu \lambda}(-\theta)\right)  {\text{Abs}}_s \bar T^J_{\lambda_1 \lambda_2 \lambda_3 \lambda_4}(s) \,.
\nn
\ea
Using the Fourier series of the Wigner matrix \eqref{eqn:dJfourier}, we can write $d^J_{\mu \lambda}(\theta)+ d^J_{\mu \lambda}(-\theta)$ as a sum over $\cos (\nu \theta)$ with real coefficients
\begin{equation}
 d^J_{\mu \lambda}(\theta)+ d^J_{\mu \lambda}(-\theta) =2 e^{i \frac{\pi}{2} (\lambda-\mu)} \sum^J_{\nu=-J} d_{\lambda \nu}^J \left( \frac{\pi}{2} \right) d_{\mu \nu}^J \left( \frac{\pi}{2} \right) \cos \left( \nu \theta \right)   \, .
\end{equation}
Substituting in the discontinuity we find
\ba
\label{TFintro}
 {\text{Abs}}_s  \Tc^{+}_{\tau_1 \tau_2 \tau_1 \tau_2} (s, \theta)  = 2\big(\sqrt{-su}\big)^{\xi} {\cal S}^{S_1+S_2}  \sum_{J, \nu}   \cos \left( \nu \theta \right) F^{J\nu}_{\tau_1 \tau_2}(s)   \, ,
\ea
where \be
\label{FJnudef}
F^{J\nu}_{\tau_1 \tau_2}(s)=\sum_{\lambda_1 \lambda_2 \lambda_3 \lambda_4}  C_{\lambda_1 \lambda_2}^{\nu*} \ {\text{Abs}}_s  \bar T_{\lambda_1 \lambda_2 \lambda_3 \lambda_4}^J \ C^{\nu}_{\lambda_3 \lambda_4}  \, ,
\ee
and
\be
 C^{\nu}_{\lambda_3 \lambda_4} =  \, u^{S_1 *}_{\tau_1 \lambda_3} u^{S_2 *}_{\tau_2 \lambda_4} e^{- i \frac{\pi}{2} \mu } d_{\mu \nu}^J \left( \frac{\pi}{2} \right)  \, .
\label{eqn:C_positivity}
\ee
By unitarity, we have established in \eqref{uniT} that $ \bar T_{\lambda_1 \lambda_2 \lambda_3 \lambda_4}^J$ is a positive definite Hermitian matrix\footnote{Strictly speaking unitarity only implies that $ \bar T$ is positive semi-definite. However following Ref.~\cite{deRham:2017avq} we may use analyticity to guarantee that it is positive definite.}, and it therefore directly follows that
\be
\text{Unitarity} \;\;  \implies \;\; F^{J\nu}_{\tau_1 \tau_2}(s)   \ge 0 \, .
\ee

To proceed further we distinguish between the case of BB or FF scattering $\xi=0$, and BF scattering $\xi=1$.

\paragraph{BB or FF scattering:} For $\xi=0$ the intermediate partial waves have integer angular momenta $J$ and similarly $\nu$ is an integer. In the forward scattering limit $t=\theta=0$ we have the familiar optical theorem
\ba
&& {\text{Abs}}_s  \Tc^{+}_{\tau_1 \tau_2 \tau_1 \tau_2} (s,t=0,u) =  2  {\cal S}^{S_1+S_2}  \sum_{J=0}^{\infty} \sum^J_{\nu=-J}     F^{J\nu}_{\tau_1 \tau_2}(s)  > 0 \, ,  \, \quad \forall  \, s \ge 4m^2 \, .
\ea
On the right hand cut, this property can be extended to any numbers of $t$ derivatives by making use of the properties of the Chebyshev polynomials for  integer $\nu$. Indeed, the Chebyshev polynomials satisfy
\begin{equation}
\label{chebydef}
N_{n,\nu} =  \left. \frac{\ud^n \cos ( \nu \theta )}{\ud \cos^n \theta} \right|_{\theta=0}  =  \prod_{k=0}^{n-1} \frac{\nu^2-k^2}{2k+1} \geq 0 ,
\end{equation}
which is strictly positive for $n\leq \nu$ and vanishes for all $n>\nu$ for integer $\nu$.
We therefore have
\bal
 \frac{\partial^n}{ \partial \cos^n \theta} {\text{Abs}}_s  \Tc^{+}_{\tau_1 \tau_2 \tau_1 \tau_2} (s, \theta)   \bigg|_{\theta=0}
=2  {\cal S}^{S_1+S_2}  \sum_{J, \nu}  ~   N_{n,\nu} F^{J\nu}_{\tau_1 \tau_2}(s) >0 \, ,
\eal
or equivalently,
\begin{equation}
\frac{\partial^n}{\partial t^n} {\text{Abs}}_s  \Tc^{+}_{\tau_1 \tau_2 \tau_1 \tau_2} (s,t,u)  \bigg|_{t=0} > 0  ~~~{\rm for~~}  s \ge  4m^2 \, .
\end{equation}
This means that, since the transversity amplitudes are analytic
in $t$ \cite{Mahoux:1969um}, we can analytically continue the optical theorem for forward scattering ${\text{Abs}}_s \, \Tc_{\tau} (s, 0 ) > 0 $ to the finite positive $t$ case
\begin{equation}
{\text{Abs}}_s  \Tc^{+}_{\tau_1 \tau_2 \tau_1 \tau_2}  (s, t, u ) > 0 , \;\;\;\; 0 \leq t < m^2   \, , \;\;\;\; s \geq 4m^2 \, ,
\label{eqn:positivity_T}
\end{equation}
where we remind the reader that the absorptive part is related to the discontinuity across the real axis by ${\text{Abs}}_s  = \frac1{2i}{\text{Disc}}$.
Note that the analytic continuation cannot take us past the first pole in $t$. We have taken the first pole to be at $t=m^2$, which occurs in generic cases. In special cases the  one may be able to extend the range in $t$ even further, for instance for purely scalar scatterings, one can go as far as $t=4m^2$  \,  \cite{deRham:2017avq,deRham:2017imi}.

\paragraph{BF scattering:}
For $\xi=1$ we have to include an extra factor of $\sqrt{-su} =\sqrt{{\cal S}} \cos(\theta /2)$ in the definition of $  \Tc^+_{\tau_1 \tau_2 \tau_1 \tau_2}$. The discontinuity is then
\ba
{\text{Abs}}_s  \Tc^{+}_{\tau_1 \tau_2 \tau_1 \tau_2} (s, \theta)
&=& 2 {\cal S}^{S_1+S_2+1/2}    \sum_{J=1/2}^{\infty} \sum^J_{\nu=-J}  \cos(\theta /2)  \cos \left( \nu \theta \right) F^{J\nu}_{\tau_1 \tau_2}(s) \nn   \\
&=&  {\cal S}^{S_1+S_2+1/2} \sum_{J, \nu}    \left(  \cos \left( (\nu+1/2) \theta \right) + \cos \left( (\nu-1/2) \theta \right)\right)  F^{J\nu}_{\tau_1 \tau_2}(s) \, .\quad
\ea
It is then straightforward to see that $ {\text{Abs}}_s \Tc^{+}_{\tau_1 \tau_2 \tau_1 \tau_2}   \big|_{\theta=0} >0$ and since $\nu\pm 1/2$ is now integer,
\ba
 && \frac{\partial^n}{ \partial \cos^n \theta} {\text{Abs}}_s  \Tc^{+}_{\tau_1 \tau_2 \tau_1 \tau_2} (s, \theta)   \bigg|_{\theta=0} = \mathcal{S}^{S_1+S_2+1/2}  \sum_{J, \nu}  \left( N_{n,\nu+{1}/{2}} + N_{n,\nu-{1}/{2}} \right)   F^{J\nu}_{\tau_1 \tau_2}(s) >0 \, . \nn
\ea
Once again this implies
\be
\frac{\partial^n}{\partial t^n} {\text{Abs}}_s  \Tc^{+}_{\tau_1 \tau_2 \tau_1 \tau_2}   (s, t, u )  > 0 \, \quad \text{for }\quad 0 \leq t < m^2 \,, \;\;\;\; s \geq 4m^2 \, ,  \quad {\rm and }\ \forall \ n\ge 0 \, .
\ee

Following an analogous procedure for $\Tc^{-}_{\tau_1 \tau_2 \tau_1 \tau_2} (s, \theta)$ it is straightforward to show that
\ba
{\text{Abs}}_s  \Tc^{-}_{\tau_1 \tau_2 \tau_1 \tau_2} (s, \theta)
%
 = -\frac{1}{\sqrt{s}}   \big(\sqrt{-su}\big)^{\xi} {\cal S}^{S_1+S_2+1}     \sum_{J,  \nu}    \sin \theta \sin \left( \nu \theta \right)  F^{J\nu}_{\tau_1 \tau_2}(s)  \, .\quad
\ea
Even though we are still dealing with a function of $\theta$, we now have the difference of two cosines to deal with, $2 \sin \theta \sin \left( \nu \theta \right) =  \cos \left( (\nu+1) \theta \right) - \cos \left( (\nu-1) \theta \right)$ and so we cannot infer the same positivity properties of either the discontinuity or its derivatives. Nevertheless, the above expression will be important in determining the discontinuity along the left hand cut.

\subsection{Crossing and the Left Hand Cut}
\label{sec:lhc}

Following the standard S-matrix paradigm, $\Tc^{s+}_{\tau_1 \tau_2 \tau_1 \tau_2}(s,t) $ is analytically continued to the whole complex Mandelstam plane in such a way as to ensure full crossing symmetry is respected. Since our positivity bounds will arise from fixed $t$ dispersion relations, it is sufficient to consider the properties of the transversity amplitudes in the complex $s$ plane, for fixed real $t$, accounting for $s \leftrightarrow u$ crossing symmetry. Once the kinematical singularities have been removed, the remaining physical singularities are the poles associated with physical particles, in this case a single pole at $s=m^2$ and the right hand (RH) branch cut $s\ge 4m^2$ associated with multi-particle states. $s \leftrightarrow u$ crossing symmetry requires that there is a second pole at $u=m^2$ as well as a  second left-hand (LH) branch cut for $u =\mu \ge 4m^2$ which corresponds to $s=4m^2-t-\mu$. In the vicinity of the LH cut the scattering amplitude can be determined by crossing symmetry
\ba
\Tc^{s+}_{\tau_1 \tau_2 \tau_1 \tau_2}(s,t, u)
&=&  (\sqrt{-su})^{\xi} {\cal S}^{S_1+S_2} \left( \Tc^s_{\tau_1 \tau_2 \tau_1 \tau_2}(s,t,u)+\Tc_{-\tau_1 -\tau_2 -\tau_1 -\tau_2}^s (s,t,u) \right)  \nn \\
&=&   (\sqrt{-su})^{\xi} {\cal S}^{S_1+S_2} \left(e^{-i \chi_u \sum_i \tau_i } \Tc^u_{-\tau_1 -\tau_2 -\tau_1 -\tau_2}(u,t,s)+e^{+i \chi_u \sum_i \tau_i } \Tc^u_{\tau_1 \tau_2 \tau_1\tau_2}(u,t,s) \right)   \nn \\
&=&  c_+ \Tc^{u+}_{\tau_1 \tau_2 \tau_1 \tau_2}(u,t,s) +c_-  \Tc^{u-}_{\tau_1 \tau_2 \tau_1 \tau_2}(u,t,s) \, ,
\label{eq:Tsp}
\ea
where
\ba
\label{eq:crossingfactors1}
&& c_+   =  \frac{{\cal S}^{S_1+S_2}}{{\cal U}^{S_1+S_2}} \cos \big( \chi_u \sum_i \tau_i  \big) \, ,  \\
\label{eq:crossingfactors2}
&& c_-  =   -  \frac{{\cal S}^{S_1+S_2}}{{\cal U}^{S_1+S_2}}  \frac{1}{\sqrt{stu}} \sin \big( \chi_u \sum_i \tau_i  \big)   ,
\ea
and where ${\cal S}$ and ${\cal U}$ are defined in \eqref{eq:CalS}.
Note that $\sin \big( \chi_u \sum_i \tau_i  \big)/\sqrt{stu}$ is even under $\theta_s \rightarrow - \theta_s$ and hence contains no branch cut.  This is as it should be since $\Tc^{s+}_{\tau_1 \tau_2 \tau_1 \tau_2}(s,t, u)$ is an even function of $\theta_s$ by construction. \\

Now, defining the $u$-channel scattering angle $\theta_u$ via
\be
t=-\frac12 (u-4m^2)(1-\cos \theta_u),
\ee
then $\Tc^{u\pm}_{\tau_1 \tau_2 \tau_1 \tau_2}(u,t,s)$
has the same analyticity properties in terms of $u$ and $\theta_u$ as $\Tc^{ s \pm}_{\tau_1 \tau_2 \tau_1 \tau_2}(s,t,u)$ has in terms  of $s$ and $\theta_s$.
Similarly as in \eqref{eq:PWTs}, $\Tc^{u\pm}_{\tau_1 \tau_2 \tau_1 \tau_2}(s,t,u)$ also has a partial wave expansion in terms of partial wave amplitudes $\tilde T_{\lambda_1 \lambda_2 \lambda_3 \lambda_4}^{u,J} $ describing the fixed total angular momentum $J$ scattering process $A + \bar D  \rightarrow C + \bar B$.
Since all kinematical singularities have been removed, following the same argument as before, the remaining discontinuities can only arise  from the partial wave scattering amplitudes themselves and so across the LH cut,
\ba
 {\text{Abs}}_{u}  \Tc^{s+}_{\tau_1 \tau_2 \tau_1 \tau_2}(s,t,u) &=& \frac{1}{2i} \left(\Tc^{s+}_{\tau_1 \tau_2 \tau_1 \tau_2}(s-i \epsilon,t)-\Tc^{s+}_{\tau_1 \tau_2 \tau_1 \tau_2}(s+i \epsilon,t) \right)  ~~\quad s\le -t\nn \\
&=& c_+\,   {\text{Abs}}_{u} \Tc^{u+}_{\tau_1 \tau_2 \tau_1 \tau_2} (u,t,s) +c_- \, {\text{Abs}}_{u} \Tc^{u-}_{\tau_1 \tau_2 \tau_1 \tau_2}(u,t,s) ~~\quad u \ge 4m^2\,. \nn
\ea
Here we have defined the $u$-channel absorptive part ${\text{Abs}}_{u}$ as the discontinuity of the $s$-channel amplitude across the LH cut, which by crossing symmetry is related to the RH cut discontinuity of the $u$-channel amplitude ${\text{Abs}}_{u} \Tc^{u+}_{\tau_1 \tau_2 \tau_1 \tau_2}(u,t)= (\Tc^{u+}_{\tau_1 \tau_2 \tau_1 \tau_2}(u+i \epsilon,t)- \Tc^{u+}_{\tau_1 \tau_2 \tau_1 \tau_2}(u-i \epsilon,t))/(2i)$. Concretely,
\ba
{\text{Abs}}_{u}  \Tc^{s+}_{\tau_1 \tau_2 \tau_1 \tau_2}(s,t)
& = & 2 (\sqrt{-su})^{\xi} {\cal U}^{S_1+S_2} \sum_{J,  \nu}    \left[   c_+    \cos \left( \nu \theta_u \right)  - c_-  \frac{\mathcal{U} }{2\sqrt{u}}  \sin \theta_u  \sin \left( \nu \theta_u \right) \right]  F^{u,J\nu}_{\tau_1 \tau_2}(u) \nn \\
& = & 2  ( \sqrt{-su})^{\xi} {\cal S}^{S_1+S_2}   \sum_{J,  \nu}    \cos \left( \nu \theta_u -\chi_u \sum_i \tau_i  \right)  F^{u,J\nu}_{\tau_1 \tau_2}(u) \, ,
\ea
where $F^{u,J\nu}_{\tau_1 \tau_2}(u)>0$ is the $u$ channel equivalent of $F^{J\nu}_{\tau_1 \tau_2}(s)$ defined in Eq.~\eqref{FJnudef}.

\paragraph{BB or FF scattering:} To proceed, we first consider $\xi=0$. Let us consider the combination
\ba
\label{eq:comb1}
\sqrt{{\cal S}}  e^{ \pm i \chi_u } = \frac{1}{2}\sqrt{{\cal U}} + \frac{(\sqrt{u} \pm 2m)\sqrt{{\cal U}}}{4 \sqrt{u}}e^{i \theta_u }+\frac{(\sqrt{u} \mp 2m)\sqrt{{\cal U}}}{4 \sqrt{u}}e^{-i \theta_u } \, .
\ea
It is clear that this is a sum of  positive functions for $u>4m^2$ times $e^{i p \theta_u}$ and the same can be said for any positive integer power of this quantity\footnote{That is $\sqrt{{\cal S}}  e^{ \pm i \chi_u } =\sum_p c^{\pm}_p(u) e^{ip \theta_u}$, with $c_p(u)>0$ for $u>4m^2$.}. Furthermore
\be
{\cal S}=s(s-4m^2)=(u-4m^2) (1+\cos \theta_u) (u+4m^2+(u-4m^2) \cos\theta_u)/4,
\ee
is similarly a sum of  positive functions for $u>4m^2$ times $e^{i p \theta_u}$. Since $ \Tc^{s+}_{\tau}= \Tc^{s+}_{-\tau}$ we can without loss of generality focus on $\tau_1+\tau_2 \ge 0$. From this we conclude that
\ba
\hspace{-10pt}
2{\cal S}^{S_1+S_2} \cos \left( \nu \theta_u -\chi_u \sum_i \tau_i  \right) &=&{\cal S}^{S_1+S_2-\tau_1-\tau_2} {\cal S}^{\tau_1+\tau_2}   \left( e^{i\nu \theta_u} e^{-i \chi_u \sum_i \tau_i  }+  e^{-i\nu \theta_u} e^{i \chi_u \sum_i \tau_i  } \right)  \nn  \\
&=& {\cal S}^{S_1+S_2-\tau_1-\tau_2}\left( e^{i\nu \theta_u} \left(\sqrt{{\cal S}}  e^{- i \chi_u }\right)^{2 (\tau_1+ \tau_2)}+  e^{-i\nu \theta_u} \left(\sqrt{{\cal S}}  e^{ i \chi_u }\right)^{2( \tau_1+ \tau_2)} \right)  \nn  \\
&=&\sum_{p=-2(\tau_1+\tau_2)-\nu}^{2(\tau_1+\tau_2)+\nu}C_{\nu,p}(u) e^{i p \theta_u}  \, ,
\label{eq:cosdiff}
\ea
where $C_{\nu,p}(u) > 0$ for $u> 4m^2$ and $C_{\nu,-p}(u)= C_{\nu,p}(u)$. Using the latter property, then finally we have
\be
{\text{Abs}}_{u}  \Tc^{s+}_{\tau_1 \tau_2 \tau_1 \tau_2}(s,t,u)= 2 \sum_{J=0}^{\infty} \sum^J_{\nu=-J} \sum_{p=0}^{2(\tau_1+\tau_2)+\nu}   C_{\nu,p}(u)  \cos(p\theta_u)   F^{u,J\nu}_{\tau_1 \tau_2}(u)  \, .
\ee
It is then straightforward to see that
\ba
&& {\text{Abs}}_{u}  \Tc^{s+}_{\tau_1 \tau_2 \tau_1 \tau_2}(s,0) >0 \, , \quad u \ge 4m^2 \, , \\
&& \frac{\partial^n}{ \partial \cos^n \theta_u} {\text{Abs}}_{u}  \Tc^{s+}_{\tau_1 \tau_2 \tau_1 \tau_2}(s,t) \bigg|_{\theta_u=0} >0  \, , \quad u \ge 4m^2 \, , \quad  \forall \; n \ge 0 \, , \\
&& \frac{\partial^n}{\partial t^n}\Big|_u {\text{Abs}}_{u}  \Tc^{s+}_{\tau_1 \tau_2 \tau_1 \tau_2}(s,t) \bigg|_{t=0} >0  \, , \quad u \ge 4m^2 \, \quad \forall \; n \ge 0 \, ,
\ea
which given the analyticity of the amplitude in $t$ may in turn be extended to
\be
 \frac{\partial^n}{\partial t^n} \Big|_u \;  {\text{Abs}}_{u}  \Tc^{s+}_{\tau_1 \tau_2 \tau_1 \tau_2}(s,t,u)   > 0  \, , \quad u > 4m^2 \, \quad 0 \le t < m^2 \, , \quad  \forall \ n \ge 0 \, .
\ee

\paragraph{BF scattering:}
In the case $\xi=1$, we include the additional factor $\sqrt{-su}=  \sqrt{{\cal U}}  \cos( \theta_u/2)$, and in this case $\nu$ and $J$ are half integers, so
\ba
 {\text{Abs}}_{u}  \Tc^{s+}_{\tau_1 \tau_2 \tau_1 \tau_2}(s,t,u) & = &
 2 \sqrt{ {\cal U} } {\cal S}^{S_1+S_2}      \sum_{J=1/2}^{\infty} \sum^J_{\nu=-J}    \cos\left( \frac{\theta_u}{2} \right) \cos \left( \nu \theta_u -2\chi_u (\tau_1+\tau_2)  \right)  F^{u,J\nu}_{\tau_1 \tau_2}(u)   \nn\\
&   = &  \sqrt{{\cal U}} {\cal S}^{S_1+S_2} \sum_{J, \nu}   \bigg[   \cos \left( (\nu+\tfrac{1}{2} ) \theta_u -2 \chi_u (\tau_1+\tau_2)  \right)     \nn \\
&&\phantom{\sqrt{{\cal U}} {\cal S}^{S_1+S_2} \sum_{J, \nu}}
+  \cos \left( (\nu- \tfrac{1}{2} ) \theta_u -2\chi_u (\tau_1+\tau_2)  \right)  \bigg]  F^{u,J\nu}_{\tau_1 \tau_2}(u)   \, .
\ea
Following the same arguments as previously we can express
\be
{\cal U}^{1/2}{\cal S}^{\tau_1+\tau_2}  \cos \left( (\nu\pm1/2) \theta_u -2 \chi_u (\tau_1+\tau_2)  \right) = \sum_p d^{\pm}_{\nu,p}(u) e^{ip \theta_u}\,,
\ee
where $d^{\pm}_{\nu,p}(u)  \ge 0$ for $u\ge 4m^2$ and $d^{\pm}_{\nu,p}(u) = d^{\pm}_{\nu,-p}(u)$ and since we have already shown this same property holds for ${\cal S}^{S_1+S_2-(\tau_1+\tau_2)}$ then
\be
{\text{Abs}}_{u}  \Tc^{s+}_{\tau_1 \tau_2 \tau_1 \tau_2}(s,t,u)  = \sum_{J,\nu} \sum_{p=0} D_{\nu,p}(u) \cos(p\theta_u)  F^{u,J\nu}_{\tau_1 \tau_2}(u)  \, ,
\ee
where $D_{\nu,p}(u) \ge 0$ for $ u \ge 4m^2$. Once again we infer that
\be
 \frac{\partial^n}{\partial t^n} \Big|_u {\text{Abs}}_{u}   \Tc^{s+}_{\tau_1 \tau_2 \tau_1 \tau_2}   (s,t,u) >0  \, , \quad {\rm for}\quad u \ge 4m^2 \, \quad 0 \le t < m^2 \, , \quad  \forall  \; n \ge 0 \, .
\ee

\subsection{Dispersion Relation}
\label{sec:disperse}

Following the famous result of Ref.~\cite{Martin:1965jj}, if the scattering amplitude satisfies a dispersion relation\footnote{This follows from the properties of the retarded Green's functions in a  QFT  and we shall assume goes through for arbitrary spins.} for $t_0 < t \le 0$, has an absorptive part which is analytic inside the Lehmann ellipse, is analytic in some small neighbourhood of $s$ and $t$ and if it satisfies positivity of all the derivatives of its absorptive parts along the LH and RH cuts, then the analyticity region in $t$ may be extended to $|t| < R$ for finite $R$. In the case of a single mass we can push to $R=4m^2$. Assuming polynomial boundedness and analyticity in Martin's extended region, it will then follow from the partial wave expansion that the scattering amplitude will satisfy a $t$ dependent extension of the Froissart bound \cite{Jin:1964zza}
\be
| \Tc_{\tau_1 \tau_2 \tau_1 \tau_2}(s,t)|_{|s| \rightarrow \infty} < |s|^{1+\epsilon(t)} \,  \quad \Rightarrow \quad  | \Tc^+_{\tau_1 \tau_2 \tau_1 \tau_2}(s,t)|_{|s| \rightarrow \infty} < |s|^{1+\epsilon(t)+N_S-2}  \, ,
\ee
where
\be
\label{NSnum}
N_S=2+2(S_1+S_2)+\xi    ,
\ee
the additional powers arising from the $(\sqrt{-su})^{\xi} {\cal S}^{S_1+S_2}$ prefactor. In the previous sections we have shown that the appropriately regularized transversity amplitudes $\Tc^+_{\tau_1 \tau_2 \tau_1 \tau_2}(s,t)$ satisfy positivity of all derivatives along the LH and RH cuts, and so we may conclude that they satisfy identically the same analyticity properties as the scalar amplitudes \cite{Mahoux:1969um}. Furthermore, following the arguments of Jin and Martin of Ref.~\cite{Jin:1964zza} when these same conditions hold, the scattering amplitude at fixed $t$ ($0 \le  t <R$) satisfies a dispersion relation with the same number of subtractions as for $t=0$ which in turn implies $\epsilon(t)<1$ and so
\be
| \Tc^+_{\tau_1 \tau_2 \tau_1 \tau_2}(s,t)|_{|s| \rightarrow \infty} < |s|^{N_S} \, .
\ee
We thus conclude that the regularized amplitudes $\Tc^+_{\tau_1 \tau_2 \tau_1 \tau_2}(s,t)$ satisfy a dispersion relation in $s$ for $0 \le t<4m^2$ with $N_S$ subtractions. This condition is of course implicit in the usual assumption of `maximal analyticity', however we have seen that the result is much stronger since we have proven the relevant positivity criteria which are the crucial ingredient in Martin's extensions of the Lehmann ellipse, the Froissart bound and the extension of the regime of the dispersion relation. \\

The transversity amplitude $ \Tc^+_{\tau_1 \tau_2 \tau_1 \tau_2}(s,t)$ contains a simple $s$-channel pole at $s=m^2$ and $u$-channel pole at $s=3m^2-t$ (or $u=m^2$), which appear already at tree level (although their residues are affected by loops). It proves useful to define an associated `pole-subtracted' transversity amplitude $\tilde\Tc_\tau (s,t)$ with these two poles removed. That is, we consider
\be
 \tilde\Tc^+_{\tau_1 \tau_2 \tau_1 \tau_2}(s,t)
= \Tc^+_{\tau_1 \tau_2 \tau_1 \tau_2}(s,t) -  \frac{\text{Res}\Tc^+_{\tau_1 \tau_2 \tau_1 \tau_2}(s=m^2,t)}{s-m^2} - \frac{\text{Res}\Tc^+_{\tau_1 \tau_2 \tau_1 \tau_2} (s=3m^2-t,t)}{s+t-3m^2}\,,
\ee
where $\text{Res}$ denotes the residue. In the scalar case it proved convenient to also subtract the $t$-channel pole \cite{deRham:2017avq,deRham:2017imi} although this subtraction  did not actually play a significant role. For general spins, such a subtraction would not be convenient since the residue of the $t$-channel pole is itself a function of $s$, and subtracting it can modify the behaviour of the amplitude\footnote{The concern is that the tree-level or finite loop residue may already violate the Froissart bound, and so subtracting it modifies the analyticity arguments which rely on the assumption of the Froissart bound in determining the overall number of subtractions.} at large $s$. \\

\begin{figure}
\centering
\begin{tikzpicture}

\draw [thick, gray] (-4.1,0) -- (4.1,0);
\draw [thick, gray] (0,-4.4) -- (0,4.4);
\draw [ultra thick, red] (2,0) -- (4.1,0);
\draw [ultra thick, red] (-0.5,0) -- (-4.1,0);

\node at (-2.8,3.5) {\scriptsize $C'$};
\draw [thick, ->] (4.01,-0.3) -- (3,-0.3); \draw [thick] (3.1,-0.3) -- (2,-0.3);
\draw [thick] (2,-0.3) arc (270:90:0.3);
\draw [thick, ->] (2,0.3) -- (3,0.3); \draw [thick] (2.9,0.3) -- (4.01,0.3);
\draw [thick, ->] (4,0.3) arc (0:130:4);
\draw [thick] (0,4.3) arc (90:180:4);
\draw [thick, ->] (-4.01,0.3) -- (-2,0.3); \draw [thick] (-2.1,0.3) -- (-0.5,0.3);
\draw [thick] (-0.5,0.3) arc (90:-90:0.3);
\draw [thick, ->] (-0.5,-0.3) -- (-2,-0.3); \draw [thick] (-1.9,-0.3) -- (-4.01,-0.3);
\draw [thick, ->] (-4,-0.3) arc (180:310:4);
\draw [thick] (0,-4.3) arc (270:360:4);

\filldraw [red] (2,0) circle (1.5pt);
\node at (2.2,-0.15) {\tiny $4m^2$};

\filldraw [red] (-0.5,0) circle (1.5pt);
\node at (-0.55,-0.175) {\tiny $-t$};

\node at (0.4,-0.2) {\tiny $m^2$};
\filldraw [red] (0.5,0) circle (1.5pt);

\node at (1.05,0.2) {\tiny $3m^2\!\!-\!t$};
\filldraw [red] (1,0) circle (1.5pt);

\node at (1.2,0.9) {\scriptsize $C$};
\draw [thick] (1.5,0) arc (0:90:0.75);
\draw [thick, ->] (1.5,0) arc (0:60:0.75);
\draw [thick] (0.75,0.75) arc (90:360:0.75);

\node at (3.7,3.7) {\small $s$};
\draw [thick]  (3.5,3.9)  -- (3.5,3.5) --  (3.9,3.5);

\end{tikzpicture}
\caption{The scattering amplitude can be analytically continued to the entire complex $s$ plane, with the poles at $s=m^2$ and $3m^2-t$ and branch cuts along the real axis from $-t$ to $-\infty$ and from $4m^2$ to $\infty$. \label{fig:1} }
\end{figure}
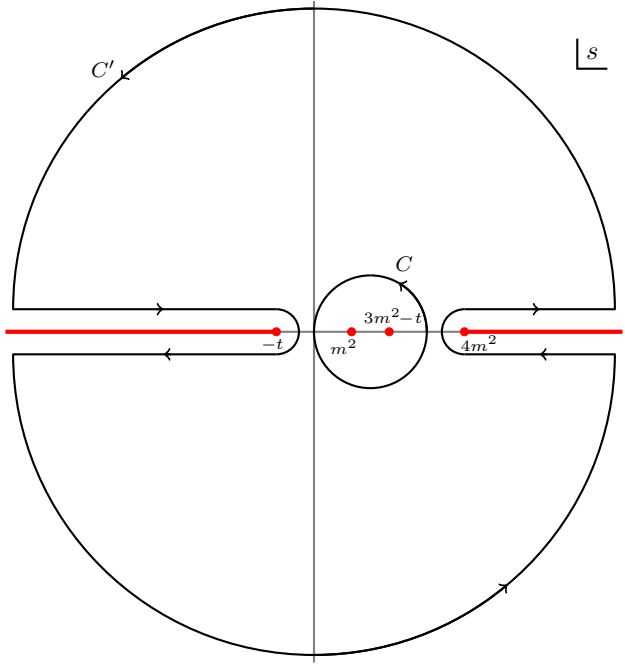

Consider a contour $C$ for $\tilde\Tc^+_\tau (s,t)$ in the complex $s$ plane, which encircles the poles at $s'=m^2$ and $s'=3m^2-t$ as well as a generic point $s$, as shown in Figure \ref{fig:1}. By Cauchy's integral formula, we have
\be
\tilde\Tc^+_{\tau_1 \tau_2 \tau_1 \tau_2}(s,t) = \frac{1}{2\pi i} \oint_C \ud s' \; \frac{ \tilde\Tc^+_{\tau_1 \tau_2 \tau_1 \tau_2}(s',t)}{(s'-s)}\,.
 \ee
We can deform this contour so that it runs around the branch cuts and closes with circular arcs at infinity (contour $C'$). We emphasize that even when we are considering higher spins, a Froissart bound still applies \cite{Mahoux:1969um} and $| \Tc^+_{\tau_1 \tau_2 \tau_1 \tau_2}(s,t)|_{|s| \rightarrow \infty} < |s|^{N_S}$. This allows us to neglect the arcs at infinity by performing a sufficient number of subtractions. We can then obtain the following dispersion relation:
\ba
\label{disrel1}
\tilde\Tc^+_{\tau_1 \tau_2 \tau_1 \tau_2}(s,t) =  \sum_{n=0}^{N_S-1} \! a_n(t) s^n
\!+\! \frac{s^{N_S}}{\pi}  \int_{4m^2}^\infty \ud \mu  \frac{ {\text{Abs}}_s \Tc^+_{\tau_1 \tau_2 \tau_1 \tau_2}(\mu,t) }{ \mu^{N_S} (\mu - s) }  \nn  \\
+\frac{u^{N_S}}{\pi}  \int_{4m^2}^\infty \ud \mu  \frac{  {\text{Abs}}_u \Tc^+_{\tau_1 \tau_2 \tau_1 \tau_2}(4m^2-t-\mu,t) }{ \mu^{N_S} ( \mu - u) } \, ,
\ea
where $N_S$ is given by Eq.~\eqref{NSnum}. \\

The subtraction functions $a_n(t)$ in the dispersion relation are undetermined by analyticity and depend on the detailed information of the particular theory involved. To eliminate them, we simply take $N_s$ derivatives and  consider the quantity
\ba
\label{faltdef}
 \hspace{-20pt} f_{\tau_1 \tau_2 } (s, t) & =&  \frac{1}{N_S!}\frac{\ud^{N_S}}{\,\,\ud s^{N_S}} \tilde\Tc^+_{\tau_1 \tau_2 \tau_1 \tau_2}(s,t) \, , \\
   &=&  \frac{1}{2\pi i} \oint_C \ud s' \; \frac{ \tilde\Tc^+_{\tau_1 \tau_2 \tau_1 \tau_2}(s',t) }{(s'-s)^{N_S+1 } }  \,,\\
 &=&  \frac{1}{\pi}  \int_{4m^2}^\infty \ud \mu  \frac{ {\text{Abs}}_s \Tc^+_{\tau_1 \tau_2 \tau_1 \tau_2}(\mu,t) }{  (\mu - s)^{N_S+1 } }+    \frac{1}{\pi}  \int_{4m^2}^\infty \ud \mu  \frac{  {\text{Abs}}_u \Tc^+_{\tau_1 \tau_2 \tau_1 \tau_2}(4m^2-t-\mu,t) }{ ( \mu - u)^{N_S+1} } \, . 
 \ea
Since we have already established that the absorptive parts are positive on both the RH and LH cuts in section \ref{sec:rhc} and \ref{sec:lhc}, then our first positivity bounds is the simple statement that
\be
\label{eq:leadingbound}
  f_{\tau_1 \tau_2 } (s, t)> 0 \, , \quad -t < s < 4m^2, \quad 0 \le t < m^2 \, ,
\ee
where the range of real $s$ is determined by the requirement that $\mu-s >0$, $\mu-u>0$ given $\mu \ge 4m^2$. This is the direct generalization, for general spins, of the bound given in \cite{Adams:2006sv} applied in a larger range of $s$ and $t$.

%
%
%
%
%
%
%
%
%
%

\subsection{Positivity Bounds for Particles with Spin}

In a recent work \cite{deRham:2017avq}, we extended the simple positivity bounds of the form \eqref{eq:leadingbound} for scalar theories, to an infinite number of bounds on the $t$ and $s$ derivatives of the scattering amplitude. The interpretation of these bounds depends somewhat on the context. In the case of an assumed weakly coupled UV completion, these infinite number bounds may be directly applied to the tree level scattering amplitudes of the EFT and put constraints on the coefficients in the effective Lagrangian. If we do not assume a weakly coupled UV completion, then the bounds may be applied to the full quantum scattering amplitudes and the contribution of the light loops may be further subtracted off to strengthen the bounds, as discussed in the example of the massive Galileon \cite{deRham:2017imi}. \\

To begin with we will derive the bounds on the exact all-loop scattering amplitude. The procedure is identical to that discussed in  \cite{deRham:2017avq} which we refer to for more details. The only distinction in the case of particles with spin is that we must use the regularized pole subtracted amplitudes, $\tilde\Tc^+_{\tau_1 \tau_2 \tau_1 \tau_2}(s,t)$, we have a larger number of subtractions $N_S \ge 2$, and the contributions from the LH and RH cut are not identical. Before getting to the general case, we can get a feel for how the bounds work by considering the first $t$ derivative of \eqref{faltdef}. Defining new variables $s=2m^2-t/2+v$, so that
\be
\label{faltdef2}
 f_{\tau_1 \tau_2 } (v, t)
 =  \frac{1}{\pi}  \int_{4m^2}^\infty \ud \mu  \frac{ {\text{Abs}}_s \Tc^+_{\tau_1 \tau_2 \tau_1 \tau_2}(\mu,t) }{  (\mu-2m^2+t/2 - v)^{N_S+1 } }+    \frac{1}{\pi}  \int_{4m^2}^\infty \ud \mu  \frac{  {\text{Abs}}_u \Tc^+_{\tau_1 \tau_2 \tau_1 \tau_2}(4m^2-t-\mu,t) }{ ( \mu -2m^2+t/2+v)^{N_S+1} } \, ,
 \ee
then differentiating with respect to $t$ gives
\ba
\frac{\partial}{\partial t}  f_{\tau_1 \tau_2 } (v, t) &=&-\frac{(N_S+1)}{2\pi}  \int_{4m^2}^\infty \ud \mu  \frac{ {\text{Abs}}_s \Tc^+_{\tau_1 \tau_2 \tau_1 \tau_2}(\mu,t) }{  (\mu -2m^2+t/2-v)^{N_S+2 } }
\\
&& -  \frac{(N_S+1)}{2\pi}  \int_{4m^2}^\infty \ud \mu  \frac{  {\text{Abs}}_u \Tc^+_{\tau_1 \tau_2 \tau_1 \tau_2}(4m^2-t-\mu,t) }{ ( \mu -2m^2+t/2+v)^{N_S+2} } \nn \\
&& + \frac{1}{\pi}  \int_{4m^2}^\infty \ud \mu  \frac{\partial_t   {\text{Abs}}_s \Tc^+_{\tau_1 \tau_2 \tau_1 \tau_2}(\mu,t) }{  (\mu-2m^2+t/2 - v)^{N_S+1 } }
\nn\\
&& +  \frac{1}{\pi}  \int_{4m^2}^\infty \ud \mu  \frac{\partial_t   {\text{Abs}}_u \Tc^+_{\tau_1 \tau_2 \tau_1 \tau_2}(4m^2-t-\mu,t)}{ ( \mu -2m^2+t/2+v)^{N_S+1} } \, .\nn
\ea
Defining
\ba
{\cal M}^2 = {\rm Min}_{ \mu \ge 4m^2}[\mu-2m^2+t/2]=2m^2+t/2\,,
\ea
 and using the integral inequality that for any positive definite function $\rho(\mu) > 0$
\be
\frac{1}{{\cal M}^2 }\int_{4m^2}^{\infty} \frac{\rho(\mu)}{(\mu-2m^2+t/2)^N}  \, \d \mu > \int_{4m^2}^{\infty} \frac{\rho(\mu)}{(\mu-2m^2+t/2)^{N+1}} \, \d \mu  \, ,
\ee
and evaluating at $v=0$ we then infer that,
\ba
 \frac{\partial}{\partial t}  f_{\tau_1 \tau_2 } (0, t) +\frac{N_S+1}{2 {\cal M}^2}  f_{\tau_1 \tau_2 } (0, t) &>& \frac{1}{\pi}  \int_{4m^2}^\infty \ud \mu  \frac{\partial_t   {\text{Abs}}_s \Tc^+_{\tau_1 \tau_2 \tau_1 \tau_2}(\mu,t) }{  (\mu-2m^2+t/2 )^{N_S+1 } }  \\
 &+&   \frac{1}{\pi}  \int_{4m^2}^\infty \ud \mu  \frac{\partial_t    {\text{Abs}}_u \Tc^+_{\tau_1 \tau_2 \tau_1 \tau_2}(4m^2-t-\mu,t)}{ ( \mu -2m^2+t/2)^{N_S+1} } > 0 \, .  \nn
\ea
Thus our second non-trivial bound is
\be
 \frac{\partial}{\partial t}  f_{\tau_1 \tau_2 } (0, t) +\frac{N_S+1}{2 {\cal M}^2}  f_{\tau_1 \tau_2 } (0, t) > 0 \, ,  \quad 0 \le t < m^2 \, .
 \ee
In practice, the above form of this bound is not so interesting since we have in mind ${\cal M}^2 \sim m^2$ and so this will be dominated by the second term.
Since $f_{\tau_1 \tau_2 } (0, t) $ is already positive from the lower bound, then there is little new content in this new bound. The situation is very different however if we imagine that the EFT has a weakly coupled UV completion. In this case, we expect the scattering amplitude already computed at tree level to satisfy all of the properties that we have utilized, specifically the Froissart bound. Given this, the above bound can be applied directly to the tree level scattering amplitudes. These amplitudes by definition do not include loops from the light fields and the branch cut will no longer be at $\mu=4m^2$ but rather at $\mu=\Lambda_{\rm th}^2$ where the threshold energy $\Lambda_{\rm th}$ is defined as the mass of the lightest state that lies outside of the regime of validity of the EFT. Assuming $\Lambda_{\rm th} \gg m$ then the bound on the tree level scattering amplitude becomes
 \be
 \frac{\partial}{\partial t}  f^{\rm tree}_{\tau_1 \tau_2 } (0, t) +\frac{N_S+1}{2 \Lambda_{\rm th}^2}  f^{\rm tree}_{\tau_1 \tau_2 } (0, t) > 0 \,,  \quad 0 \le t < m^2 \, .
 \ee
 In a typical EFT, both of these terms are of comparable order with the first potentially dominating, and so the bound becomes meaningfully independent of the existing requirement $f^{\rm tree}_{\tau_1 \tau_2 } (0, t)>0 $. \\

 Even if we do not have a weakly coupled UV completion, we can use our knowledge of the light loops in the regime in which perturbation theory is valid (e.g. $|k| \ll \varepsilon \Lambda_{\rm th}$ with $\varepsilon \ll 1$) to subtract off their contribution from the amplitudes, thus removing part of the branch cut. This is achieved by defining
 \ba
  f^{\varepsilon \Lambda_{\rm th}}_{\tau_1 \tau_2 } (v, t)  =  \frac{1}{N_S!}\frac{\ud^{N_S}\tilde\Tc^+_{\tau_1 \tau_2 \tau_1 \tau_2}(s,t)}{\,\,\ud s^{N_S}} & - & \frac{1}{\pi}  \int_{4m^2}^{\varepsilon^2 \Lambda_{\rm th}^2} \ud \mu  \frac{ {\text{Abs}}_s \Tc^+_{\tau_1 \tau_2 \tau_1 \tau_2}(\mu,t) }{  (\mu-2m^2+t/2 - v)^{N_S+1 } }  \\
   &-& \frac{1}{\pi}  \int_{4m^2}^{\varepsilon^2 \Lambda_{\rm th}^2} \ud \mu  \frac{ {\text{Abs}}_u \Tc^+_{\tau_1 \tau_2 \tau_1 \tau_2}(4m^2-t-\mu,t)}{ ( \mu -2m^2+t/2+v)^{N_S+1} }  \, , \nn
  \ea
  where the RHS is computed using the knowledge of the light loops in the EFT.
  Assuming again $\varepsilon \Lambda_{\rm th} \gg m$ we may then impose
  \be
   \frac{\partial}{\partial t}  f^{\varepsilon \Lambda_{\rm th}}_{\tau_1 \tau_2 } (0, t) +\frac{N_S+1}{2 \varepsilon^2 \Lambda_{\rm th}^2 }  f^{\varepsilon \Lambda_{\rm th}}_{\tau_1 \tau_2 } (0, t) > 0 \, ,  \quad 0 \le t < m^2 \, .
  \ee
  Provided the hierarchy $\varepsilon \Lambda_{\rm th} \gg m$ is sufficiently great, then this will impose a bound independent of the leading order one $f^{\varepsilon \Lambda_{\rm th}}_{\tau_1 \tau_2 } (0, t) > 0$.
  We see that in all three cases, the form of the bound is the same, and the only distinction is the choice of ${\cal M}^2$. \\

  \subsection{General Higher Order Positivity Bounds}

As discussed in the scalar case in  \cite{deRham:2017avq} we can generalize the previous procedure to put bounds on all $t$ derivatives and all even $v$ derivatives of the scattering amplitude, provided the subtraction functions do not enter. In the scalar case, the triviality of the crossing relation implied that all odd $v$ derivatives of the amplitude were zero.
Following the notation of \cite{deRham:2017avq} and defining
\be
\tilde B_{\tau_1 \tau_2}(v,t) = \tilde\Tc^+_{\tau_1 \tau_2 \tau_1 \tau_2}(s=2m^2-t/2+v,t)
\ee
and further defining
\be
B_{\tau_1 \tau_2}^{(2N,M)}(t) = \frac{1}{M!} \partial_v^{2N} \partial_t^M \tilde B_{\tau_1 \tau_2}(v,t)  \Big|_{v=0} \, ,
\ee
then provided $N \ge N_S/2$ so that the subtraction functions drop out, the $B_{\tau_1 \tau_2}^{(2N,M)}(t)$ can be given in terms of the positive definite integrals
\ba
I^{(q,p)}_{\tau_1 \tau_2}(t) = \frac{q!}{p!} \frac{1}{\pi} \int_{4m^2} ^{\infty} \d \mu \frac{\left[\partial_t^p \( {\text{Abs}}_s \Tc^+_{\tau_1 \tau_2 \tau_1 \tau_2}(\mu,t)\)+ \partial_t^p \({\text{Abs}}_u \Tc^+_{\tau_1 \tau_2 \tau_1 \tau_2}(4m^2-t-\mu,t) \)\right]}{(\mu+t/2 -2m^2)^{q+1}} >0\,,\nn
\ea
in the form
\be
B_{\tau_1 \tau_2}^{(2N,M)}(t) =\sum_{k=0}^M \frac{(-1)^k}{k!2^k} I_{\tau_1 \tau_2}^{(2N+k,M-k)} \, , \quad \forall \ \, N \ge N_S/2 \, , \, M>0 \, .
\ee
Then following \cite{deRham:2017avq} verbatim, the general positivity bounds are
\be
Y^{(2N,M)}_{\tau_1 \tau_2}(t)>0 \, , \quad \forall\ \, N \ge N_S/2 \, , \, M>0 \, , \quad 0 \le t < 4m^2 \, ,
\ee
where the $Y^{(2N,M)}_{\tau_1 \tau_2}$ are determined by the recursion relation
\ba
\label{eqn:Y}
\hspace{-20pt} Y_{\tau_1 \tau_2}^{(2N,M)}(t) &=& \sum_{r=0}^{M/2} c_r B_{\tau_1 \tau_2}^{(2N+2r,M-2r)}(t) \nn\\
&& ~~~~~~ +\frac{1}{{\cal M}^2} \sum_{ \text{even}\, k =0}^{(M-1)/2} (2N+2k+1)  \beta_k   Y_{\tau_1 \tau_2}^{(2N+2k ,M-2k-1)}(t)> 0\,,
\ea
with $Y_{\tau_1 \tau_2}^{(2N,0)}=B_{\tau_1 \tau_2}^{(2N,0)}$ and $c_r$ and $\beta_k$ are determined by
\ba
&& c_0 = 1 ~\quad{\rm and}\quad~ c_k = - \sum_{r=0}^{k-1} \, \frac{ 2^{2 (r-k)} c_r}{ (2k-2r)! }\,, \quad \forall \quad k\ge 1\,. \\
&& \beta_k= (-1)^k \sum_{r=0}^k\; \frac{2^{2(r-k)-1} c_r }{ (2k-2r+1)! }\,.
\ea
In the present case the scale ${\cal M}^2 $ is the minimum of $\mu+t/2 -2m^2$ which is $ 2m^2+t/2$ or for the tree-level bounds applicable for a weakly coupled UV completion $\Lambda_{\rm th}^2+t/2 -2m^2\approx \Lambda_{\rm th}^2$.

\section{Extensions}
\label{sec:different}

In previous sections, we considered the simple case where the mass of all the particles were the same $m_1=m_2=m_3=m_4$. Having laid out the strategy in that case, we can now move to more general scenarios.

\subsection{Two Mass Eigenstates}

We now consider a theory with two available mass states.
Without loss of generality, we can order $m_1 > m_2$.
In order to use the optical theorem at $t=0$, we should consider transition amplitudes to final states with $m_3 = m_1$, $m_4=m_2$, $S_3=S_1$ and $S_4=S_2$. Note that in order for the heavy particle to remain stable against decay into two light particles, we further assume $m_1 < 2 m_2$. This condition is not necessary when considering the tree level positivity bounds, since any associated branch cut arises at loop order. 
Defining the positive mass difference
\be
\Delta = m_1^2 - m_2^2  ,
\ee
we can generalize the analytic combinations ${\cal S}$ and ${\cal U}$ defined in \eqref{eq:CalS} as follows
\ba
\mathcal{S} &=& ( s - (m_1 - m_2)^2 ) (s - (m_1+m_2)^2 ) \,, \\
\mathcal{U} &=& ( u - (m_1 - m_2 )^2 ) ( u - (m_1+m_2)^2 ) \,,
\ea
which are positive on both the RH and the LH cuts. We also define the following quantity
\ba
\Psi = -su + \Delta^2\,,
\ea
in terms of which we have a scattering angle given by
\ba
\cos^2 \frac{\theta}{2} = \frac{ \Psi }{ \mathcal{S} }\quad  {\rm and} \quad   \sin^2 \frac{\theta}{2} = \frac{ - s t }{  \mathcal{S} } .
\ea

\paragraph{Transversity amplitudes:}
We now define $\Tc_{\tau_1 \tau_2 \tau_3 \tau_4}$ as the scattering amplitude between eigenstates of the transversity operator,
\be
\tau_i= - \frac{1}{m_i} w_\mu W^\mu(k_i)  \;\;\;\; \text{with} \;\;\;\; w_\mu = \frac{- 2 \epsilon_{\mu \nu \rho \sigma} k_1^\nu k_2^\rho k_3^\sigma  }{\sqrt{  - t \Psi  } } ,
 \ee
exactly analogous to \eqref{eqn:transversity_operator}, but with an overall normalization of $\sqrt{-t \Psi}$ in place of $\sqrt{stu}$. Such an amplitude obeys the crossing relation (assuming $S_3=S_1$ and $S_4=S_2$, \ie assuming that the out-going particles are the same as the in-going ones but without necessarily the same transversity),
\ba
\Tc^s_{\tau_1 \tau_2 \tau_3 \tau_4} (s,t) =  (-1)^{  2S_1 + 2S_2 } e^{i \pi \sum_i \tau_i}  e^{-i \chi_1 (\tau_1 + \tau_3)} e^{-i \chi_2 (\tau_2 + \tau_4)} \Tc^u_{-\tau_1 -\tau_4 -\tau_3 -\tau_2} (u,t)\,,
\ea
with $u$ channel crossing angles
\ba
\label{eq:chi1}
 \cos \chi_1 = \frac{ - (s + \Delta) (u + \Delta ) + 4 m_1^2 \Delta }{ \sqrt{  \mathcal{S} \mathcal{U} } } \,, & \quad &
 \sin \chi_1 = \frac{ - 2 m_1 \sqrt{ - t \Psi  }  }{ \sqrt{ \mathcal{S} \mathcal{U} } }  ,
 \\
 \label{eq:chi2}
 \cos \chi_2 =  \frac{ -(s - \Delta) (u - \Delta ) - 4 m_2^2 \Delta }{ \sqrt{  \mathcal{S} \mathcal{U} } } \,,
& \quad &
 \sin \chi_2 = \frac{ - 2 m_2 \sqrt{ - t \Psi  }  }{ \sqrt{ \mathcal{S} \mathcal{U} } }  .
  \ea
which clearly reduces to \eqref{chianglesame} when the mass difference $\Delta \to 0$.  Note that to avoid cluttering notations we have omitted the $u$ channel label for $\chi_i$.

\begin{figure}
\centering

\begin{tikzpicture}

\node at (6,-0.2) {Re$\,s$};
\draw [thick, gray, ->] (-6,0) -- (6,0);
\draw [ultra thick, red] (4,0) -- (6,0);
\draw [ultra thick, red] (-3,0) -- (-6,0);

\node at (4.,0.2) {\scriptsize $(m_1\!\!+\!m_2)^2$};
\filldraw [red] (4,0) circle (1.5pt);

\node at (-3,-0.2) {\scriptsize $(m_1\!\!-\!m_2)^2\!-\!t$};
\filldraw [red] (-3,0) circle (1.5pt);

\node at (-1,-0.2) {\scriptsize $m_2^2$};
\filldraw [red] (-1.1,0) circle (1.5pt);

\node at (0.1,-0.2) {\scriptsize $m_1^2$};
\filldraw [red] (0,0) circle (1.5pt);

\node at (1.05,0.2) {\scriptsize $m_1^2\!\!+\!2 m_2^2\!\!-\!t$};
\filldraw [red] (1,0) circle (1.5pt);

\node at (2.55,-0.2) {\scriptsize $2m_1^2\!\!+\!2m_2^2\!\!-\!t$};
\filldraw [red] (2.5,0) circle (1.5pt);

\end{tikzpicture}

\caption{The scattering amplitude on the real $s$ axis in a theory with two massive states. We can draw integration contours analogous to $C$ and $C'$ from Figure \ref{fig:1}, providing that none of the poles overlap with the branch cuts. \label{fig:2}}
\end{figure}
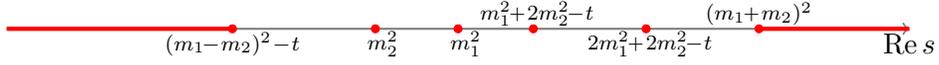

\paragraph{Physical singularities:}
In the complex $s$ plane, by unitarity the scattering amplitude must have poles at $s=m_1^2$ and $s=m_2^2$, and a branch cut from $s=(m_1+m_2)^2$ to infinity. The crossing relation then tells us that there are also poles at $s= 2m_1^2 + m_2^2-t$ and $s= m_1^2 + 2m_2^2 - t$, and a branch cut from $s=(m_1-m_2)^2-t$ to infinity. Note that since $m_1<2 m_2$ the poles and branch cuts are separated.  We can hence proceed as before, providing that we have
\begin{equation}
0 \leq t < m_2^2 \leq m_1^2 < 4 m_2^2  .
\end{equation}
The corresponding $\text{Re}\, s$ axis (at fixed $t$) is shown in Figure~\ref{fig:2}.

\paragraph{Kinematical singularities:}
Just as in the case of a single mass, the transversity amplitudes may possess different types of kinematical singularities or branch points at:
\begin{itemize}

\item  $\mathcal{S}=0$: When writing $\cos \theta$ and $\sin \theta$ in terms of $s,t$, one introduces spurious singularities in $1/\mathcal{S}$. These are factorizable, and can be removed with a prefactor of $\mathcal{S}^{S_1 + S_2}$. This corresponds to removing the $s-4m^2$ singularities in the equal mass limit.

 \item  $\sqrt{-t \Psi} =0$: When writing $\sin \theta$ in terms of $s,t$, one further introduces an unphysical branch point at $-t \Psi = 0$.
     Since $\sqrt{-t \Psi}$ is odd under $\theta \rightarrow - \theta$ this branch point is removed by taking the following even combinations,
\begin{equation}
\Tc_{\tau_1 \tau_2 \tau_3 \tau_4} ( s, \theta) + \Tc_{\tau_1 \tau_2 \tau_3 \tau_4} (s, -\theta)
\;\;\;\; \text{or}  \;\;\;\;
\sqrt{-t \Psi} \left[  \Tc_{\tau_1 \tau_2 \tau_3 \tau_4} ( s, \theta) - \Tc_{\tau_1 \tau_2 \tau_3 \tau_4} (s, -\theta) \right]\,.
\end{equation}
 These are the exact analogue to \eqref{eqn:T_sum} and \eqref{eqn:T_diff} combinations that remove the branch cut associated with $\sqrt{stu}$ in the equal mass limit.

\item  $\Psi=0$:  For boson-fermion scattering, $\cos \theta/2$ creates an unphysical branch point at $\Psi=0$, which can be removed with a prefactor of $
 \big( \sqrt{ \Psi } \big)^{\xi} $, where as before, $\xi$ is 0 for BB or FF, and 1 for BF scattering. This corresponds to removing the branch point at $u=0$ in the equal mass limit.

\end{itemize}
To summarize,  we can consider the combinations
\begin{align}
\Tc_{\tau_1 \tau_2 \tau_1 \tau_2}^+ &=  ( \sqrt{\Psi} )^\xi \mathcal{S}^{S_1 + S_2} \left[ \Tc_{\tau_1 \tau_2 \tau_3 \tau_4} (s, \theta) + \Tc_{\tau_1 \tau_2 \tau_3 \tau_4} (s, - \theta)  \right]  , \label{eqn:Tplus}  \\
\Tc_{\tau_1 \tau_2 \tau_1 \tau_2}^- &= -i ( \sqrt{\Psi} )^\xi \mathcal{S}^{S_1 + S_2}   \sqrt{- t \Psi} \left[ \Tc_{\tau_1 \tau_2 \tau_3 \tau_4} (s, \theta) - \Tc_{\tau_1 \tau_2 \tau_3 \tau_4} (s, - \theta)  \right]  , \label{eqn:Tminus}
\end{align}
which are free of all kinematical singularities. These are both analytic in the twice-cut $s$ plane at fixed $t$, and therefore obey dispersion relations with $N_S=2 +2(S_1 + S_2) + \xi$ subtractions.

\paragraph{Positivity of the right hand cut:}
The relation between the transversity and helicity basis, as well as the partial wave expansion of the helicity amplitudes, are exactly as described earlier---even for unequal masses. Similarly, the Fourier series of the Wigner matrix \eqref{eqn:dJfourier} is unchanged. Therefore $\mathcal{T}^+_{\tau_1 \tau_2 \tau_1 \tau_2}$, as defined in \eqref{eqn:Tplus}, can be shown to obey the same positivity property \eqref{eqn:positivity_T} as its equal mass limit,
\ba
\label{eq:AbsT+diffmass}
\text{Abs}_s  \mathcal{T}^+_{\tau_1 \tau_2 \tau_1 \tau_2}  = \sum_{J, \nu}
 (\sqrt{\Psi} )^\xi \mathcal{S}^{S_1 + S_2} 2 \cos ( \nu \theta)F^{J\nu}_{\tau_1 \tau_2}(s)    \,,
\ea
and so as in the equal mass limit, using $\sqrt{\Psi} = \sqrt{\mathcal{S}} \cos (\theta/2)$, and the fact that all the Chebyshev polynomials $N_{n,\nu}$ defined in \eqref{chebydef} are positive, we can infer
all the derivatives of  \eqref{eq:AbsT+diffmass} with respect to $\cos \theta$ can be shown to be positive. Indeed, for $\xi=0$,
\ba
\frac{\partial^n}{ \partial \cos^n \theta} \text{Abs}_s  \mathcal{T}^+_{\tau_1 \tau_2 \tau_1 \tau_2} \bigg|_{\theta=0} =2  \mathcal{S}^{S_1 + S_2} \sum_{J, \nu}   F^{J\nu}_{\tau_1 \tau_2}(s)
N_{n,\nu} >0\,,
\ea
and for $\xi=1$ (\ie for BF scattering),
\ba
\frac{\partial^n}{ \partial \cos^n \theta} \text{Abs}_s  \mathcal{T}^+_{\tau_1 \tau_2 \tau_1 \tau_2} \bigg|_{\theta=0} =
\mathcal{S}^{S_1 + S_2+1/2} \sum_{J, \nu}   F^{J\nu}_{\tau_1 \tau_2}(s)
\left( N_{n,\nu+1/2} + N_{n,\nu-1/2}    \right)>0\,.
\ea
It therefore immediately follows that
\ba
\frac{\partial^n}{\partial t^n} \text{Abs}_s  \mathcal{T}^+_{\tau_1 \tau_2 \tau_1 \tau_2}   > 0 \quad \text{for} \quad 0 \leq t < m_2^2  \quad {\rm and}\quad \forall \ n \geq 0   \,.
\ea
It will also be useful to express the absorptive part of the difference as
\ba
\text{Abs}_s  \mathcal{T}^-_{\tau_1 \tau_2 \tau_1 \tau_2}  = - 2 \sqrt{-t \Psi} (\sqrt{\Psi} )^\xi  \mathcal{S}^{S_1 + S_2} \sum_{J, \nu}
  \sin ( \nu \theta)  F^{J\nu}_{\tau_1 \tau_2}(s)   \,,
\ea
which is not necessarily positive, but will be useful in what follows.

\paragraph{Positivity of the left hand cut:} Having defined the appropriate quantity \eqref{eq:AbsT+diffmass} which is positive on the RH cut, we can now, exactly as before, establish the scattering amplitude on the LH cut by crossing symmetry.
We proceed as in the equal mass limit and first relate the $s$ channel transversity amplitude with the $u$ channel one as in \eqref{eq:Tsp}
\be
\mathcal{T}^{s+}_{\tau_1 \tau_2 \tau_1 \tau_2} (s,t) = c_+  \mathcal{T}^{ u+}_{\tau_1 \tau_2 \tau_1 \tau_2} (u,t) + c_-  \mathcal{T}^{u-}_{\tau_1 \tau_2 \tau_1 \tau_2} (u,t)    ,
\ee
with
\begin{align}
c_+  &=  \frac{ \mathcal{S}^{S_1 + S_2} }{ \mathcal{U}^{S_1 + S_2} }  \cos \left( 2 \chi_1 \tau_1 + 2 \chi_2 \tau_2 \right)  ,
 \\
c_-  &=   -\frac{ \mathcal{S}^{S_1 + S_2} }{ \mathcal{U}^{S_1 + S_2} }  \frac{ \sin \left( 2 \chi_1 \tau_1 + 2 \chi_2 \tau_2 \right)  }{\sqrt{-t \Psi} }   \,.
\end{align}
and so
\ba
\label{eq:AbsTsp2}
\text{Abs}_u  \mathcal{T}^{s+}_{\tau_1 \tau_2 \tau_1 \tau_2} (s, t)  &=2 ( \sqrt{ \Psi})^\xi \mathcal{S}^{S_1+S_2}   \sum_{J,  \nu}  \cos \left( \nu \theta_u - 2 \chi_1 \tau_1 - 2 \chi_2 \tau_2 \right)  F^{u,J\nu}_{\tau_1 \tau_2}(u)\,.
\ea
 Next we re-write this expression as a sum over $\cos (p \theta_u)$, after replacing $\sqrt{\Psi} = \sqrt{\mathcal{U}} \cos (\theta_u/2)$.
To express the previous relation as a sum of $\cos$, we derive an exact analogue to \eqref{eq:comb1}, where the only subtlety to keep track off is the fact that we have now two different $u$ channel angles $\chi_1$ and $\chi_2$ (defined in \eqref{eq:chi1} and \eqref{eq:chi2}) as opposed to only one $\chi_u$ in the equal mass limit. This implies the existence of two separate sets of coefficients $\sqrt{\cal S} e^{\pm i \chi_i}= \sum_p c^{\pm}_{i , p}(u) e^{i p \theta}$ as opposed to the only one set in  \eqref{eq:comb1}, but it nevertheless remains that $c^{\pm}_{i , p}(u) >0$ for $ u > (m_1 + m_2)^2$.    \\

Noting again that $ \mathcal{S}$ is still a sum of $\cos(p \theta_u)$ with positive coefficients when $u > (m_1 + m_2)^2$, more precisely,
\ba
 \mathcal{S} =
 \frac{\cal U}{4u^2} &\big[&\left(u-({m_1}+{m_2})^2\right) \cos \theta _u + u+({m_1}+{m_2})^2\, \big] \nn \\
\times &\big[&\left(u-({m_1}-{m_2})^2\right) \cos \theta _u + u+({m_1}-{m_2})^2\, \big]\,,
\ea
then we are led to precisely the same relation as in \eqref{eq:cosdiff} with now
\ba
\label{eq:SC2}
{\cal S}^{S_1+S_2}  \cos \left( \nu \theta_u -2\chi_1  \tau_1-2 \chi_2 \tau_2   \right) &=&{\cal S}^{S_1+S_2-\tau_1-\tau_2}   {\cal S}^{\tau_1+\tau_2} \cos \left( \nu \theta_u -2\chi_1  \tau_1-2 \chi_2 \tau_2   \right)  \nn \\
&=& \sum_{p=-2(\tau_1+\tau_2)-\nu}^{2(\tau_1+\tau_2)+\nu}C_{\nu,p}(u) e^{i p \theta_u} \, ,
\ea
where $C_{\nu,p} (u)$ is analytic in $u$, and strictly positive for $u> (m_1+m_2)^2$.\\

Using this expression in \eqref{eq:AbsTsp2} and taking derivatives of the Chebyshev polynomials at $\theta_u=0$, we finally have
\ba
&& \frac{\partial^n}{\partial \cos^n \theta_u} \text{Abs}_u  \mathcal{T}^{s +}_{\tau_1 \tau_2 \tau_1 \tau_2} (s, t)   \bigg|_{\theta_u=0}
\nn\\
&&= \sum_{J, \nu,  p}
F^{u,J\nu}_{\tau_1 \tau_2}(u)   C_{\nu, p} (u) \begin{cases}
2 N_{n,p}  \;\;\;\;& \text{if } \xi = 0 \\
\sqrt{\mathcal{U}} \left( N_{n,p+1/2} + N_{n,p-1/2}   \right)
 \;\;\;\;& \text{if } \xi = 1   ,
\end{cases}
\ea
where the coefficients $C_{\nu, p} (u)$ are all positive in the region $u > (m_1 + m_2)^2$ and so are the Chebyshev polynomials $N_{n,p}$. This proves the  following positivity relation:
\be
 \frac{\partial^n}{\partial t^n}  {\text{Abs}}_{u}   \Tc^{s+}_{\tau_1 \tau_2 \tau_1 \tau_2}   (s,t) >0  \, , \quad {\rm for } \quad u > (m_1 + m_2)^2 \,, \quad 0 \le t < m_2^2 \, , \quad  \forall  \; n \ge 0 \, .
\ee\\

\paragraph{Positivity bounds:} We can now derive the dispersion relation to work out the positivity bounds. The only subtlety to bear in mind is the existence of additional poles (associated with the existence of different masses). However, those are to be subtracted as in the single mass case and we define the pole subtracted amplitude as
\begin{equation}
\tilde{\mathcal{T}}_{\tau_1 \tau_2 \tau_1 \tau_2}^+ = \mathcal{T}^+_{\tau_1 \tau_2 \tau_1 \tau_2}  - \frac{ \text{Res}_{s=m_1^2} }{s-m_1^2} - \frac{ \text{Res}_{s=m_2^2} }{s-m_2^2} - \frac{ \text{Res}_{u=m_1^2} }{u-m_1^2} - \frac{ \text{Res}_{u=m_2^2} }{u-m_2^2}   \,.
\end{equation}
We then have a dispersion relation with $N_S = 2+2S_1+2S_2+\xi$ subtractions given by
\ba
\tilde\Tc^+_{\tau_1 \tau_2 \tau_1 \tau_2}(s,t) = \! \sum_{n=0}^{N_S-1} \! a_n(t) s^n
\!&+&\! \frac{s^{N_S}}{\pi}  \int_{4m^2}^\infty \ud \mu  \frac{ {\text{Abs}}_s \Tc^+_{\tau_1 \tau_2 \tau_1 \tau_2}(\mu,t) }{ \mu^{N_S} (\mu - s) }  \nn  \\
&+&\frac{u^{N_S}}{\pi}  \int_{4m^2}^\infty \ud \mu  \frac{  {\text{Abs}}_u \Tc^+_{\tau_1 \tau_2 \tau_1 \tau_2}(2m_1^2+2m_2^2-t-\mu,t) }{ \mu^{N_S} ( \mu - u) } \, .
\ea
Then, we can follow the same steps as the equal mass case discussed in section \ref{sec:disperse}  (see the steps following Eq.~(\ref{disrel1})) to derive positivity bounds.

\subsection{Multiple Mass Eigenstates}

We can even generalize the previous procedure further and suppose we had $n$ possible intermediate states, $m_i$, which we order $m_1 \geq m_2 \geq ... \geq m_n > 0$. Then all we need in order to construct positivity bounds as we have done is to require
\begin{align}
m_1^2  <  4 m_n^2  \,,
\end{align}
\ie all of the particles are kinematically stable against two-body decay (once again this is not necessary when considering the tree level bounds). Then all of the poles lie within the $s$ and $u$ channel branch points, and so are contained within a contour like $C$ in Fig.~\ref{fig:1}. In this case, we are led to the same positivity bound, providing $t$ is restricted to the region
\begin{equation}
0 \leq t < m_n^2   \,.
\end{equation}
The proof is identical to the preceding subsection, with $m_1$ and $m_2$ replaced with any desired pair of particle masses $m_i$ and $m_j$.

\section{Discussion}
\label{sec:conc}

The utility and universality of effective field theories is both a blessing and a curse. They successfully describe the low energy world, and decoupling ensures that predictions from a low energy EFT can be made without any precise knowledge of its explicit high energy completion.
On the other hand this suggests that all EFTs are from a theoretical point of view equal, and only experiments/observations can distinguish them.  In certain fields, for example in constructing EFTs for inflation, dark energy or physics beyond the standard model, observations and experiments are sufficiently limited that there remain a large wealth of candidate EFTs that can describe the same data. In this article we have discussed theoretical tools by which this impasse may be partly broken. We have seen that the very existence of a standard UV completion leads to powerful constraints on the scattering amplitudes of the low-energy EFTs. Violating any of these constraints directly implies an obstruction of the EFT from ever admitting a standard UV completion. Establishing the IR implications of UV completion is thus essential in segregating between different types of EFTs. For 2-to-2 elastic scattering amplitude, these constraints were previously known in the forward scattering limit for arbitrary particles. In this work we have derived an infinite number of constraints that span beyond the forward scattering limit for particles of arbitrary spins. \\

To this aim, given the 2-to-2 elastic scattering amplitude for particles of arbitrary spins, first we have shown how to construct a regularized transversity amplitude which is free of kinematic singularities, has the same analyticity structure as a scalar scattering amplitude, and has a positive discontinuity along both its left hand and right hand cuts. This generalizes an approach given for a special case in \cite{Mahoux:1969um} (for helicity amplitudes) to the case of general spins. This result is far from straightforward, due in particular to the complex analyticity structure for fermion scattering.
Crucially, in order to achieve positivity along both cuts we must work with the transversity amplitudes and not helicity amplitudes. In appendix \ref{app:multispinor} we have given an independent derivation of the $s-u$ crossing formula for scattering amplitudes of arbitrary spin away from the forward scattering limit which is central to these arguments. This uses the multispinor framework \cite{Bargmann:1948ck,Schwinger:1966zz,PhysRev.161.1316,Schwinger:1978ra} where particles of general spin are viewed as being made up of tensor products of spin 1/2 states.
\\

Once this has been done, the positivity bounds that have been derived for scalars \cite{Adams:2006sv} and in particular their non-forward limit extensions \cite{deRham:2017avq,deRham:2017imi} can immediately be passed over to particles of general spin, with the only caveat being that the number of subtractions increases with the number of spins for the regularized amplitudes, \ie those combinations which are free of kinematic singularities. \\

These bounds will apply to scattering of massive states in any low energy effective theory that arises from an analytic, Lorentz invariant UV completion. We also expect them to apply in the massless limit for tree level scattering amplitudes, since the usual obstruction to the massless limit comes from the branch cut that begins at $s=4m^2$, but this branch cut only arises when light loops are included. The massless pole is itself harmless since it can be subtracted out. This is true provided that the Froissart bound holds. For massless particles, Froissart could be violated; nevertheless we would expect that the fixed $t$ dispersion relations remain bounded by a polynomial, and this is sufficient to derive positivity bounds with the only difference being that the number of subtractions $N_S$ should be increased sufficiently to account for the growth of fixed $t$ amplitude at large $s$. \\

In general, the bounds apply away from the forward scattering limit; however, whether they will be stronger away from this limit will depend on the model. At least the example of Galileon EFTs given in \cite{deRham:2017imi} is a proof of principle that in certain cases the non-forward limit bounds are stronger. We will give explicit examples of the application of these bounds to particular classes of EFTs elsewhere \cite{dRMTZ}.
\vskip 30pt

\section*{Acknowledgments}

We would like to thank Brando Bellazzini for useful comments.
CdR thanks the Royal Society for support at ICL through a Wolfson Research Merit Award.
CdR is supported in part by the European Union's Horizon 2020 Research Council grant 724659 MassiveCosmo ERC-2016-COG. SM is funded by the Imperial College President's Fellowship. AJT thanks the Royal Society for support at ICL through a Wolfson Research Merit Award.

\appendix

\section{Analyticity and Causality}
\label{app:AnalyticCausal}

The connection between the analyticity of the scattering amplitude for fixed momenta transfer, and causality is well established, but sufficiently forgotten that we will review here the essential details of the proof without dwelling on the lengthy mathematical subtleties. We shall do so for scalar particles, but this can easily be extended to general spin by utilizing the appropriate helicity wavefunctions and accounting for statistics.
Denoting $a_A^{\dagger}$ the creation operators for a particle of type $A$ and $a_{\bar A}^{\dagger}$ that for the associated anti-particle with relativistic normalization $[a_A(k),a^{\dagger}_B(k')] = (2 \pi)^3 \delta^3(\vec k - \vec k') 2 \omega_k \delta_{AB}$, then the $s$-channel $A+B \rightarrow C+D$ scattering amplitude is ($\hat S = 1+ i \hat T$)
\ba
\langle 0 | a_C(k_3) a_D(k_4)\hat T a^{\dagger}_B(k_2) a^{\dagger}_A(k_1) | 0 \rangle &=& \langle k_3 |  a_D(k_4)\hat T a^{\dagger}_B(k_2) | k_1 \rangle \nn  \\
&=&-i  \langle k_3 |  a_D  (k_4 )[ \hat S , a^{\dagger}_B (k_2 )] | k_1 \rangle  \, ,
\ea
where in the last step we have used the fact that $\langle k_3 |  a_D(k_4)a^{\dagger}_B(k_2)=0$ unless $B$ is identical to $C$ or $D$ in which case it will correspond a non-scattering process which drops out of $\hat T$. Denoting the (in general complex) free fields by
\be
\phi_A(x) = \int \frac{\d^3 k}{(2\pi)^3 2 \omega_k} \left(e^{ik.x} a_A(k) +  e^{-ik.x} a_{\bar A}^{\dagger}(k) \right)
\ee
then it is straightforward to show as a consequence of Wick's theorem
\be
[ \hat S , a^{\dagger}_A(k)]  = \int \d^4 x e^{ik.x} \frac{\delta \hat S}{\delta \phi_A(x)} \, ,
\ee
which just follows from the elementary fact that
\be
[ \phi_B(y) , a^{\dagger}_A(k)]  = \int \d^4 x e^{ik.x} \frac{\delta \phi_B(y)}{\delta \phi_A(x)}  = \int \d^4 x e^{ik.x} \delta_{AB} \delta^4(x-y) \, ,
\ee
and the fact that the commutator respects the Leibniz rule. Then using the property of the stability of the one-particle states $\hat S | k \rangle = | k  \rangle=\hat S^{\dagger} | k \rangle$
\ba
\langle 0 | a_C (k_3) a_D(k_4)\hat T a^{\dagger}_B(k_2) a^{\dagger}_A(k_1) | 0 \rangle &=& - i \int \d^4 x e^{ik_2.x} \langle k_3 |  a_D(k_4)\frac{\delta \hat S}{\delta \phi_B(x)} S^\dagger  | k_1 \rangle  \nn \\
 &=& -  \int \d^4 x e^{ik_2.x} \langle k_3 |  a_D(k_4) \hat J_B(x)  | k_1 \rangle
\ea
where we have defined the current operator $ \hat J_B(x) = i \frac{\delta \hat S}{\delta \phi_B(x)} \hat S^\dagger = - i \hat S \frac{\delta \hat S^\dagger}{\delta \phi_B(x)}   $. Similarly using
\be
[  a_A(k), \hat S]  = \int \d^4 x e^{-ik.x} \frac{\delta \hat S}{\delta \phi_{ \bar A}(x)} \, ,
\ee
we have
\be
\langle 0 | a_C(k_3) a_D(k_4)\hat T a^{\dagger}_B(k_2) a^{\dagger}_A(k_1) | 0 \rangle = -  \int \d^4 x \int \d^4 y e^{-i k_4.y+ik_2.x} \langle k_3 |  \frac{\delta \hat J_B(x)}{\delta \phi_{\bar D}(y)}  | k_1 \rangle \, .
\ee
Causality is encoded in the statement that (this is expounded on in \cite{bogoliubov1959introduction}, section 17.5)
\be
\frac{\delta \hat J_B(x)}{\delta \phi_{\bar D}(y)} =0 \, \quad \text{if } \quad y^0<x^0 \quad  \text{and/or } \quad (y-x)^2>0 \, .
\ee
Roughly speaking, the response of the $S$-matrix to the fluctuation at $x$ described by $\hat J_B (x)$ cannot influence fields outside of the future lightcone of $x$.
Since
\be
\frac{\delta \hat J_B(x)}{\delta \phi_{\bar D}(y)} = i \( \frac{\delta^2}{\delta \phi_{\bar D}(y) \delta \phi_B(x)} \hat S\) \hat S^{\dagger} + i \frac{\delta \hat S}{\delta \phi_B(x)} \frac{\delta \hat S^{\dagger}}{\delta \phi_{\bar D}(y)}
\ee
then we infer
\be
 \( \frac{\delta^2}{\delta \phi_{\bar D}(y) \delta \phi_B(x)} \hat S\) \hat S^{\dagger}  = -  \frac{\delta \hat S}{\delta \phi_B(x)} \frac{\delta \hat S^{\dagger}}{\delta \phi_{\bar D}(x)} = -\hat J_B(x) \hat J_{\bar D}(y) \, \,  \text{if } \, \,  y^0<x^0 \, \,   \text{and/or } \, \, (y-x)^2>0 \, .
\ee
Using the commutativity of functional derivatives $\frac{\delta^2}{\delta \phi_{\bar D}(y) \delta \phi_B(x)} \hat S=\frac{\delta^2}{\delta \phi_{B}(x) \delta \phi_{\bar D}(y)} \hat S$ then we infer
\be
 \( \frac{\delta^2}{\delta \phi_{\bar D}(y) \delta \phi_B(x)} \hat S\) \hat S^{\dagger}  = -\hat J_{\bar D}(y) \hat J_{B}(x) \, \,  \text{if } \, \,  x^0<y^0 \, \,   \text{and/or } \, \, (y-x)^2>0 \, .
\ee
which implies the more familiar statement of causality, that operators commute outside of the lightcone
\be
[\hat J_B(x) , \hat J_{\bar D}(y) ] =0 \, , \quad \text{for} \, \, (y-x)^2>0 \, .
\ee
Putting this together we have
\be
\frac{\delta \hat J_B(x)}{\delta \phi_{\bar D}(y)} =0 = \theta(y^0-x^0) [\hat J_B(x),\hat J_{\bar D}(y)] + \text{contact terms}
\ee
where the contact terms vanish for $x^0 \neq y^0$, \ie they are (derivatives of) delta functions. The precise form of these contact terms cannot be determined by causality (since they are instantaneous) or unitarity. In momentum space, the contact terms correspond to polynomial functions of energy/momenta and so their addition is equivalent to modifying the subtraction terms in the dispersion relation. Consequently it is sufficient to derive a dispersion relation assuming no subtractions are needed (\ie no contact terms), and include the subtractions at the end of the calculation. With this in mind the scattering amplitude can be taken to be
\be
\langle 0 | a_C(k_3) a_D(k_4)\hat T a^{\dagger}_B(k_2) a^{\dagger}_A(k_1) | 0 \rangle = -  \int \d^4 x \int \d^4 y e^{-i k_4.y+ik_2.x} \theta(y^0-x^0) \langle k_3 | [\hat J_B(x),\hat J_{\bar D}(y)]  | k_1 \rangle \, .
\ee

Using the translation properties of momentum eigenstates to remove an overall momentum conserving delta function, the stripped $s$-channel scattering amplitude is
\be
A^{A+B \rightarrow C+D}(k_1,k_2;k_3,k_4) =  \int \d^4 x  \, e^{-i(k_2+k_4).x/2} \theta(x^0) \langle k_3 | [\hat J_{\bar D}(x/2),\hat J_B(-x/2)]  | k_1 \rangle \, .
\ee
An identical calculation for the $u$-channel amplitude would give
\ba
&~& A^{A+\bar D \rightarrow C+\bar B}(k_1,-k_4;k_3,-k_2)
\nn\\
&=&   \int \d^4 x  e^{i(k_2+k_4).x/2} \theta(x^0) \langle k_3 | [\hat J_{B}(x/2),\hat J_{\bar D}(-x/2)]  | k_1 \rangle  \nn \\
&= &  - \int \d^4 x e^{-i(k_2 + k_4).x/2} \theta(-x^0) \langle k_3 | [\hat J_{\bar D}(x/2),\hat J_{B}(-x/2)]  | k_1 \rangle .
\ea
In the latter form, it is clear that the only difference between the $u$-channel scattering amplitude and the $s$-channel is the choice of retarded $\theta(x^0)$ versus advanced $-\theta(-x^0)$ boundary conditions (up to a sign).
As a consequence their difference is
\bal
&\qquad A^{A+B \rightarrow C+D}(k_1,k_2;k_3,k_4) -A^{A+\bar D \rightarrow C+\bar B}(k_1,-k_4;k_3,-k_2)
\nn\\
& = \int \d^4 x e^{-i(k_2+k_4).x/2} \langle k_3 | [\hat J_{\bar D}(x/2),\hat J_{B}(-x/2)]  | k_1 \rangle \, .
\eal
Inserting a complete set of positive energy multi-particle momentum eigenstates $|p_n \rangle$ this can be written as
\bal
 &\qquad A^{A+B \rightarrow C+D}(k_1,k_2;k_3,k_4) -A^{A+\bar D \rightarrow C+\bar B}(k_1,-k_4;k_3,-k_2)
  \nn\\
&= \nn  (2\pi)^4 \sum_n  \bigg[ \langle k_3 | \hat J_{\bar D}(0) |p_n \rangle \langle p_n |\hat J_{B}(0) | k_1 \rangle \delta^4(q+(k_1+k_3)/2-p_n)
\nn\\
&\qquad\qquad\qquad  - \langle k_3 | \hat J_B (0) |p_n \rangle \langle p_n |\hat J_{\bar D}(0) | k_1 \rangle \delta^4(q-(k_1+k_3)/2-p_n)\bigg]
\eal
where $q=(k_2+k_4)/2$. As a function of $q$, this is only non-zero if either $(q\pm(k_1+k_3)/2)^2 \ge m_L^2$ where $m_L$ is the mass of the lightest particle (corresponding to the lightest $p_n$) or if the pole terms corresponding to single particle intermediate states are subtracted out, then $4 m_L^2$. In terms of Mandelstam variables this is $s \ge 4 m_L^2$ or $u \ge 4 m_L^2$.
The region where this is not satisfied is precisely the Mandelstam triangle and so we conclude that the $s$-channel and $u$-channel functions are identical in the analyticity window of the Mandelstam triangle and are elsewhere analytic continuations of each other. The full rigorous proof of this is lengthy \cite{bogoliubov1959introduction,Hepp_1964,Bremermann:1958zz} but relies only on the above physical considerations. \\

Focussing on elastic scattering, $m_1 = m_3$ and $m_2 = m_4$ we may choose the Breit coordinate system
\ba
&& k_1 = (\sqrt{\vec p{\,}^2+m_1^2},\vec p)\, , \quad k_3=(\sqrt{\vec p{\,}^2+m_1^2},-\vec p) , \nn \\
&&  k_2=(E,-\vec p+\lambda \vec e) \, , \quad k_4=( E, \vec p + \lambda \vec e)
\ea
where $\vec e.\vec p=0$ and $|\vec e|=1$. In terms of Mandelstam variables we have $t = - 4 \vec p{\,}^2$ and $E=\sqrt{\vec p{\,}^2+\lambda^2+m_2^2}= (s+t/2-m_1^2-m_2^2)/(\sqrt{4m_1^2-t})$.
Then the $s$-channel amplitude is
\ba
A^{A+B \rightarrow C+D}(k_1,k_2;k_3,k_4)
=  \int \d^4 x  \, e^{iE x^0 - i \vec e.\vec x \sqrt{E^2-m_2^2+t/4} } \theta(x^0) \langle k_3 | [\hat J_{\bar D}(x/2),\hat J_B(-x/2)]  | k_1 \rangle \, .\nn
\ea
Analyticity in $s$ at fixed $t < 4m_1^2$ corresponds to analyticity in $E$ at fixed $t$. From the above expression, to define an analytic continuation we must carefully deal with the convergence of the integral associated with the analytic continuation of the square root $\sqrt{E^2-m_2^2+t/4}$. However, if we focus purely on the high energy regime we may approximate this as
\be
A^{A+B \rightarrow C+D}(k_1,k_2;k_3,k_4) \approx  \int \d^4 x  \, e^{iE (x^0 -\vec e.\vec x)} \theta(x^0) \langle k_3 | [\hat J_{\bar D}(x/2),\hat J_B(-x/2)]  | k_1 \rangle \, .
\ee
Since the integrand vanishes for $|\vec x|^2> (x^0)^2$ and $x^0<0$, we conclude that the domain of integration is $x^0 -\vec e.\vec x  \ge 0$ and so the $s$-channel scattering amplitude may be extended to an analytic function of $E$ in the upper half complex energy plane since
\ba
&& e^{iE (x^0 -\vec e.\vec x)} \theta(x^0 -\vec e.\vec x) \rightarrow 0,  \text{ as } x^0 - \vec e. \vec x \rightarrow \infty \, , \text{ for fixed } {\rm Im}(E) >0 \, .
\ea

Similarly at high energies the $u$-channel amplitude at high energies takes the form
\ba
A^{A+\bar D \rightarrow C+\bar B}(k_1,-k_4;k_3,-k_2) \approx  -\int \d^4 x  \, e^{iE (x^0 -\vec e.\vec x)} \theta(-x^0) \langle k_3 | [\hat J_{\bar D}(x/2),\hat J_B(-x/2)]  | k_1 \rangle \, , \nn
\ea
where now the integrand has support for $x^0 -\vec e.\vec x \le 0$ and so may be taken as an analytic function of $E$ in the lower half complex energy plane. Once these results are extended to low energies as well (e.g. \cite{bogoliubov1959introduction,Hepp_1964,Bremermann:1958zz}) modulo poles, then using the property that the $s$-channel and $u$-channel amplitudes coincide in the Mandelstam region, we infer that the scattering amplitude is analytic in the whole complex $E$ plane, at fixed $t$ modulo poles and the branch cuts along the real axis gapped by the Mandelstam triangle. Since $E= (s+t/2-m_1^2-m_2^2)/(\sqrt{4m_2^2-t})$ this is equivalent to analyticity in $s$ at fixed real $t<4m_2^2$. The original rigorous proofs \cite{bogoliubov1959introduction,Hepp_1964} only applied for $t \le 0$ or some intermediate positive value \cite{Bremermann:1958zz}, however the results of Martin which make additional use of unitarity of the scattering amplitude (that do not immediately follow from the above integral representation) extend these to $t<(m_1+m_2)^2$ \cite{Martin:1965jj} where $m_L$ is the lightest particle mass.

\section{Crossing Relations from Multispinors}

\label{app:multispinor}

In this section we will derive the crossing formula by calculating tree level scattering amplitudes for arbitrary spin particles,
\ba
\mathcal{T}^{A+B \to C+D}_{\tau_1 \tau_2 \tau_3 \tau_4} (k_1, k_2 ; k_3, k_4 )\, .
\ea
 To do this we will make significant use of transversity spinors, and so we first set notation.
Throughout we will consider scattering in the $xz$-plane. Using Lorentz invariance to fix the total $3$-momentum to be zero, on-shell conditions to fix the energies in terms of the $3$-momenta, and overall momentum conservation, one can write the scattering  amplitude as a function of just two variables namely the 3-momentum $|\mathbf{p}|=p$ and the scattering angle $\theta$. The four-momentum of each particles is then given by
\ba
k_i = \left( \sqrt{s}/2 , p \sin \theta_i, 0 , p \cos \theta_i  \right) \,, \quad {\rm with }\quad p =  \tfrac{1}{2} \sqrt{ s - 4 m^2 }\,,
\label{eqn:standard}
\ea
with a respective angle for each particles in the scattering process $A+B \to C+ D$:
\ba
\label{eq:theta_i}
\theta_1=0\,,\quad \theta_2=\pi\,, \quad \theta_3=\theta \quad {\rm and }\quad \theta_4=\pi+\theta\,.
\ea
Before giving any explicit amplitudes, it will be useful to derive precisely the transversity spinor states and polarizations.  For that we start in the standard helicity basis and convert to transversity.

\subsection{Transversity states}

The helicity spinors $\tilde u_\lambda$ that satisfy the Dirac equation $[-i \partialslash +m]\tilde u_\lambda e^{i p.x} =0 $ are
\ba
\label{eq:helicitySpinors}
\tilde u_+=\frac{1}{\sqrt{2m (m+E)}}\(\begin{array}{c}
(E+m)\cos (\theta/2)\\
(E+m)\sin (\theta/2)\\
p \cos (\theta/2)\\
p \sin (\theta/2)
\end{array}\)\,,\
\tilde u_-=\frac{1}{\sqrt{2m (m+E)}}\(\begin{array}{c}
- (E+m)\sin (\theta/2)\\
(E+m)\cos (\theta/2)\\
p \sin (\theta/2)\\
- p \cos (\theta/2)
\end{array}\)\,.\qquad
\ea
The helicity anti-spinor states $\tilde v_\lambda$ can then be derived by charge conjugation,
\ba
\label{eq:antispinor}
\bar {\tilde v}_\lambda  = \tilde u^T_\lambda C\,, \quad  {\rm or }\quad \bar {\tilde u}_\lambda = \tilde v^T_\lambda C\,,
\ea
with the charge conjugation matrix $C$ given by $C=-i \gamma^0 \gamma^2$ and where we  work in the standard Dirac convention for the $\gamma$ matrices. \\

The transversity spinors $u_\tau$ are simply a superposition of the helicity ones
\ba
u_\tau=\sum_{\lambda} u^{1/2}_{\tau \lambda} \tilde u_\lambda\,,
\ea
where the Wigner matrix $u^{1/2}_{\tau \lambda}$ is given by
\ba
u^{1/2}_{\tau \lambda} = \frac{1}{\sqrt{2}} \left( \begin{array}{c c}
1 & i \\
i & 1
\end{array}  \right)\,,
\ea
hence leading to the transversity spinors $u_{1/2}=\frac{1}{\sqrt{2}}\(\tilde u_++i \tilde u_-\)$ and $u_{-1/2}=\frac{1}{\sqrt{2}}\(i\tilde u_++\tilde u_-\)$, or more explicitly,
\ba
u_\tau(\theta)=\frac{e^{i \pi/4}}{\sqrt{4m(m+E)}}\(\begin{array}{c}
(E+m)e^{-i  \tau (\theta+\pi/2)}\\
(E+m)e^{-i  \tau (\theta-\pi/2)}\\
pe^{i  \tau (\theta-\pi/2)}\\
-pe^{i  \tau (\theta+\pi/2)}
\end{array}\)\,.
\ea
Now for the vector polarizations, those can be constructed out of the spinors directly in transversity as follows,
\ba
\epsilon_{\tau=\pm 1}^\mu = -\frac{1}{\sqrt{2}}\bar v_{\tau/2}\ \gamma^\mu\ u_{\tau/2}\,, \qquad
\epsilon_{\tau=0}^\mu = -\frac{1}{2}\(\bar v_{1/2}\ \gamma^\mu \ u_{-1/2}+\bar v_{-1/2}\ \gamma^\mu\ u_{1/2}\)\,,
\ea
or more explicitly,
\ba
\epsilon_{\tau = \pm 1}^\mu(\theta)
&=& \frac{i}{\sqrt{2} m} \(  p , \;\;  E \sin \theta  \pm i m \cos \theta  , \;\; 0 ,  E \cos \theta \mp i m \sin \theta  \)  \\
\epsilon_{\tau=0}^\mu (\theta)  &= &  \( 0 , \;\; 0 , \;\;  1  ,\;\; 0   \) \,,
\ea
which satisfy the correct transversity relations, $p_\mu \epsilon^\mu_\tau=0$, $\eta\mn \epsilon^\mu_\tau (\epsilon^\nu_{\tau'})^*=\delta_{\tau \tau'}$ and $\sum_\tau \epsilon ^\mu_\tau \, (\epsilon^\nu_\tau)^*=\eta^{\mu\nu}{+}p^\mu p^\nu/m^2$, with $\epsilon_{\tau}=(-1)^\tau\, \epsilon_{-\tau}^*$. Following the same logic, we could go ahead and construct the transversity states for particles of any spin.

\subsection{Multispinors}

Although our interest is in the scattering of particles which are irreducible representations of the Poincar\'e group, the crossing formula apply equally for scattering of unphysical states which sit in reducible representations\footnote{Although reducible representations will necessarily contain ghost states, e.g. $1/2 \otimes 1/2 = 3 \oplus 1 $(ghost) these states will still have to respect crossing symmetry in the same way and so this need not concern us in deriving the crossing relations.}, specifically those that are given as tensor products of spin 1/2 representations. This is because all that is relevant is the transformation of states under complex Lorentz transformations. Since it is significantly easier to derive the crossing formula for arbitrary reducible spin states we shall do so here. The wavefunction for such a state of maximal spin $2S$ will be given by a multispinor with $2S$ components \cite{Bargmann:1948ck,Schwinger:1966zz,Scadron:1968zz,PhysRev.161.1316,Schwinger:1978ra}
\be
\psi_{\alpha_1 \dots \alpha_{2S}}^{\tau_1 \dots \tau_{2S}}  = \Pi_{i=1}^{2S} u^{\tau_i}_{\alpha_i}(x) \, ,
\ee
and the associated quantum field for particle $A$ takes the form
\be
\Psi^A_{\alpha_1 \dots \alpha_{2S}} = \sum_{\tau_i} \int \frac{\d^3 k }{(2 \pi)^3  2 \omega_k} \left[   \left( \Pi_{i=1}^{2S} u^{\tau_i}_{\alpha_i}(k) \right) \hat a_{A,\tau_1 \dots \tau_{2 S}}(k) + \left(   \Pi_{i=1}^{2S} v^{\tau_i}_{\alpha_i}(k) \right) \hat a^{\dagger}_{\bar A, \tau_1 \dots \tau_{2 S}}(k)  \right] \, .
\ee
Here $\hat a_{A,\tau_1 \dots \tau_{2 S}}(k) $ annihilates a particle $A$ with transversity string $\tau_1 \dots \tau_{2S}$ corresponding to a reducible representation of total transversity $\tau = \tau_1 + \dots \tau_{2S}$, and $\hat a^{\dagger}_{\bar A,\tau_1 \dots \tau_{2 S}}(k) $ creates the associated anti-particle $\bar A$ and the canonical normalization is such that
\be
[\hat a_{A,\tau_1 \dots \tau_{2 S}}(k), \hat a^{\dagger}_{B,\tau_1' \dots \tau_{2 S}'}({ k'})] = 2 \omega_k \delta_{AB} \Pi_{i=1}^{2 S} \delta_{\tau_i \tau_i'} { (2\pi)^3 \delta^3(k-k')} \, .
\ee

Consider an interaction process $A+B \rightarrow C+D$ in which each particle is represented by a multispinor field. By angular momentum conservation, $S_1+S_2=S_3+S_4+\text{integer}$. Let us assume for now that the overall integer is positive, so that there are more incoming spinor indices than outgoing. In this case, to construct a scalar interaction we will need to contract incoming indices, which is naturally achieved with the charge conjugation matrix since
\be
u^{\tau_1}_{\alpha}(k_1) C_{\alpha \beta} u^{\tau_2}_{\beta}(k_2) = \bar v^{\tau_1}(k_1) u^{\tau_2}(k_2)
\ee
To begin with, let us consider the case $S_3 \ge S_1$, $S_2 \ge S_4$ and define $L=S_1+S_2-S_3-S_4$. With this choice, we can consider the following interaction Lagrangian

\ba
{\cal L} &=& \sum_I  \left( \bar \Psi_C^{\alpha_1 \dots \alpha_{2S_1-L} \gamma_1 \dots \gamma_{2S_3-2S_1+L}} \hat O^I \Psi^A_{\alpha_1 \dots \alpha_{2S_1}} \right) \\
&&\times  C^{\alpha_{2S_1-L+1} \rho_1} \dots C^{\alpha_{2S_1} \rho_L} \left( \bar \Psi_D^{\beta_1 \dots \beta_{2S_4} } \hat O_I \Psi^B_{\beta_1 \dots \beta_{2S_4} \gamma_1 \dots \gamma_{2S_3-2S_1+L} \rho_1 \dots \rho_L}\right) \, .\nn
\ea
Here $\bar \Psi^{\alpha_1 \alpha_2 \dots } = \Psi^{\dagger}_{\beta_1 \beta_2 \dots} (\gamma^0)^{\alpha_1 \beta_1 }(\gamma^0)^{\alpha_2 \beta_2 } \dots$ is the multispinor generalization of the Dirac conjugate \cite{Doughty:1986ya}, and the operators $\hat O$ are tensor products of operators acting in the ${\rm Min}(2S_1,2S_4)$ dimensional spinor space. The simplest choice $\hat O=1$ will preserve total transversity $\tau_1+\tau_2=\tau_3+\tau_4$ and so to be more general we will consider operators of the form
\be
\hat O_I = \gamma_{\mu_1} \otimes \gamma_{\mu_2} \dots 1 \otimes 1 \, , \quad \hat O^I = \gamma^{\mu_1} \otimes \gamma^{\mu_2} \dots 1 \otimes 1 \,
\ee
and $\sum_I = \sum_{\mu_1,\mu_2 \dots}$ and we will assume the number of $\gamma$ factors is $\kappa \le {\rm Min}(2S_1,2S_4)$. \\

The s-channel scattering amplitude is schematically given by (we take the initial state in the form $\hat a^{\dagger}_B \hat a^{\dagger}_A|0 \rangle $ and the final state $\hat a^{\dagger}_D \hat a^{\dagger}_C|0 \rangle $)
\ba
T_{\tau_1\tau_2\tau_3\tau_4}^s(s,t,u) &=&  \eta_{DC} \eta_{DA}  \left( \bar u(k_3) \gamma_{\mu} u(k_1)\bar u(k_4) \gamma^{\mu} u(k_2) \right)^{\kappa} \times \\
&& \left( \bar u(k_3) u(k_1) \right)^{2S_1-\kappa-L} \left( \bar u(k_3) u(k_2) \right)^{2(S_3-S_1)+L} \left(  \bar u(k_4) u(k_2) \right)^{2S_4-\kappa} (\bar v(k_1) u(k_2))^L\, , \nn
\ea
where $ \eta_{DC} \eta_{DA} $ are the usual statistics factors arising from reordering creation and annihilation operators ($\eta_{ij}=-1$ if both particles $i$ and $j$ are fermions and $+1$ otherwise).
More precisely
\be
 \eta_{DC} \eta_{DA} = \langle 0 | \hat a_C \hat a_D  \hat a^{\dagger}_C \hat a_A  \hat a^{\dagger}_D \hat a_B   \hat a^{\dagger}_B\hat a^{\dagger}_A | 0 \rangle \, .
\ee
In writing this short hand, we understand $\left( \bar u(k_3) u(k_1) \right)^{2S_1-\kappa-L} $ to mean a product of $2S_1-\kappa-L$ factors in which each $u$ and $\bar u$ has a distinct transversity, similarly for the other factors. The total transversity for particle $A$ is the sum of all the individual $\tau_1$'s for each element $u_{\tau_1}(k_1)$ in the string, and similarly for the other particles $\tau_i$.
 \\

The $u$-channel amplitude for the same process is
\ba
T^u_{\tau_1\tau_2\tau_3\tau_4}(s,t,u) &=& \eta_{AB} \eta_{BD} \eta_{BC} \left( \bar u(k_3) \gamma_{\mu} u(k_1)\bar v(k_2) \gamma^{\mu} v(k_4) \right)^{\kappa} \times  \\
&& \left( \bar u(k_3) u(k_1) \right)^{2S_1-\kappa-L} \left( \bar u(k_3) v(k_4) \right)^{2(S_3-S_1)+L} \left(  \bar v(k_2) v(k_4) \right)^{2S_4-\kappa} (\bar v(k_1) v(k_4))^L \, , \nn
\ea
in the same shorthand notation where the statistics factor is now
\be
 \eta_{AB} \eta_{BD} \eta_{BC}= \langle 0 | \hat a_C \hat a_{\bar B}  \hat a^{\dagger}_C \hat a_A  \hat a_{\bar D} \hat a^{\dagger}_{\bar B}   \hat a^{\dagger}_{\bar D}\hat a^{\dagger}_A | 0 \rangle \, .
\ee

Consider then the ratio
\be
\frac{T_{\tau_1\tau_2\tau_3\tau_4}^s(s,t,u) }{T^u_{-\tau_1-\tau_4-\tau_3-\tau_2}(u,t,s)} =\eta_{DC} \eta_{DA} \eta_{AB} \eta_{BD} \eta_{BC} \, F_1^{\kappa} F_2^{2S_1-\kappa-L} F_3^{2(S_3-S_1)+L} F_4^{2S_4-\kappa} F_5^L \, ,
\ee
where being more explicit about indices for individual factors
\ba
&& F_1 = \frac{\bar u_{\tau_3}(k_3) \gamma_{\mu} u_{\tau_1}(k_1)\bar u_{\tau_4}(k_4) \gamma^{\mu} u_{\tau_2}(k_2)}{\left[\bar u_{-\tau_3}(k_3) \gamma_{\mu} u_{-\tau_1}(k_1)\bar v_{-\tau_4}(k_2) \gamma^{\mu} v_{-\tau_2}(k_4) \right] |_{s \leftrightarrow u}} \, ,\\
&& F_2 = \frac{\bar u_{\tau_3}(k_3)  u_{\tau_1}(k_1)}{\left[\bar u_{-\tau_3}(k_3) u_{-\tau_1}(k_1) \right] |_{s \leftrightarrow u}}  \, ,\\
&& F_3 = \frac{\bar u_{\tau_3}(k_3)  u_{\tau_2}(k_2)}{\left[\bar u_{-\tau_3}(k_3) v_{-\tau_2}(k_4) \right] |_{s \leftrightarrow u}} \, , \\
&& F_4 =  \frac{\bar u_{\tau_4}(k_4)  u_{\tau_2}(k_2)}{\left[\bar v_{-\tau_4}(k_2) v_{-\tau_2}(k_4) \right] |_{s \leftrightarrow u}} \, , \\
&& F_5 = \frac{\bar v_{\tau_1}(k_1) u_{\tau_2}(k_2)}{\left[ \bar v_{-\tau_1}(k_1) v_{-\tau_2}(k_4)\right]  |_{s \leftrightarrow u}} \, .
\ea
The last 4 factors are easiest to determine as they are diagonal. We note that given our conventions
\be
\bar u_{\tau_3} (k_3) u_{\tau_1}(k_1) =\sqrt{\frac{s+u}{4m^2}} e^{i \tilde \chi (\tau_1+\tau_3)} \delta_{\tau_1 \tau_3}
\ee
where
\be
e^{i \tilde \chi(s,t,u)} = \frac{\sqrt{-su} + \frac{is}{2m}\sqrt{-t}}{\sqrt{s(s-4m^2)} \sqrt{\frac{s+u}{4m^2}}} \, .
\ee
Similarly
\be
\bar u_{\tau_4} (k_4) u_{\tau_2}(k_2) =\sqrt{\frac{s+u}{4m^2}} e^{i \tilde \chi (\tau_2+\tau_4)} \delta_{\tau_2 \tau_4} \, , \quad
\bar v_{\tau_4} (k_4) v_{\tau_2}(k_2) =-\sqrt{\frac{s+u}{4m^2}} e^{i \tilde \chi (\tau_2+\tau_4)} \delta_{\tau_2 \tau_4} \, .
\ee
Since
\be
e^{i \tilde \chi(s,t,u)+ i \tilde \chi(u,t,s)} = \frac{\sqrt{u(u-4m^2)}}{\sqrt{s(s-4m^2)}} \left(\frac{\sqrt{-su} + \frac{is}{2m}\sqrt{-t}}{\sqrt{-su} - \frac{iu}{2m}\sqrt{-t}} \right) = e^{-i \chi_u(s,t,u)}
\ee
then we infer that
\be
F_2 =- e^{i \pi (\tau_1+\tau_{ 3})} e^{-i \chi_u(\tau_1+\tau_3)}  \delta_{\tau_1 \tau_3}\, , \quad F_4  = e^{i \pi (\tau_2+\tau_4)}  e^{-i \chi_u(\tau_2+\tau_4)}  \delta_{\tau_2 \tau_4}\, .
\ee
To determine $F_3$ we need
\be
\bar u_{\tau_3} (k_3) u_{\tau_2}(k_2) = \sqrt{1-\frac{u}{4m^2}} \left[ \frac{\sqrt{\frac{-t}{s-4m^2}}-\frac{i\sqrt{s}}{2m} \sqrt{\frac{-u}{s-4m^2}}}{\sqrt{1-\frac{u}{4m^2}}}\right]^{\tau_3+\tau_2} \delta_{\tau_3 \tau_2} \, ,
\ee
and
\be
\bar u_{-\tau_3} (k_3) v_{-\tau_2}(k_4) = \frac{i}{2m} \sqrt{s-4m^2} \, e^{\frac{i \pi}{2} (\tau_2+\tau_3)}  \delta_{\tau_2 \tau_3}=- \frac{1	}{2m} \sqrt{4m^2-s} \, e^{\frac{i \pi}{2} (\tau_2+\tau_3)}  \delta_{\tau_2 \tau_3}
\ee
where in the last step we have analytically continued from $s>4m^2$ to $s<4m^2$ via an anticlockwise contour which avoids right hand branch cut corresponding to $\sqrt{s-4m^2}  \rightarrow i \sqrt{4m^2-s}$.
Then
\ba
F_3 &=& - e^{\frac{-i \pi}{2} (\tau_2+\tau_3)} \left[ \frac{\sqrt{\frac{-t}{s-4m^2}}-\frac{i\sqrt{s}}{2m} \sqrt{\frac{-u}{s-4m^2}}}{\sqrt{1-\frac{u}{4m^2}}}\right]^{\tau_3+\tau_2}\delta_{\tau_2 \tau_3}  \, , \\
&=& - e^{i \pi(\tau_3+\tau_2)} e^{-i \chi_u (\tau_3+\tau_2)} \delta_{\tau_2 \tau_3} \, .
\ea
To compute $F_5$ we need
\be
\bar v_{\tau_1}(k_1) u_{\tau_2}(k_2) = - 2 \tau_1 \frac{1}{2 m}\sqrt{s-4m^2}\delta_{\tau_1 \tau_2}
\ee
and
\be
\bar v_{-\tau_1}(k_1) v_{-\tau_2}(k_4) =  2 \tau_1  \left(  \frac{(2 \tau_1) \sqrt{-t}+ \frac{i}{2m} \sqrt{-su}}{\sqrt{s-4m^2}}\right)  \delta_{\tau_1 \tau_2}\,.
\ee
Again analytically continuing this into the region $s<4m^2$ along an anticlockwise contour that avoids the right hand branch cut then
\be
F_5 =  \frac{\sqrt{s-4m^2}}{2 i m \left( \frac{(2 \tau_1) \sqrt{-t}+ \frac{i}{2m} \sqrt{-su}}{\sqrt{4m^2-u}}\right) } \delta_{\tau_1 \tau_2} = e^{i \pi (\tau_1+\tau_2)} e^{-i (\tau_1+\tau_2) \chi_u} \delta_{\tau_1 \tau_2} \, .
\ee
Finally after some algebra (checking all $2^4$ combinations)
\be
F_1 = -e^{i \pi (\tau_1+\tau_2+\tau_3+\tau_4)} e^{-i (\sum_i \tau_i ) \chi_u} \delta_{(\tau_1+\tau_4)(\tau_2 +\tau_3)} \, .
\ee
In the above we have calculated each ratio for a given spinor factor. The full amplitude is a product of such factors, however we have written each factor so that the product is straightforward to take\footnote{For instance the factor $e^{i \pi (\tau_1+\tau_2+\tau_3+\tau_4)}$ in $F_1$ is always unity, but including it allows us to combine $F_1$ into the general answer.} with the result that
\ba
F_1^{\kappa} F_2^{2S_1-\kappa-L} F_3^{2(S_3-S_1)+L} F_4^{2S_4-\kappa} F_5^L  &=& (-1)^{\kappa} (-1)^{2S_1-\kappa-L} (-1)^{2(S_3-S_1)+L} \ e^{i \pi \sum_i \tau_i }e^{-i (\sum_i \tau_i ) \chi_u}  \nn \\
&=&(-1)^{2S_3}e^{i \pi \sum_i \tau_i }e^{-i (\sum_i \tau_i ) \chi_u} \, ,
\ea
where $\tau_i $ now denotes the total transversity which is a sum of all the $\tau_i$ transversities for each element in the string. \\

Thus the crossing relation takes the form
\be
T_{\tau_1\tau_2\tau_3\tau_4}^s(s,t,u)  =\eta_u' e^{i \pi \sum_i \tau_i }e^{-i  \chi_u \sum_i \tau_i} T^u_{-\tau_1-\tau_4-\tau_3-\tau_2}(u,t,s) \, ,
\label{eq:crossingm}
\ee
where
\be
\eta_u'=\eta_{AB}\eta_{BC}  \eta_{CD} \eta_{DA}  \eta_{BD}  (-1)^{2S_3} \, .
\label{eq:etaup}
\ee
Although this was derived under the assumption $S_3 \ge S_1$, $S_2 \ge S_4$ and $S_1+S_2 \ge S_3+S_4$, the final result cannot depend on this fact, which can be demonstrated by recomputing for interactions in the opposite case following the same procedure.
Focusing on the elastic (for spins but not necessarily transversities) case $S_1=S_3$ and $S_2=S_4$ this becomes simply
\be
T_{\tau_1\tau_2\tau_3\tau_4}^s(s,t,u) =(-1)^{2(S_1+S_2)} e^{i \pi \sum_i \tau_i }e^{-i  \chi_u  \sum_i \tau_i  } T^u_{-\tau_1-\tau_4-\tau_3-\tau_2}(u,t,s)\, .
\ee
Finally for elastic transversities as well $\tau_1=\tau_3$, $\tau_2=\tau_4$ this is
\be
T_{\tau_1\tau_2\tau_1 \tau_2}^s(s,t,u) =e^{-i  \chi_u \sum_i \tau_i  } T^u_{-\tau_1-\tau_2-\tau_1-\tau_2}(u,t,s)\, .
\ee
which is the result needed in the main text.

\section{Explicit Examples}
\label{app:examples}

In this Appendix we give some of the simplest scattering amplitudes between spin-0, spin-1/2 and (now irreducible!) spin-1 particles, and show that they behave exactly as expected with regards to their kinematic singularities and crossing properties. We consider different types of four-point interactions and confirm their analyticity and crossing relations.

\subsection{Scalar-Scalar}

The simplest four-scalar interaction in this context is a $\lambda \phi^4$ interaction, which gives a trivial tree-level scattering amplitude $A^s (s,t,u) = \lambda$.
The  analyticity and crossing property are then manifestly trivial,
\ba
A^{s} ( s, t, u) = A^t ( t, s, u  )  = A^u ( u, t , s ) \,.
\ea

\subsection{Scalar-Spinor}

Next we turn to scalar-spinor interactions and start with the four-point of the form  $\lambda \psi \bar \psi \phi \phi  $ interaction, where for simplicity we consider scattering between four distinct particles (\ie the two scalars are distinct -- even if they carry the same mass -- and so are the two fermions). \\

\noindent {\it $\bullet$ $\psi  \phi  \to \psi \phi $ scattering}: Let us consider the following interaction in the Lagrangian,
\ba
\L_{\rm int}= \lambda \bar \psi_C \psi_A \phi_B \phi_D^{\dagger}\,.
\ea
Then  the $s$ channel transversity amplitudes for the scattering process $\psi_A \phi_B \to  \psi_C \phi_D$ is
\ba
\mathcal{T}^s_{\tau_1 0 \tau_3 0}(s,t,u)=\mathcal{T}^{\psi \phi \to \psi \phi}_{\tau_1 0 \tau_3 0} = \lambda \bar u_{\tau_3}(\theta_3) u_{\tau_1}(\theta_1)=
\frac{\lambda s}{\sqrt{-s u}\sqrt{\Sc}}\(-u +i\tau_1 \frac{\sqrt{s t u}}{ m}\)\, \delta_{\tau_1 \tau_3}
\,,
\ea
where the angles $\theta_i$ are given in \eqref{eq:theta_i}. \\

\paragraph{Analyticity:}
First we can clearly see that the combinations $\mathcal{T}^{\pm}$ defined in \eqref{eqn:T+} and \eqref{eqn:T-} (with $\xi=1$ since we are dealing with BF scattering) are explicitly analytic in $s, t$ and $u$ and free of all kinematical singularities as argued in section~\ref{sec:trfm}. Indeed for the $u$ channel scattering process $\psi \phi \to \psi \phi$,
\ba
\mathcal{T}^{+}_{\tau_1 0 \tau_3 0}(s,t,u)&=&-2\lambda s u  \, \delta_{\tau_1 \tau_3}\\
\mathcal{T}^{-}_{\tau_1 0 \tau_3 0}(s,t,u)&=&\frac{2\tau_1 \lambda}{m} \,  s^2 t u  \, \delta_{\tau_1 \tau_3}\,.
\ea

\paragraph{Crossing:}
Next, when it comes to the crossing relation, we can see that  the $u$ channel transversity amplitude is identical to the $s$ channel,
\ba
\mathcal{T}^u_{\tau_1 0 \tau_3 0}(s,t,u) = \mathcal{T}^s_{\tau_1 0 \tau_3 0}(s,t,u)\,,
\ea
and so for any $\tau_1=\pm 1/2$ and any $\tau_3=\pm 1/2$ the following relation is identically satisfied,
\ba
\mathcal{T}^s_{\tau_1 0 \tau_3 0}(s,t,u) = e^{- i \chi_u (\sum \tau_i) }  \mathcal{T}^u_{-\tau_1 0 -\tau_3 0}(u,t,s)\,,
\ea
with
\ba
e^{-i \chi_u}=\frac{\mathcal{T}^s_{+ 0 + 0}(s,t,u)}{\mathcal{T}^u_{- 0 - 0}(u,t,s)}
=\frac{s}{u}\sqrt{\frac{\Uc}{\Sc}}\frac{u-\frac{i}{2m}\sqrt{s t u}}{s + \frac{i}{2 m}\sqrt{s t u}}
=\frac{1}{\sqrt{\Sc \Uc }}\(- s u +2 i m \sqrt{s t u}\)\,,
\ea
which is precisely the crossing relation \eqref{eq:crossingm} derived in appendix~\ref{app:multispinor} with in this case, $\eta_u'=(-1)^{2S_3}=-1$, $e^{i \pi \sum \tau_i}=e^{2i \pi \tau_1}=-1$ and with  the angle $\chi_u$ precisely as in \eqref{eq:anglephi} or \eqref{chianglesame} \,.\\

\noindent {\it $\bullet$ $\phi \phi \to \bar \psi  \psi$ scattering}: We now consider the interaction $\L_{\rm int}=\lambda \phi_A \phi_B \bar \psi_C \psi_{\bar D}$.
The $s$ channel transversity amplitudes for the process $\phi_A \phi_B \to  \psi_C \psi_D$ is then
\ba
\mathcal{T}^s_{0 0 \tau_3 \tau_4}(s,t,u)=
\mathcal{T}^{\phi \phi \to \bar \psi  \psi}_{0 0 \tau_3 \tau_4} &=&
\lambda\ \langle 0 |a_C a_D
\(\bar \psi_C \psi_{\bar D} \phi_A \phi_B\)
a_B^\dagger a_A^\dagger|0\rangle \\
&=&\lambda \bar u_{\tau_3}(\theta_3) v_{\tau_4}(\theta_4)\
\langle 0 |a_C a_D
a_C^\dagger a_D^\dagger |0\rangle \\
&=& -  \lambda\bar u_{\tau_3}(\theta_3) v_{\tau_4}(\theta_4) \\
&=& - \frac{ \lambda \tau_3}{m}\sqrt{\frac{\Sc}{s}}\, \delta_{\tau_3,  \tau_4}\,.
\ea
The combinations $\mathcal{T}^{\pm}$ are again manifestly analytic (note that since we are not dealing with an elastic process one should use the prefactor introduced in \eqref{eq:pref4m2} instead of that in \eqref{eq:pref4m22}),
\ba
\mathcal{T}^{+}_{0 0 \tau_3 \tau_4}=-\frac{2\lambda \tau_3}{m} (s-4m^2) \delta_{\tau_3, \tau_4} \qquad{\rm and }\qquad
\mathcal{T}^{-}_{0 0 \tau_3 \tau_4}=0\,.
\ea
The corresponding $u$ channel $\phi_A \psi_{\bar D} \to \psi_{C} \phi_B$ transversity amplitude is
\ba
\mathcal{T}^u_{0 \tau_2 \tau_3 0}(s,t,u)=\mathcal{T}^{\phi \bar \psi \to \bar \psi \phi}_{0 \tau_2 \tau_3 0}&=& \lambda\
\langle 0 |a_C a_B
\(\bar \psi_C \psi_{\bar D} \phi_A \phi_B\)
a_{\bar D}^\dagger a_A^\dagger|0\rangle \\
&=&  \lambda  \bar u_{\tau_3}(\theta_3) u_{\tau_2}(\theta_2) \\
&=&\frac{-i \lambda }{\sqrt{\Sc}\sqrt{u}}\(\sqrt{stu}- i \tau_3 \frac{su}{m}\)\, \delta_{\tau_3,  \tau_2}\,,
\ea
where we have used the analytic continuation appropriate for this case\footnote{To be more specific, we define $\sqrt{-u}$ as $\sqrt{-u}=\sqrt{s-x}$, with $x=4m^2-t>0$. The physical region being defined for $s>4m^2$, as the limit from the upper half of the complex plane for $s$. Then defining $s=x+\rho e^{i \theta}$, we have $\sqrt{s-x}=\sqrt{\rho} e^{i \theta/2}$. Then analytically continuing from $\theta=0^+$ to $\theta=\pi^-$, we have $\sqrt{-u}=\sqrt{s-x}=i \sqrt{\rho}=i \sqrt{x-s}=i \sqrt{u}$.}, $\sqrt{-u}=i \sqrt{u}$.
Then the crossing relation gives,
\ba
\mathcal{T}^s_{0 0 \tau_3 \tau_4}(s,t,u)= - e^{-i \chi_u (\tau_3+\tau_4)} \mathcal{T}^u_{0 -\tau_4 -\tau_3 0}(u,t,s)\,,
\ea
which precisely matches again the crossing relation \eqref{eq:crossingm} where this time we have $\eta_u'=\eta_{CD}(-1)^{2S_3}=1$ and $e^{i \pi \sum \tau_i}=-1$. \\

\subsection{Spinor-Spinor}

Next, we may look at spinor-spinor four-point interactions. The simplest one is of the form  $\lambda (\bar \psi \psi)^2$ where we consider again distinct spinors more specifically,
\ba
\L_{\rm int}= \lambda \, \(\bar \psi_C \psi_A\) \(\bar \psi_D \psi_B\)\,.
\ea
The scattering process $\psi_A \psi_B \to \psi_C \psi_D$  has amplitude,
\ba
\mathcal{T}^s_{\tau_1 \tau_2 \tau_3 \tau_4}(s,t,u)&=& \lambda\ \langle 0 |a_C a_D
 \(\bar \psi_C \psi_A\) \(\bar \psi_D \psi_B\)
a_B^\dagger a_A^\dagger|0\rangle \\
&=& \lambda \(\bar u_{\tau_3}(\theta_3) u_{\tau_1}(\theta_1)\)\(\bar u_{\tau_4}(\theta_4) u_{\tau_4}(\theta_4)\)\\
&=&
\lambda \, \delta_{\tau_1, \tau_3}\delta_{\tau_2,\tau_4}\left\{
\begin{array}{ccc}
\frac{1}{\Sc}\(\sqrt{-s u}+\frac{i \tau_1}{m}s\sqrt{-t}\)^2 & \quad& {\rm if } \ \tau_1=\tau_2\\
\frac{s+u}{4m^2} & \quad& {\rm if } \ \tau_1\ne \tau_2\\
\end{array}
\right.\,.
\ea
In this case we manifestly have $\eta_u'=1$ and when $\tau_1\ne \tau_2$ we have $\sum \tau_i=0$ (since the amplitude is proportional to $ \delta_{\tau_1, \tau_3}\delta_{\tau_2,\tau_4}$) so the crossing and analyticity relations are trivially satisfied in that case.

On the other hand, when $\tau_1=\tau_2$, then $\sum \tau_i=4\tau_1=\pm 2$, and we have
\ba
\mathcal{T}^s_{\pm \pm \pm \pm }(s,t,u)&=& \frac{1}{\Sc \Uc} \(s u \mp 2i m \sqrt{s t u}\)^2  \mathcal{T}^u_{\mp \mp \mp \mp}(u,t,s) \\
&=& e^{-\pm 2 i \chi_u }\mathcal{T}^u_{\mp \mp \mp \mp}(u,t,s)\,,
\ea
as found in appendix~\ref{app:multispinor}. \\

To make things slightly more interesting, we can also consider the following  interaction $ (\bar \psi_C \gamma^\mu \psi_A) (\bar \psi_D \gamma_\mu \psi_B)$, where again we consider all the fields to be distinct even though they carry the same mass and spin.
Such an interaction then leads to an $s$ channel transversity amplitude which is non-trivial when $\{\tau_3, \tau_4\}\ne \{\tau_1, \tau_2\}$. Using the expression of the anti-spinor in terms of the spinors \eqref{eq:antispinor} $\bar \psi_D \gamma_\mu \psi_B = - \bar \psi_{\bar B} \gamma_\mu \psi_{\bar D}$, we can easily see that the transversity amplitude in the $s$ and $u$ channels associated to the scattering process $\psi_A  \psi_B \to \psi_C \psi_D$ are simply related as
\ba
\mathcal{T}^s_{\tau_1 \tau_2 \tau_3 \tau_4}(s,t,u)=(\bar u_{\tau_3}(\theta_3). \gamma^\mu . u_{\tau_1}(\theta_1)) (\bar u_{\tau_4}(\theta_4). \gamma_\mu . u_{\tau_2}(\theta_2)) =-\mathcal{T}^u_{\tau_1 \tau_2 \tau_3 \tau_4}(s,t,u)\,.
\ea
\begin{itemize}
\item Starting with the cases where $\sum \tau_i=0$, the simplest one is when $\tau_1=-\tau_2=\tau_3=-\tau_4$. This case we have
    \ba
    \mathcal{T}^s_{\pm \mp \pm \mp}(s,t,u)=\frac{u-s}{4m^2}\,,
    \ea
so the associated combination $\mathcal{T}^+$ given in \eqref{eqn:T+} is clearly analytic, while the $\mathcal{T}^-$ vanishes. Crossing symmetry is also trivial in that case, and satisfies \eqref{eqn:T24} with $\eta_u'=1$,
\ba
\mathcal{T}^s_{\pm \mp \pm \mp}(s,t,u)=-\mathcal{T}^u_{\mp \pm \mp \pm}(s,t,u)=\mathcal{T}^u_{\mp \pm \mp \pm}(u,t,s)\,.
\ea
\item Still considering the case where $\sum \tau_i=0$, one can look at $\tau_1=-\tau_2=-\tau_3=\tau_4$ or the crossing equivalent $\tau_2 \leftrightarrow  \tau_4$  which simply leads to
\ba
\mathcal{T}^s_{\pm \mp \mp \pm}(s,t,u)=-\mathcal{T}^s_{\pm \pm \mp \mp}(s,t,u)=\frac{t}{4m^2}\,.
\ea
The analyticity of the associated combinations  $\mathcal{T}^\pm$ defined in (\ref{eqn:T+},\ref{eqn:T-}) are trivially satisfied and so are the crossing relations derived in \eqref{eqn:T24},
\ba
\mathcal{T}^s_{\pm \mp \mp \pm}(s,t,u)&=&\mathcal{T}^u_{\mp \mp \pm \pm}(u,t,s)\\
\mathcal{T}^s_{\pm \pm \mp \mp}(s,t,u)&=&\mathcal{T}^u_{\mp \pm \pm \mp}(u,t,s)\,.
\ea
\item In the case where $\sum \tau_i=\pm 1$, all the scattering amplitudes vanish so we are just left with the case where $\sum \tau_i=\pm 2$,
\ba
\mathcal{T}^s_{\pm \pm \pm \pm}(s,t,u)=-\frac{s-u}{4m^2}-\frac{2 s t}{\Sc}\mp \frac{i s}{m \Sc}\sqrt{s t u}\,,
\ea
and we clearly see that both combinations $\mathcal{T}^\pm$ defined in (\ref{eqn:T+},\ref{eqn:T-}) are analytic. Moreover, the $u$ channel is given by
\ba
\mathcal{T}^u_{\mp \mp \mp \mp}(u,t,s)=-\mathcal{T}^s_{\mp \mp \mp \mp}(u,t,s)=
-\frac{s-u}{4m^2}+\frac{2 u t}{\Uc}\mp \frac{i u}{m \Uc}\sqrt{s t u}\,,
\ea
and the crossing relation \eqref{eqn:T24} is again confirmed
\ba
\mathcal{T}^u_{\mp \mp \mp \mp}(s,t,u)=e^{-2 i \chi_u} \mathcal{T}^u_{\mp \mp \mp \mp}(u,t,s)\,.
\ea
This concludes all the different cases one can consider for the scattering amplitude associated with the 4-spinor interaction $(\bar \psi \gamma_u \psi)^2$.
\end{itemize}

\subsection{Scalar-Vector}

Finally, we shall consider a scalar-vector example. The simplest four-point scalar-vector interaction in this context is $ {A_C^\mu}^{\dagger} A^A_\mu \phi_D^{\dagger} \phi_B$, where again we consider all particles to be distinct and the two vectors $A_\mu$ carry the same mass $m$. \\

\noindent {\it $A_A \phi_B \to A_C \phi_D$ scattering:} The $s$ channel transversity amplitudes for this process are
\ba
\mathcal{T}^s_{\tau_1 0 \tau_3 0} (s,t,u)&=&  \epsilon_{\mu}^{\tau_1}(\theta_1) \({\epsilon^{\mu}_{\tau_3}(\theta_3)}\)^* \\
 &=&  \( \begin{array}{c c c}
\frac{1}{\Sc}\(\sqrt{- s u}-\frac{i s}{2m}\sqrt{-t}\)^2 &\quad  0 \quad & \frac{t}{4m^2} \\
0 & 1 & 0 \\
\frac{t}{4m^2} & 0 & \frac{1}{\Sc}\(\sqrt{- s u}+\frac{i s}{2m}\sqrt{-t}\)^2
\end{array} \)\,.
\label{eqn:Tscalarvector}
\ea
The corresponding $u$ channel obviously satisfies $\mathcal{T}^u_{\tau_1 0 \tau_3 0} (s,t,u)=\mathcal{T}^s_{\tau_1 0 \tau_3 0} (s,t,u)$.

\begin{itemize}
\item Starting again with the case where $\sum \tau_i=0$, the non-trivial amplitudes are either for $\tau_1=\tau_3=0$ (for which $\mathcal{T}^s=\mathcal{T}^u=1$) or  $\tau_1=-\tau_3=\pm1$ (for which $\mathcal{T}^s=\mathcal{T}^u=t/4m^2$). In either cases the analyticity and crossing relations are trivially satisfied.
\item On the other hand, when  $\sum \tau_i =\pm 2$, \ie when $\tau_1=\tau_3=\pm 1$, then
\ba
\mathcal{T}^u_{\pm 0 \pm 0} (u,t,s)=\frac{1}{\Uc}\(\sqrt{- s u}\pm\frac{i  u}{2m}\sqrt{-t}\)^2\,,
\ea
and so
\ba
\mathcal{T}^s_{\pm 0 \pm 0} (s,t,u)= e^{\mp 2 i \chi_u}\mathcal{T}^u_{\mp 0 \mp 0} (u,t,s)\,,
\ea
as it should be.
\end{itemize}

\noindent {\it $A_A A_B \to \phi_C \phi_D$ scattering:} Now considering an interaction  $ A_B^\mu A^A_\mu \phi_D^{\dagger} \phi_C^{\dagger}$, the $s$ channel amplitude is then
\ba
\mathcal{T}^s_{\tau_1 \tau_2 0 0} (s,t,u) = \epsilon^{\mu}_{\tau_1}(\theta_1)  \epsilon_{\mu}^{\tau_2}(\theta_2) =\left( \begin{array}{c c c}
\frac{\Sc}{4m^2 s} & \quad 0 \quad & \frac{s}{4m^2}  \\
0 & 1 & 0 \\
\frac{s}{4m^2}  & 0 & \frac{\Sc}{4m^2 s}
\end{array} \right)\,,
\ea
while the corresponding $u$ channel  $A \phi \to \phi A$ is given by
\ba
\mathcal{T}^u_{\tau_1 0 0 \tau_2} (s,t,u) =\epsilon^{\mu}_{\tau_1}(\theta_1)   \(\epsilon_{\mu}^{\tau_2}(\theta_4)\)^* = \left( \begin{array}{c c c}
\frac{\Uc}{4 m^2 u}e^{-2 i \chi_u} & \quad 0 \quad & \frac{u}{4m^2}  \\
0 & 1 & 0 \\
\frac{u}{4m^2}  & 0 & \frac{\Uc}{4 m^2 u}e^{2 i \chi_u}
\end{array} \right)\,.
\ea
In this case $S_3=0$ and since we are only dealing with bosons, $\eta_u'=1$. Moreover since $\mathcal{T}^s_{\pm 1 0 0 0}=\mathcal{T}^s_{ 0 \pm 1 0 0}=0,$ for the relevant cases we always have $e^{i \pi \sum \tau_i}=1$.  These transversity amplitudes therefore  satisfying the appropriate crossing relations derived in \eqref{eq:crossingm}.

\section{Crossing Relations from Lorentz Rotations}
\label{app:crossing}

To make contact with previous analyses, we now compare the transversity crossing relations derived in Appendix \ref{app:multispinor} with the `historical' approach found in the literature \cite{Trueman:1964zzb,cohen-tannoudji_kinematical_1968,Hara:1970gc,Hara:1971kj}. In these works the crossing relations are given in terms of a complex Lorentz transformation which for the helicity amplitudes takes the form
\begin{align}
\mathcal{H}^{s}_{\lambda_1 \lambda_2 \lambda_3 \lambda_4} ( s,t,u ) &= \eta_t \sum_{\lambda_i'} e^{i \pi ( \lambda_1' - \lambda_4' ) } \times \qquad      \label{eqn:H23} \\
& \,  d^{S_1}_{\lambda_1' \lambda_1} ( \chi_t ) \ d^{S_3}_{\lambda_3' \lambda_3} (  \pi - \chi_t )\
 \mathcal{H}^{t}_{\lambda_1' \lambda_3' \lambda_2' \lambda_4'} ( t,s,u ) \
d^{S_2}_{\lambda_2' \lambda_2} ( - \pi + \chi_t ) \ d^{S_4}_{\lambda_4' \lambda_4} ( - \chi_t )\,,
\nn \\
\mathcal{H}^{s}_{\lambda_1 \lambda_2 \lambda_3 \lambda_4} ( s,t,u ) &= \eta_u \sum_{\lambda_i'} e^{i \pi ( \lambda_1' - \lambda_3' ) } \times \qquad   \label{eqn:H24}
  \\
& \,  d^{S_1}_{\lambda_1' \lambda_1} ( \chi_u ) \ d^{S_4}_{\lambda_3' \lambda_3} (  \pi - \chi_u )\
 \mathcal{H}^{u}_{\lambda_1' \lambda_4' \lambda_3' \lambda_2'} ( u,t,s ) \
 d^{S_3}_{\lambda_4' \lambda_4} ( - \chi_u ) \  d^{S_2}_{\lambda_2' \lambda_2} ( - \pi + \chi_u )  \, ,\nn
\end{align}
where $\eta_t$ and $\eta_u$ are  statistics factors which are difficult to determine \cite{Hara:1970gc,Hara:1971kj}. Their value depends on the choice of branches for the phase of the arguments of each of the Wigner matrices (which are only periodic in $4\pi$ for fermions). \\

It is apparent from the above equations for the helicity amplitudes that the crossing relations do not relate sign definite amplitudes to other sign definite amplitudes, simply because the Wigner matrices will contain negative signs terms. They are therefore of limited interest when it comes to establishing positivity bounds. 
In the remainder of this appendix, we shall first describe how to recast the helicity crossing relations \eqref{eqn:H23} and \eqref{eqn:H24} in the more useful transversity basis (see also section~\ref{sec:trfm}), and then provide a review of the historical derivation of \eqref{eqn:H23} and \eqref{eqn:H24}.

\subsection{Transversity Crossing Relations}

As can be seen from equations \eqref{eqn:H23} and \eqref{eqn:H24} for helicity amplitudes, the crossing relations are highly non-trivial and positivity of the amplitude of the $s$ channel on the right hand cut implies no particular information on the sign of the $u$ channel amplitude to be evaluated on the left hand cut. To be able to derive the relevant positivity bounds, we should therefore work in terms of scattering amplitudes which are (semi)-diagonal under crossing. This is possible by working in the transversity basis. Indeed the Wigner (small) $d$ matrices that enter the crossing relations for definite helicities are related to the Wigner $D$ matrices arising from the rotation operator, $D_{ab}^S(\alpha,\beta,\gamma)=e^{-i a \alpha} d^S_{ab}(\beta) e^{-i b \gamma}$ and the $u^S_{ab}$ defined as
\ba
u^S_{ab}=D^S_{ab}\(\frac{\pi}{2}, \frac{\pi}{2}, -\frac{\pi}{2}\)
\ea
are precisely what transforms a helicity state into a transversity state.\\

From the properties of these matrices (see appendix \ref{app:dJ}), namely,
\ba
 \mathcal{T}^s_{\tau_1 \tau_2 \tau_3 \tau_4} (s,t,u) &=& \sum_{\lambda_i} u^{S_1}_{\tau_1 \lambda_1} u^{S_2}_{\tau_2 \lambda_2} u^{S_3 *}_{\tau_3 \lambda_3} u^{S_4 *}_{\tau_4 \lambda_4} \mathcal{H}^s_{\lambda_1 \lambda_2 \lambda_3 \lambda_4} (s,t,u)   \\
&=&     \eta_t \sum_{\lambda_i'} \Bigg[ e^{i \pi ( \lambda_1' - \lambda_4' ) }\
 \mathcal{H}^{t}_{\lambda_1' \lambda_3' \lambda_2' \lambda_4'} ( t,s,u )  \nn   \\
&& \times \(\sum_{\lambda_i}  u^{S_1}_{\tau_1 \lambda_1} u^{S_2}_{\tau_2 \lambda_2} u^{S_3 *}_{\tau_3 \lambda_3} u^{S_4 *}_{\tau_4 \lambda_4}  d^{S_1}_{\lambda_1' \lambda_1} ( \chi_t ) \ d^{S_3}_{\lambda_3' \lambda_3} (  \pi - \chi_t ) d^{S_2}_{\lambda_2' \lambda_2} ( - \pi + \chi_t ) \ d^{S_4}_{\lambda_4' \lambda_4} ( - \chi_t )  \)\Bigg]\nn\\
&=& \eta_t \,  (-1)^{S_1 - \tau_1} (-1)^{S_4 - \tau_4} e^{i \pi (S_2-S_3)}  e^{- i \chi_t \sum_i \tau_i} e^{i \pi ( \tau_2 + \tau_3 )} \nn\\
&& \times \sum_{\lambda'_i}    u_{-\tau_1 \lambda_1'}^{S_1}   u_{-\tau_2 \lambda_2'}^{S_2*}  \mathcal{H}^t_{\lambda_1' \lambda_3' \lambda_2' \lambda_4'} (t,s,u)  u_{-\tau_3 \lambda_3'}^{S_3} u_{-\tau_4 \lambda_4'}^{S_4*}
  \, ,\nn
\ea
which becomes
\be
\mathcal{T}^s_{\tau_1 \tau_2 \tau_3 \tau_4} (s,t,u)  =   \eta_t' \, e^{i \pi \sum_i \tau_i} \; e^{-i \chi_t \sum_i \tau_i} \mathcal{T}^t_{-\tau_1 -\tau_3 -\tau_2 -\tau_4} (t, s, u)  \,   ,  \label{eqn:T23}
\ee
with the overall sign
\begin{equation}
\eta_t' =  (-1)^{\sum_i S_i }  (-1)^{2 S_2}  \eta_t .
\end{equation}

Similarly, we can apply \eqref{eqn:H24} and find
\ba
\mathcal{T}^s_{\tau_1 \tau_2 \tau_3 \tau_4} (s,t,u) &=& \eta_u   (-1)^{S_1 - \tau_1} (-1)^{S_3 - \tau_3} e^{i \pi S_2}  e^{-i \pi S_4 } e^{- i \chi_u \sum_i \tau_i} e^{i \pi ( \tau_2 + \tau_4 )} \\
&& \times \sum_{\lambda_i'}  u_{-\tau_1 \lambda_1'}^{S_1}  u_{-\tau_2 \lambda_2'}^{S_2*}  \mathcal{H}^u_{\lambda_1' \lambda_4' \lambda_3' \lambda_2'} (u,t, s)
 u_{-\tau_3 \lambda_3'}^{S_3*}    u_{-\tau_4 \lambda_4'}^{S_4}     , \nn
\ea
which becomes
\be
\mathcal{T}^s_{\tau_1 \tau_2 \tau_3 \tau_4} (s,t,u)  =   \eta_u'  e^{i \pi \sum_i \tau_i} \;\; e^{-i \chi_u \sum_i \tau_i} \mathcal{T}^u_{-\tau_1 -\tau_4 -\tau_3 -\tau_2} (u, t, s)  \, ,       \label{eqn:T24}
\ee
with the overall sign
\begin{equation}
\eta_u' = (-1)^{ \sum_i S_i } \, (-1)^{2 S_2}    \, \eta_u .
\end{equation}
We have now specified all four crossing relations (\ref{eqn:H23}, \ref{eqn:H24}, \ref{eqn:T23}, \ref{eqn:T24}), up to two undetermined signs $\eta_t$ and $\eta_u$.

To completely determine the crossing relations we need to specify $\eta_t$, $\eta_u$ or equivalently $\eta_u'$. This turns out to be quite difficult by this method\footnote{See for instance the critical discussion in \cite{Hara:1971kj} which notes that the result given in \cite{cohen-tannoudji_kinematical_1968} is ambiguous due to an unclear specification of the branches of the arguments of the Wigner matrices.}. In \cite{Hara:1971kj} this is achieved by using closure, CPT and particle exchange properties, as well as some example amplitudes. However, since these coefficients only depend on the spins, by far the simplest way is to directly determine $\eta_u'$ by computing explicit tree level examples, as we have done in the previous appendices, and in particular in the multispinor approach \eqref{eq:etaup}. Once known for explicit examples, then they remain the same for any scattering amplitude with the same spins, to all loops.

In the remainder of this appendix, we will review the derivation of our starting point: the helicity crossing relations \eqref{eqn:H23} and \eqref{eqn:H24}.

\subsection{Helicity Crossing Relations}
\label{sec:helicitycrossing}

To make connection with older literature \cite{Trueman:1964zzb,cohen-tannoudji_kinematical_1968,Hara:1970gc,Hara:1971kj} we will now summarize the steps in deriving the $s-t$ and from it the $s-u$ crossing relations.
\subsubsection*{$B \leftrightarrow C$ Crossing}

The scattering amplitude $A+B \to C+D$ depends on the $4$-momenta of the external states, and can be defined in any reference frame,
\ba
A^{A+B \to C+D}_{\lambda_1 \lambda_2 \lambda_3 \lambda_4} (k_1,k_2 ; k_3, k_4 )\, .
\ea
Similarly as in appendix \ref{app:multispinor}, we fix the four-momenta in terms of the 3-momentum and the scattering angle $\theta$, leading to the expressions for the 4-momenta \eqref{eqn:standard}, with respective angles $\theta_i$ given in \eqref{eq:theta_i}. \\

Introducing the `polarization', $\varepsilon_{\lambda_i}^{\mu_i} (k_i)$, where $\mu$ denotes a collection of tensor/spinor indices appropriate for particle $i$, we can write the amplitude as
\begin{align}
A^{A+B \to C+D}_{\lambda_1 \lambda_2 \lambda_3 \lambda_4} ( k_1 , k_2  ; k_3, k_4 ) &= \varepsilon_{\lambda_1}^{\mu_1} ( k_1 ) \varepsilon_{\lambda_2}^{\mu_2} ( k_2 ) \bar \varepsilon_{\lambda_3}^{\mu_3} ( k_3 ) \bar \varepsilon_{\lambda_4}^{\mu_4} ( k_4 ) M_{\mu_1 \mu_2 \mu_3 \mu_4}^{A+B \to C+D} ( k_1 , k_2 , k_3 , k_4  )
\end{align}
where $M_{\mu_1 \mu_2 \mu_3 \mu_4}^{A+B \to C+D} ( k_1 , k_2 , k_3 , k_4  )$ is the vertex one would calculate using Feynman rules in the usual way.
The crossing of particles $B$ and $C$ is implemented via the rearrangement
\ba
A^{A+B \to C+D}_{\lambda_1 \lambda_2 \lambda_3 \lambda_4} ( k_1 , k_2  ; k_3, k_4 ) &=&  \eta \varepsilon_{\lambda_1}^{\mu_1} ( k_1 )  \bar \varepsilon_{\lambda_3}^{\mu_3} ( k_3 )  \varepsilon_{\lambda_2}^{\mu_2} ( k_2 ) \bar \varepsilon_{\lambda_4}^{\mu_4} ( k_4 )  M_{\mu_1 \mu_2 \mu_3 \mu_4}^{A + \bar C \to \bar B + D} ( k_1 , -k_3 ,  - k_2 , k_4  ) \nn  \\
&=&  \eta e^{i \pi (\lambda_2 - \lambda_3) } A^{A + \bar C \to \bar B + D}_{\lambda_1 \lambda_3 \lambda_2 \lambda_4} ( k_1, -k_3 ; -k_2, k_4 ) \,,
\ea
where $\eta$ is an overall sign which depends on the statistics of the particles, and the factor of $e^{i \pi (\lambda_2 - \lambda_3 )}$ comes from the polarizations, which obey\footnote{
This is required by $\mathtt{CPT}$ invariance, up to an overall $\mathtt{CPT}$ phase which is independent of the helicity, and is conventionally set to unity.
},
\begin{equation}
\varepsilon_{\lambda}^\mu (k) = e^{i \pi \lambda} \; \bar \varepsilon_{\lambda}^\mu ( -k )\,.
\end{equation}
For example, for standard spin-1/2 and spin-1 polarizations
\ba
\tilde u_\lambda ( - k) = e^{i \pi \lambda} \tilde v_{\lambda} (k) , \;\;\;\; \tilde \epsilon^\mu_\lambda (-k) = (-1)^\lambda ( \tilde \epsilon^\mu_{\lambda} (k ) )^*\,.
\ea
The helicity spinors are defined explicitly in Appendix \ref{app:multispinor} (see \eqref{eq:helicitySpinors}), and can be used to construct canonical polarizations for any desired spin.

\paragraph{Return to center of mass frame:}

The amplitude of the new channel $A +\bar C \to \bar B + D$ is no longer evaluated in the center of mass frame. To remedy this, we perform a Lorentz transformation $\hat L$ given by
\ba
\hat L\, k_0^s=k_0^t\,,\quad -\hat L\, k_{\theta_s}^s=k_{\pi}^t\,, \quad - \hat L\,  k_{\pi}^s=k_{-\theta_t}^t\quad{\rm and}\quad \hat L\, k_{\theta_s+\pi}^s=k_{\pi -\theta_t}^t\,.
\ea
remembering that in the $s$-channel
\be
k_1=k_0^s \, , \quad k_2 = k_{\pi}^s \, , \quad k_3 = k_{\theta_s}^s \, \quad  \text{and } k_4=k^s_{\theta_s+\pi}
 \ee
 with the four-momenta
 \be
 k^s_{\theta}= (\sqrt{s}/2,p \sin\theta ,0 , p \cos \theta) \, \quad p = \frac{1}{2} \sqrt{s-4m^2}\, .
 \ee
  and similarly for the $t$-channel. \\

In the scattering plane, there are only three independent Lorentz generators,  so any three of these relations uniquely specifies $\hat L$ (and the fourth is guaranteed by momentum conservation). This transformation brings us to the $t$-channel center of mass frame, which differs from the $s$-channel frame by a reversal of the normal to the scattering plane. The explicit form for this Lorentz transformation is well-known, and it can be written \cite{Trueman:1964zzb,cohen-tannoudji_kinematical_1968,Hara:1970gc}
\ba
\label{eq:Rot}
\hat L = B_0^t  R_{\tilde\chi_1} \left( B_0^s \right)^{-1} = B_{\pi}^t  R_{\tilde\chi_2} \left( B_{\theta_s}^s \right)^{-1} = B_{-\theta_t}^t  R_ {\tilde\chi_3} \left( B_{\pi}^s \right)^{-1} = B_{\pi - \theta_t}^t  R_{\tilde\chi_4} \left( B_{\pi+\theta_s}^s \right)^{-1}\quad
\ea
where $B_\theta^s$ is a boost along $\vec{p}{\,}_\theta^s$ of magnitude $| \vec{p}{\,}_\theta^s|$, where the spatial part of $p^s_\theta$ is given in \eqref{eqn:standard}, \ie $B_\theta^s  \, (m,0,0,0)^T = k_\theta^s$. $R_{\tilde\chi}$ is a rotation of angle $\tilde\chi$ about the axis perpendicular to the scattering plane with
\ba
\label{eq:AngleChoice}
{ \tilde\chi_1 = \tilde\chi_4 = \pi - \tilde\chi_2 = \pi - \tilde\chi_3 = \chi_t \,,}
\ea
where,
\ba
\label{eq:chi}
\cos \chi_t =  - \frac{s t}{ \sqrt{\mathcal{ST}}}, \;\;\;\; \sin \chi_t = {+} \frac{2 m \sqrt{stu} }{ \sqrt{\mathcal{ST}}} \,,
\ea
and care has been taken to ensure that the square roots are well-defined in the $s$ and $t$ physical regions, in which
\ba
\mathcal{S T} =  s ( s-4m^2)  \; t ( t-4m^2)  > 0  , \;\;\;\; stu > 0  \,.
\ea
It is then unambiguous that $\cos \chi_t >0$ and $\sin \chi_t < 0$, which implies that the range for the angle $\chi_t$ is ${ 0 \le  \chi_t \le \pi/2}$ in the physical $s$ and $t$ regions.  \\

This tensorial Lorentz transformation only determines the angles $\chi_i$ up to a shift of $2\pi$. Note that when fermions are involved, one should also consider the spinorial Lorentz transform, generated by $- \tfrac{1}{4} [ \gamma^\mu, \gamma^\nu ]$, in place of the usual Lorentz generators. However, note that shifting any of the $\chi_i$ by $2\pi$ will only introduce an overall sign, $(-1)^{2S_i}$, which can be absorbed into the overall statistics prefactor $\eta$.
\\

\paragraph{Reversed scattering plane normal:}

We have almost transformed back to the standard kinematic setup \eqref{eqn:standard}, however now the normal to the scattering plane has the opposite direction. We can account for this by noting that
\begin{equation}
A^{A + \bar C \to \bar B + D}_{\lambda_1' \lambda_3' \lambda_2' \lambda_4'} ( k_0^t,  k_{\pi}^t ; k_{-\theta_t}^t , k_{\pi - \theta_t}^t )  =  ( -1 )^{\lambda_1' - \lambda_3' - \lambda_2' + \lambda_4'} A^{A + \bar C \to \bar B + D}_{\lambda_1' \lambda_3' \lambda_2' \lambda_4'} ( k_0^t,  k_{\pi}^t ; k_{\theta_t}^t , k_{\pi + \theta_t}^t )    ,
\end{equation}
which follows from the $d^J$ partial wave expansion of the $t$-channel amplitude  (see Eq.~(\ref{ddef2})). \\

Overall, the crossing relation can then be written in terms of the rotation angle $\chi_t$ \cite{Hara:1970gc}
\ba
\label{eq:stcrossing}
&& A^{A+B\to C+D}_{\lambda_1 \lambda_2 \lambda_3 \lambda_4} ( k_{0}^s , k_{\pi}^s  ; k_{\theta_s}^s, k_{\theta_s+\pi}^s ) = \eta_t \sum_{\lambda_i'} e^{i \pi ( \lambda_1' - \lambda_4' ) } \times \qquad  \\
&& ~~~~~~~ d^{S_1}_{\lambda_1' \lambda_1} ( \chi_t ) \ d^{S_3}_{\lambda_3' \lambda_3} (  \pi - \chi_t )\
 A^{A\bar C \to \bar B D}_{\lambda_1' \lambda_3' \lambda_2' \lambda_4'} ( k_0^t,  k_{\pi}^t ; k_{\theta_t}^t , k_{\pi + \theta_t}^t ) \
d^{S_2}_{\lambda_2' \lambda_2} ( - \pi + \chi_t ) \ d^{S_4}_{\lambda_4' \lambda_4} ( - \chi_t )\,,\nn
\ea
where the Wigner matrices arise from the spin rotations of the states, and $\eta_t$ represents an overall sign which depends only on the statistics of the particles.

\paragraph{Going to Mandelstam variables:}

Replacing the 4-momenta with the Mandelstam invariants is not always trivial, because for particles with spin the amplitude is not an analytic function of $s,t,u$. Fortunately, the non-analyticities can be factorized  (see Eq.~(\ref{ddef2}))
\ba
 A^{A+B \to C+D}_{\lambda_1 \lambda_2 \lambda_3 \lambda_4} ( k_{0}^s , k_{\pi}^s  ; k_{\theta_s}^s, k_{\theta_s+\pi}^s )  =  \left(  \cos \frac{\theta_s}{2} \right)^{| \lambda + \mu |}  \left(  \sin \frac{\theta_s}{2} \right)^{| \lambda - \mu |} R_{\lambda_1 \lambda_2 \lambda_3 \lambda_4} ( s, \cos \theta_s ) ,
 \label{eq:non-analyticfactor}
\ea
where $\lambda=\lambda_1-\lambda_2, \mu = \li_3 - \li_4$, and $R$ is an analytic function of $s$ and $\cos \theta_s$. Recall that the scattering angle is replaced with Mandelstam invariants according to
\begin{equation}
\cos \frac{\theta_s}{2} =  \frac{ \sqrt{-su} }{ \sqrt{ \mathcal{S} } } , \;\;\;\; \sin \frac{\theta_s}{2} =  \frac{ \sqrt{-s t} }{ \sqrt{ \mathcal{S} } }  \;\;\;\; \left( \text{Physical } s \text{ region} \right)\,,
\end{equation}
which is sufficient to unambiguously define the $s$-channel amplitude in its physical region
\ba
\mathcal{H}^s_{\lambda_1 \lambda_2 \lambda_3 \lambda_4} (s,t,u) &=& A_{\lambda_1 \lambda_2 \lambda_3 \lambda_4}^{A+B\to C+D} ( k_{0}^s , k_{\pi}^s  ; k_{\theta_s}^s, k_{\theta_s+\pi}^s ) .
\ea
Similarly, making the replacement
\begin{equation}
\cos \frac{\theta_t}{2} =  \frac{ \sqrt{-t u} }{ \sqrt{ \mathcal{T} } } , \;\;\;\; \sin \frac{\theta_t}{2} =  \frac{ \sqrt{-s t} }{ \sqrt{ \mathcal{T} } }   \;\;\;\; \left( \text{Physical } t \text{ region} \right)\,,
\end{equation}
we can analogously define
\ba
\mathcal{H}^t_{\lambda_1 \lambda_3 \lambda_2 \lambda_4} (t, s, u) &=& A_{\lambda_1 \lambda_3 \lambda_2 \lambda_4}^{A+\bar C\to \bar B + D} ( k_{0}^t , k_{\pi}^t  ; k_{\theta_t}^t, k_{\theta_t+\pi}^t ) .
\ea
In terms of these functions, the relation \eqref{eq:stcrossing} becomes the expression announced in \eqref{eqn:H23} that describes the crossing of $B \leftrightarrow C$. Although the left and right hand functions have been constructed in the physical $s$ and $t$ regions respectively, the resulting equality can now be analytically continued to any common values of $s,t,u$.

\subsubsection*{$B \leftrightarrow D$ Crossing}

Now, consider evaluating the amplitude at $\theta_s + \pi$, with particles $C$ and $D$ relabelled
\begin{align}
A^{A+B \to D+C}_{\lambda_1 \lambda_2 \lambda_4 \lambda_3} ( k_{0}^s , k_{\pi}^s  ; k_{\theta_s+\pi}^s, k_{\theta_s+2\pi}^s ) =  \eta_{CD} (-1)^{2 S_3}  A^{A+B \to C+D}_{\lambda_1 \lambda_2 \lambda_3 \lambda_4} ( k_{0}^s , k_{\pi}^s  ; k_{\theta_s}^s, k_{\theta_s+\pi}^s )\,,
\end{align}
where replacing $k_{\theta+2\pi}$ with $k_{\theta}$ incurs a sign if the particle is a fermion, and $\eta_{CD}$ is $-1$ if particles C and D are both fermions, and $+1$ otherwise.
Similarly, in the $t$-channel (\ie for the amplitude on the right hand side of \eqref{eq:stcrossing}) we can evaluate the amplitude at $\pi + \theta_t$ with $D$ and $\bar B$ interchanged
\begin{align}
A^{A+\bar C \to  D + \bar B }_{\lambda_1 \lambda_3  \lambda_4 \lambda_2} ( k_0^t,  k_{\pi}^t ; k_{\pi+\theta_t}^t , k_{2\pi + \theta_t}^t )
  =  \eta_{BD} (-1)^{2S_2}
A^{A+\bar C \to \bar B+ D}_{\lambda_1 \lambda_3 \lambda_2 \lambda_4} ( k_0^t,  k_{\pi}^t ; k_{\theta_t}^t , k_{\pi + \theta_t}^t )\,.
\end{align}
Substituting the previous expressions into the $B \leftrightarrow C$ crossing relation \eqref{eq:stcrossing}, we find
\ba
&& A^{A+B \to D+C}_{\lambda_1 \lambda_2 \lambda_4 \lambda_3} ( k_{0}^s , k_{\pi}^s  ; k_{\theta_s+\pi}^s, k_{\theta_s+2\pi}^s ) = (-1)^{2S_2+2S_3} \eta_{CD} \eta_{BD}  \eta_t  \sum_{\lambda_i'} e^{i \pi ( \lambda_1' - \lambda_4' ) } \times \quad  \\
&&\quad~~~     d^{S_1}_{\lambda_1' \lambda_1} ( \chi_t )   d^{S_3}_{\lambda_3' \lambda_3} ( \pi - \chi_t )
A^{A\bar C \to D \bar B}_{\lambda_1' \lambda_3'  \lambda_4' \lambda_2'} ( k_0^t,  k_{\pi}^t ; k_{\pi + \theta_t}^t , k_{2 \pi + \theta_t}^t )
 d^{S_4}_{\lambda_4' \lambda_4} ( - \chi_t )  d^{S_2}_{\lambda_2' \lambda_2} ( - \pi + \chi_t )  \, . \nn
\ea
Now, we can trivially relabel $C$ and $D$ (this simply corresponds to a shift in notation of no physical significance), and this gives a relation between $A^{A+B \to C+D}$ and $A^{A+\bar D \to C +\bar B}$, as desired. \\

\paragraph{Going to Mandelstam variables:}

We see from \eqref{eq:non-analyticfactor} that
\begin{align}
A^{A+B \to C+D}_{\lambda_1 \lambda_2 \lambda_3 \lambda_4} ( k_{0}^s , k_{\pi}^s  ; k_{\theta_s+\pi}^s, k_{\theta_s+2\pi}^s ) = (-1)^{\lambda + \mu} \mathcal{H}^s_{\lambda_1 \lambda_2 \lambda_3 \lambda_4} (s, u, t )\,,
\end{align}
since $\cos \tfrac{\pi+\theta_s}{2} = - \sin \tfrac{\theta_s}{2} $, $\sin \tfrac{\pi + \theta_s}{2}= \cos \tfrac{\theta_s}{2} $ (\ie although the momenta are exchanged $k_3 \leftrightarrow k_4$, this particular swap corresponds to $\sqrt{-u} \leftarrow - \sqrt{-t}$, $\sqrt{-t} \to \sqrt{-u}$. Simply taking $t \leftrightarrow u$ is insufficient because of the non-analyticities in the amplitude).
Similarly, defining the $u$-channel amplitude
\begin{align}
\mathcal{H}^u_{\lambda_1 \lambda_2 \lambda_3 \lambda_4} (u, t, s ) = A^{A + \bar D \to C+ \bar B}_{\lambda_1 \lambda_2 \lambda_3 \lambda_4} ( k_{0}^u , k_{\pi}^u  ; k_{\theta_u}^u, k_{\pi + \theta_u}^u )\,,
\end{align}
where,
\begin{equation}
\cos \frac{\theta_u}{2} =  \frac{ \sqrt{-s u} }{ \sqrt{ \mathcal{U} } } , \;\;\;\; \sin \frac{\theta_u}{2} =  \frac{ \sqrt{-u t} }{ \sqrt{ \mathcal{U} } }   \;\;\;\; \left( \text{Physical } u \text{ region} \right)\,,
\end{equation}
we can write the right hand amplitude in the crossing relation as
\begin{align}
A^{A + \bar D \to C+ \bar B}_{\lambda_1' \lambda_4' \lambda_3' \lambda_2'} ( k_{0}^t , k_{\pi}^t  ; k_{\pi + \theta_t}^t, k_{2\pi + \theta_t}^t ) = (-1)^{\lambda_1' - \lambda_4' + \lambda_3' - \lambda_2'}  \mathcal{H}^u_{\lambda'_1 \lambda'_4 \lambda'_3 \lambda'_2} (t, u, s )
\end{align}
which gives
\ba
&&  \mathcal{H}^{s}_{\lambda_1 \lambda_2 \lambda_3 \lambda_4} (  s, u, t ) =  (-1)^{2S_2 + 2 S_4} \eta_{CD} \eta_{BC}  \left( \eta_t \Big|_{C \leftrightarrow D} \right)  \sum_{\lambda_i'} e^{i \pi ( \lambda_1' - \lambda_3' ) }   (-1)^{\sum_i \lambda_i - \lambda_i' }  \times \quad  \\
&& ~~~~~~~~   d^{S_1}_{\lambda_1' \lambda_1} ( \chi_t )   d^{S_4}_{\lambda_4' \lambda_4} ( \pi - \chi_t )  \mathcal{H}^{u}_{\lambda_1' \lambda_4'  \lambda_3' \lambda_2'} ( t,u,s )
d^{S_3}_{\lambda'_3 \lambda_3} ( - \chi_t ) d^{S_2}_{\lambda_2' \lambda_2} ( - \pi + \chi_t )  \,.\nn
\ea
This can be written as \eqref{eqn:H24} by relabelling $t$ and $u$ (at this stage they are just complex arguments of a function; note that this does not include changing the subscripts of $\eta_t$ and $\chi_t$ from $t$ to $u$), and absorbing the $(-1)^{\lambda_i-\lambda_i'}$ into the Wigner matrices. Explicitly, this gives us the relations
\begin{equation}
\eta_u =  \eta_{BC}  \, \eta_{CD} \, \left( \eta_t \Big|_{C \leftrightarrow D} \right) , \;\;\;\; \chi_u = - \chi_t \Big|_{t \leftrightarrow u}\,.
\end{equation}
That is, the $\chi_u$ crossing angle is defined in both the $s$ and $u$ physical regions as
\bal
\label{eq:anglephi}
 \cos \chi_u  &= -  \frac{s u}{ \sqrt{\mathcal{SU}}}\,, \qquad \sin \chi_u = { -} \frac{2 m \sqrt{stu} }{ \sqrt{\mathcal{SU}}}\,,\\
 \mathcal{S U} &= s (s-4m^2)  \; u ( u-4m^2) > 0  \, .
\eal
and has a range ${ -\pi/2  \le \chi_u \le 0}$. \\

\section{Discrete Symmetries of Helicity and Transversity Amplitudes}
\label{app:symm}

This appendix documents the properties of the amplitudes under $\mathtt{C}, \mathtt{P}$ and $\mathtt{T}$. These are operators which act on the Hilbert space, and throughout we will adopt the notation, for such an operator $X$
\begin{equation}
X :  \;\;\;\; A   = \langle f | \hat T | i \rangle \to  \langle X f | \hat T | X i \rangle = A'
\end{equation}
\ie  the amplitude $A$ is mapped under $X$ to another function $A'$, related to $A$ by the relevant operator acting on each of the incoming and outgoing states.

\subsection*{Parity}

We define our rest frame helicity states as eigenstates of $\mathtt{P}$
\begin{equation}
\mathtt{P} | 0 \lambda \rangle = \eta^P | 0 \lambda \rangle
\end{equation}
where $\eta^P$ is a phase which is independent of $\lambda$ (because $\mathtt{P}$ commutes with $J$) and must be $\pm 1$ (because $\mathtt{P}^2 =1$). Parity commutes with rotations, but not with boosts. Applying a boost in the $z$ direction gives
\begin{equation}
 \mathtt{P} | p_{0}, \lambda \rangle =  \eta^P (-1)^{S+\lambda} | p_{\pi} , -\lambda \rangle\,.
\end{equation}
The two-particle state then transforms as\footnote{
Note that this can also be written as,
\begin{align}
\mathtt{P} | p_{\theta, \phi} \lambda \rangle &=   \eta^P e^{-i \pi S}  | p_{ \pi - \theta, \pi + \phi} , - \lambda \rangle
\end{align}
where the momentum undergoes a standard inversion. We will not use this expression however, because it sends our $\phi=0$ states to $\phi=\pi$ states, and would require keeping track of the azimuthal phase.
},
\begin{align}
\mathtt{P} | p_{\theta}, \lambda \rangle &= \eta^P_{12} (-1)^{S_1 - S_2 + \lambda} | p_{\pi+\theta} , -\lambda_1 - \lambda_2 \rangle\,.
\end{align}
Therefore under a parity transformation, the helicity amplitude transforms as
\begin{equation}
\mathtt{P} : \mathcal{H}_{\lambda_1 \lambda_2 \lambda_3 \lambda_4}  \to  \eta^P_{1234} (-1)^{S_1 - S_2 - S_3 + S_4 } (-1)^{\lambda-\mu}  \mathcal{H}_{-\lambda_1 -\lambda_2 -\lambda_3 -\lambda_4} .  \label{eqn:PA}
\end{equation}
and so the transversity amplitude transforms as
\begin{equation}
\mathtt{P} : \mathcal{T}_{\tau_1 \tau_2 \tau_3 \tau_4}  \to  \eta^P_{1234} (-1)^{\tau_1+\tau_2-\tau_3-\tau_4}  \mathcal{T}_{\tau_1 \tau_2 \tau_3 \tau_4} .
\end{equation}

Finally, for an angular momentum state
\begin{equation}
\mathtt{P} | p , JM, \lambda_1 \lambda_2 \rangle = \eta^P_{12} (-1)^{J - S_1 + S_2} | p , JM,  -\lambda_1 -\lambda_2 \rangle
\end{equation}
and so the partial wave amplitude then transforms as
\begin{equation}
\mathtt{P} : T^J_{\lambda_1 \lambda_2 \lambda_3 \lambda_4} \to \eta^P_{1234} (-1)^{S_1 - S_2 -  S_3 + S_4 } T^J_{-\lambda_1 -\lambda_2 -\lambda_3 - \lambda_4}  .
\end{equation}
Utilizing the partial wave expansion, we see that this agrees with \eqref{eqn:PA}, since $d_{\lambda \mu} (\theta) = (-1)^{\lambda - \mu} d_{-\lambda -\mu} (\theta)$.

\subsection*{Time Reversal}

Time reversal is an anti-unitary operator which commutes with rotations and reverses the direction of boosts
\begin{equation*}
\mathtt{T}^{-1} J_i \mathtt{T} = - J_i , \;\;\;\; \mathtt{T}^{-1} K_i \mathtt{T} = K_i
\end{equation*}
\begin{equation*}
\implies \;\;\;\; \mathtt{T}^{-1} R_{\phi \theta \varphi} \mathtt{T} =  R_{\phi \theta \varphi} , \;\;\;\; \mathtt{T}^{-1} B_i (p) \mathtt{T} = B_i (-p)
\end{equation*}
On one particle states, $\mathtt{T}$ can be implemented by
\begin{equation}
 \mathtt{T} | p_{0} , \lambda \rangle =   \eta^T  e^{-i \pi J_y}  | p_{0} ,  \lambda \rangle
\end{equation}
where $\eta^T$ can be an arbitrary constant phase (because of antiunitarity, $\mathtt{T}^\dagger \mathtt{T}$ is guaranteed to be unity for any phase).
Therefore our two-particle states transform as
\begin{align}
\mathtt{T} | p_{\theta} \lambda_1 \lambda_2 \rangle &=  \eta^T_{12} | p_{\pi+\theta} , \lambda_1 \lambda_2 \rangle
\end{align}
where $\eta^T_{12} = \eta^T_1 \eta^T_2$ is the product of the particle phases.
The resulting action on the helicity amplitude is straightforward
\begin{equation}
\mathtt{T} :  \;\;\;\; \mathcal{H}_{\lambda_1 \lambda_2 \lambda_3 \lambda_4}^{s}  \to  \eta^T_{1234}   \mathcal{H}_{\lambda_1 \lambda_2 \lambda_3 \lambda_4}^{s}\,.
\end{equation}

Since $\mathtt{T}$ is antiunitary and antilinear, invariance under $\mathtt{T}$ requires $\mathtt{T}^\dagger S \mathtt{T} = S^\dagger$, so that
$$ \langle \mathtt{T} \beta | S | \mathtt{T} \alpha \rangle = \left( \langle \beta | \mathtt{T}^\dagger  S  \mathtt{T} | \alpha \rangle \right)^* = \left( \langle \beta | S^\dagger | \alpha \rangle \right)^* =   \langle \alpha | S | \beta \rangle $$
Therefore,
\begin{align}
\mathtt{T} \;\; \text{invariance} \;\; \implies \;\; \mathcal{H}_{\lambda_1 \lambda_2 \lambda_3 \lambda_4}^{A+B \to C+D} ( s, \theta)  &=   \eta^T_{1234}  \mathcal{H}_{\lambda_3 \lambda_4 \lambda_1 \lambda_2}^{C+ D \to A + B} \left( s, - \theta \right) \nonumber \\
&= \eta^T_{1234} (-1)^{\lambda-\mu} \mathcal{H}_{\lambda_3 \lambda_4 \lambda_1 \lambda_2}^{C+ D \to A + B} \left( s, \theta \right)  \label{eqn:TAs}
\end{align}
and so the transversity amplitudes would obey
\begin{equation}
\mathtt{T} \;\; \text{invariance} \;\; \implies \;\;  \mathcal{T}_{\tau_1 \tau_2 \tau_3 \tau_4}  \to  \eta^P_{1234} (-1)^{\tau_1-\tau_2 - \tau_3 + \tau_4}  \mathcal{T}_{\tau_1 \tau_2 \tau_3 \tau_4} .
\end{equation}

On the angular momentum states,
\begin{equation}
\mathtt{T} | p , JM, \lambda_1 \lambda_2 \rangle = \eta^T_{12} (-1)^{J -
M} | p , J -M,  \lambda_1 \lambda_2 \rangle\,.
\end{equation}
Physically, time reversal inverts both momenta and spin vectors, so although the helcity is preserved, the $J_z$ projection flips sign. Consequently,
\begin{equation}
\mathtt{T} : T^J_{\lambda_1 \lambda_2 \lambda_3 \lambda_4} \to  \eta^T_{1234}  T^{J}_{\lambda_1 \lambda_2 \lambda_3 \lambda_4}
\end{equation}
and so,
\begin{equation}
\mathtt{T} \;\; \text{invariance} \;\; \implies \;\;  T^J_{\lambda_1\lambda_2 \lambda_3 \lambda_4}  = \eta^T_{1234} T^{J}_{\lambda_3 \lambda_4 \lambda_1 \lambda_2}
\end{equation}
which confirms that \eqref{eqn:TAs} agrees with the partial wave expansion, since $d_{\lambda \mu} (\theta) = (-1)^{\lambda-\mu} d_{\mu \lambda} (\theta)$.

\subsection*{Charge Conjugation}

Under charge conjugation, the quantum numbers are conjugated, but the kinematics (momenta and helicities) are unaffected. This means it is particularly trivial to write down the action of $\mathtt{C}$ on our states. Introducing a label $a$ for the particle species (\ie all of the quantum numbers excluding spin and momentum)
\begin{equation}
\mathtt{C} | p_{\theta\phi}, \lambda , a \rangle = \eta^C | p_{\theta\phi}, \lambda , \bar a \rangle
\end{equation}
where $\bar a$ is the antiparticle of $a$, with inverse charges. $\eta^C$ is an overall phase, which must be $\pm 1$ (because $C^2=1$) and cannot depend on the helicity (because $C$ commute with $J$). $\mathtt{C}$ acts on each particle in a multi-particle state
\begin{equation}
\mathtt{C} | p_{\theta\phi}, \lambda_1 \lambda_2 , a_1 a_2 \rangle = \eta^C_{12} | p_{\theta\phi}, \lambda_1 \lambda_2 , \bar a_1 \bar a_2 \rangle
\end{equation}
and does not affect the kinematics or relative phases.
On the helicity amplitude, we therefore have a trivial replacement of particles with antiparticles
\begin{equation}
\mathtt{C} : \mathcal{H}^{A+B \to C+ D}_{\lambda_1 \lambda_2 \lambda_3 \lambda_4} \to  \eta^C_{1234}  \mathcal{H}^{\bar A + \bar B \to \bar C + \bar D}_{\lambda_1 \lambda_2 \lambda_3 \lambda_4} .
\end{equation}
The action on transversity amplitudes is analogous.

\subsection*{CPT}

Now it is simply a matter of combining the previous results. For the underlying QFT to be consistent, the amplitudes must respect $\mathtt{CPT}$ invariance.
The action of $\mathtt{CPT}$ on a 1-particle state is
\begin{equation}
\mathtt{CPT} | p_{\theta} \lambda a \rangle = \eta^{CPT} (-1)^{S-\lambda} | p_{\theta} , -\lambda, \bar a \rangle
\end{equation}
\ie both $P$ and $T$ effectively invert the momentum, $P$ inverts the helicity, and $C$ replaces particle with antiparticle. $\eta^{CPT}$ is the product $\eta^C \eta^P \eta^T$, which we can set to unity. Then, for two particles,
\begin{align}
\mathtt{CPT} | p_{\theta} , \lambda_1 \lambda_2 , a_1 a_2 \rangle =   (-1)^{S_1 \pm S_2 - \lambda} | p_{\theta} , -\lambda_1 -\lambda_2 , \bar a_1 \bar a_2 \rangle
\end{align}
So under $\mathtt{CPT}$, we find
\begin{align}
\mathtt{CPT} :  \mathcal{H}^{A + B \to C + D}_{\lambda_1 \lambda_2 \lambda_3 \lambda_4}  \to
 e^{i \pi \sum S_i} e^{i \pi \lambda_i} \mathcal{H}^{\bar A + \bar B \to \bar C + \bar D}_{-\lambda_1 -\lambda_2 -\lambda_3 -\lambda_4}
\end{align}
As an antiunitary, antilinear operator, $\mathtt{CPT}$ invariance requires
$$ \langle \mathtt{CPT} \beta | S  | \mathtt{CPT} \alpha \rangle = \left( \langle \beta |  ( \mathtt{CPT} )^\dagger S \mathtt{CPT} | \alpha \rangle \right)^* = \left( \langle \beta | S^\dagger | \alpha \rangle \right)^* = \langle \alpha | S | \beta \rangle $$
and so we arrive at,
\begin{equation}
\label{eqn:CPT}
\mathcal{H}_{\lambda_1 \lambda_2 \lambda_3 \lambda_4}
 = \pm (-1)^{S_1 - S_2 + S_3 - S_4}  \mathcal{H}_{-\lambda_3 -\lambda_4 -\lambda_1 -\lambda_2}
\end{equation}
where the sign is $+$ for $B+B\to B+B$ or $F+F\to F+F$ and $-$ otherwise, and arises from permuting the creation operators. The corresponding transversity condition is
\begin{equation}
\mathcal{T}_{\tau_1 \tau_2 \tau_3 \tau_4}
 = \pm (-1)^{2S_1 + 2 S_4}  \mathcal{T}_{\tau_1 \tau_2 \tau_3 \tau_4}
\end{equation}
The combined sign $\pm (-1)^{2S_1 + 2 S_4}$ is $+$ for $BB \to BB$, $BF \to BF$, $FB\to FB$ and $FF \to FF$, and $-$ otherwise---\ie $\mathtt{CPT}$ prevents a particle from changing its total spin.

\section{Properties of Wigner's Matrices}
\label{app:dJ}

The Wigner (small) $d$ matrices are defined as
\be
\label{defWigd}
d^J_{a b} (\beta) = \langle J a | e^{-i \beta J_y} | J b \rangle  \, ,
\ee
which are related to Wigner's $D$ matrices via $D^J_{a b} (\alpha, \beta, \gamma )  = e^{-i a \alpha } d^J_{a b} (\beta) e^{-i b \gamma }$ that furnish the $2J+1$ dimensional irreducible representations of the rotation operator $R(\alpha,\beta,\gamma)$, \ie
\ba
R(\alpha,\beta,\gamma) \left|J b \right\rangle = \sum_{a=-J}^J \, D^J_{ab}(\alpha, \beta,\gamma)\,  \left|J a \right\rangle\,.
\ea
 Since  $\langle J a | J_y | J b \rangle$  are purely imaginary, it is clear that $d^j_{ab} (\beta)$ are real quantities, which implies that
\be
\label{eqn:dJsymm1}
d_{ab}^J (\theta) = d_{ba}^J (-\theta)   .
\ee
When $\theta$ is real, the $d^J_{m' m} (\beta)$ matrix by definition is unitary (so $| d_{ab}^J (\theta \in R) | \leq 1$), and, thanks to the above properties, $d^j_{ab} (\beta)$ can be analytically continued to be real analytic: $d_{ab}^J (\theta^* ) = \left[ d_{ab}^J (\theta) \right]^*$.  Also, from the definition (\ref{defWigd}), we can see that
\be
\label{eqnthetaplus}
d^J_{ab} (\theta_1+\theta_2)  = \sum^J_{c=J} d^J_{ac} (\theta_1) d^J_{cb} (\theta_2) .
\ee

The $d^J_{ab} (\theta)$ can be represented in a number of ways:
\begin{itemize}

\item Wigner himself gave an explicit formula \cite{wigner_algebra_1931}, which is easy to evaluate for small $J$,
\be
\label{ddef1}
 d^J_{ab} (\theta) =  \sum^{\text{Min} ( J-a, J+b )}_{k= \text{Max} (0,  b-a )} w_k^{(Jab)} \left( \cos \frac{\theta}{2} \right)^{2J+b-a-2k} \left( - \sin \frac{\theta}{2} \right)^{a-b+2k}
\ee
with
\be
w_k^{(Jab)}   =   \frac{(-1)^{k}\sqrt{ (J+a)!(J-a)!(J+b)!(J-b)! }} {(J-a-k)! (J+b-k)! (a-b+k)! k!  }\,.
\ee
That is, the summation of $k$ is such that the factorials are nonnegative. This formula can be derived using the trick of decomposing the angular momentum algebra into an algebra of two uncoupled harmonic oscillators \cite{Sakurai:1167961, wigner_algebra_1931}. From Eq.~(\ref{defWigd}), it is clear that
\ba
\label{eqn:dJsymm2}
&& d^J_{ab} (-\theta) = (-1)^{a-b}  d^J_{ab} (\theta)   ,
\\
\label{eqn:dJsymm3}
&& d^J_{-a,-b} (\theta) =  d^J_{ba} (\theta)  ,
\\
\label{eqndjpi}
&& d^J_{ab}(\pi) = (-1)^{J+a} \delta_{a,-b} .
\ea
By Eq.~(\ref{eqndjpi}) and Eq.~(\ref{eqnthetaplus}), we also have
\be
d^J_{ab}(\theta+\pi) = (-1)^{J-b} d^J_{a,-b}(\theta)    .
\ee

\item For asymptotic properties (such as large $J$ or small $\theta$), the closed expression in terms of Jacobi polynomials is useful \cite{Mahoux:1969um}
\be
\label{ddef2}
 d^J_{ab} (\theta) = \sqrt{  \frac{ (J+b)! (J - b)!}{ (J + a)! (J-a)! }  } \left( \sin \frac{\theta}{2} \right)^{b-a} \left( \cos \frac{\theta}{2} \right)^{b+a} P_{J-b}^{b-a, b+a} ( \cos \theta )
 \ee
 where the Jacobi polynomials are defined as
\be
P_n^{\alpha, \beta}(z) = \frac{(-1)^n}{2^n n!} (1-z)^{-\alpha}  (1+z)^{-\beta} \frac{\ud^n}{\ud z^n} \left[  (1-z)^\alpha (1+z)^\beta (1-z^2)^n \right]   ,
\ee
for integer $n, \alpha, \beta$.  However, note that this formula is only valid for $b \geq |a|$ as the Jacobi polynomial is only defined for $b \geq |a|$, but the properties \eqref{eqn:dJsymm1}, \eqref{eqn:dJsymm2} and \eqref{eqn:dJsymm3} can be used to generate all other cases.

\item It is also useful to Fourier-expand $d^J_{ab} (\theta)$, which is a periodic function of period $2\pi$ (or $4\pi$) when $J$ is an integer (or a half integer). To achieve this, note that, by a standard Euler angle counting, we have
\be
|Ja\rangle_y   \equiv e^{i\frac{\pi}2J_x}|Ja\rangle= e^{-i\frac{\pi}2 J_z} e^{-i\frac{\pi}2J_y} e^{i\frac{\pi}2J_z} |Ja\rangle .
\ee
Then its Fourier decomposition is given by \cite{Feng:2015mqa}
\begin{equation}
 d^J_{ab} (\theta) = \sum_{\nu=-J}^J e^{-i \nu \theta} t_\nu^{(J a b)}    ,
\label{eqn:dJfourier}
\end{equation}
where
\be
t_\nu^{(J a b)}  = \langle Ja|J\nu \rangle_y\, {}_y\langle J \nu | J b\rangle  = e^{i \frac{\pi}{2} (b-a)} d_{a\nu}^J \left( \frac{\pi}{2} \right) d_{b\nu}^J \left( \frac{\pi}{2} \right)   .
\ee
This can also be written as a real series in $\cos ( \nu \theta )$ when $b-a$ is even and $\sin (\nu \theta)$ when $b-a$ is odd.
The Fourier coefficients $t_\nu^{(J a b)}$ have the following properties \cite{Tajima:2015owa, Feng:2015mqa}
\bal
 t_\nu^{(J a b)} &=   (-1)^{2J + a + b} t^{(J ab)}_{-\nu}  ,
 \\
  t_\nu^{(J a b)}  &=   t_\nu^{(J, -b, -a)} = (-1)^{b-a}  t_\nu^{(J b a)}
 \\
  \sum_{\nu} t_\nu^{(Jab)} & = \delta_{ab}
\eal

\end{itemize}

The following is a summary of some useful properties of $d^J_{ab} (\theta)$ matrices \cite{Mahoux:1969um, Tajima:2015owa, Feng:2015mqa}:
\ba
\label{eq:dProp1}
&& d_{ba}^J (\theta) = d_{-a,-b}^J (\theta) = d_{ab}^J (- \theta) =  (-1)^{b-a} d_{ab}^J (\theta)     ,
 \\
\label{eq:dProp2}
&& d^J_{ab}(\pi-\theta) = (-1)^{J+a} d^J_{a,-b}(\theta)  = (-1)^{J-b} d^J_{-ab} (\theta)   ,
\\
\label{eq:dProp3}
&& d^J_{ab}(\pi+\theta) = (-1)^{J-b} d^J_{a,-b}(\theta)  = (-1)^{J+a} d^J_{-ab} (\theta)   ,
\\
\label{eq:dProp4}
&& d^J_{ab}(\theta+2\pi) = (-1)^{2J} d^J_{ab}(\theta)
\\
&& d^J_{ab}(\pi) = (-1)^{J+a} \delta_{a,-b}   ,  ~~~ d^J_{ab}(-\pi) = (-1)^{J-a} \delta_{a,-b}    ,
\\
&&  d^J_{ab}(0) = \delta_{ab}, ~~~   d^J_{ab}(2\pi) = (-1)^{2J} \delta_{ab}  ,
\\
&&d^J_{ab} (\theta_1+\theta_2)  = \sum_{c} d^J_{ac} (\theta_1) d^J_{cb} (\theta_2)  ,
\\
\label{eq:dProp5}
&& d_{ab}^J (\theta^* ) = \left[ d_{ab}^J (\theta) \right]^*  , ~~~~~~ | d_{ab}^J (\theta \in \mathbb{R} ) | \leq 1  .
\ea

The unitary matrices that transform the helicity to transversity states can be written via the Wigner $D^S$ matrices
\be
u^S_{ab} = D^S_{ab}\(\frac{\pi}{2},\frac{\pi}{2},-\frac{\pi}{2}\) ,
\ee
where $S$ is the spin of the relevant particle. Note that $t^{(Sab)}_\nu=u^S_{a\nu} (u^{S}_{b\nu})^*$.  From the properties of $d^J_{ab} (\theta)$, it is straightforward to derive the following properties of $u^S_{ab}$ \cite{Kotanski:1965zz}:
\ba
\label{uSeq1}
&&  u^S_{ba} = u^S_{-a-b} = u^S_{ab}  ,
 \\
&&  \(u^{S}_{ab}\)^* = (-1)^{a-b} u^S_{ab} ,
\\
&& u^S_{a,-b} = e^{i \pi S} \(u^{S}_{ab}\)^*    ,
\\
&& \(u^{S}_{a,-b}\)^* = e^{-i \pi S} u^S_{ab} ,
\\
&& \sum_b u^S_{ab} u^S_{bc} = e^{i \pi S} \delta_{a,-c} ,
\\
&& \sum_b u^S_{ab} (u^{S}_{-b d})^* = \delta_{a,-d} ,
\\
&& \sum_b u^S_{ab} e^{i \pi b} u^S_{bc} = e^{i \pi c} \delta_{ac}
\\
&&  \sum_{b,c} u^S_{ab} d^S_{bc} (\theta) u^{S}_{cd}   = e^{i \pi S} e^{i a \theta } \delta_{a,-d}     .
\ea
In particular, the Wigner $d^S$ matrices can be diagonalized with the $u^S_{ab}$ matrices
\be
\label{uSeqlast}
  \sum_{b,c} u^S_{ab} d^S_{bc} (\theta) (u^{S}_{dc})^* = e^{i a \theta } \delta_{ad}     \,.
\ee
These relations are particularly relevant when deriving the relations between the helicity and transversity amplitudes.

\bibliographystyle{JHEP}
\bibliography{refs}

\providecommand{\href}[2]{#2}\begingroup\raggedright\begin{thebibliography}{10}

\bibitem{Vafa:2005ui}
C.~Vafa, \emph{{The String landscape and the swampland}},
  \href{https://arxiv.org/abs/hep-th/0509212}{{\ttfamily hep-th/0509212}}.

\bibitem{Adams:2006sv}
A.~Adams, N.~Arkani-Hamed, S.~Dubovsky, A.~Nicolis and R.~Rattazzi,
  \emph{{Causality, analyticity and an IR obstruction to UV completion}},
  \href{http://dx.doi.org/10.1088/1126-6708/2006/10/014}{\emph{JHEP} {\bfseries
  10} (2006) 014}, [\href{https://arxiv.org/abs/hep-th/0602178}{{\ttfamily
  hep-th/0602178}}].

\bibitem{deRham:2017avq}
C.~de~Rham, S.~Melville, A.~J. Tolley and S.-Y. Zhou, \emph{{Positivity Bounds
  for Scalar Theories}},  \href{https://arxiv.org/abs/1702.06134}{{\ttfamily
  1702.06134}}.

\bibitem{deRham:2017imi}
C.~de~Rham, S.~Melville, A.~J. Tolley and S.-Y. Zhou, \emph{{Massive Galileon
  Positivity Bounds}},  \href{https://arxiv.org/abs/1702.08577}{{\ttfamily
  1702.08577}}.

\bibitem{Bellazzini:2015cra}
B.~Bellazzini, C.~Cheung and G.~N. Remmen, \emph{{Quantum Gravity Constraints
  from Unitarity and Analyticity}},
  \href{http://dx.doi.org/10.1103/PhysRevD.93.064076}{\emph{Phys. Rev.}
  {\bfseries D93} (2016) 064076},
  [\href{https://arxiv.org/abs/1509.00851}{{\ttfamily 1509.00851}}].

\bibitem{Cheung:2016wjt}
C.~Cheung and G.~N. Remmen, \emph{{Positivity of Curvature-Squared Corrections
  in Gravity}},
  \href{http://dx.doi.org/10.1103/PhysRevLett.118.051601}{\emph{Phys. Rev.
  Lett.} {\bfseries 118} (2017) 051601},
  [\href{https://arxiv.org/abs/1608.02942}{{\ttfamily 1608.02942}}].

\bibitem{Cheung:2016yqr}
C.~Cheung and G.~N. Remmen, \emph{{Positive Signs in Massive Gravity}},
  \href{http://dx.doi.org/10.1007/JHEP04(2016)002}{\emph{JHEP} {\bfseries 04}
  (2016) 002}, [\href{https://arxiv.org/abs/1601.04068}{{\ttfamily
  1601.04068}}].

\bibitem{Bonifacio:2016wcb}
J.~Bonifacio, K.~Hinterbichler and R.~A. Rosen, \emph{{Positivity constraints
  for pseudolinear massive spin-2 and vector Galileons}},
  \href{http://dx.doi.org/10.1103/PhysRevD.94.104001}{\emph{Phys. Rev.}
  {\bfseries D94} (2016) 104001},
  [\href{https://arxiv.org/abs/1607.06084}{{\ttfamily 1607.06084}}].

\bibitem{Bellazzini:2016xrt}
B.~Bellazzini, \emph{{Softness and amplitudes’ positivity for spinning
  particles}}, \href{http://dx.doi.org/10.1007/JHEP02(2017)034}{\emph{JHEP}
  {\bfseries 02} (2017) 034},
  [\href{https://arxiv.org/abs/1605.06111}{{\ttfamily 1605.06111}}].

\bibitem{Jacob:1959at}
M.~Jacob and G.~C. Wick, \emph{{On the general theory of collisions for
  particles with spin}},
  \href{http://dx.doi.org/10.1016/0003-4916(59)90051-X}{\emph{Annals Phys.}
  {\bfseries 7} (1959) 404--428}.

\bibitem{Trueman:1964zzb}
T.~L. Trueman and G.~C. Wick, \emph{{Crossing relations for helicity
  amplitudes}},
  \href{http://dx.doi.org/10.1016/0003-4916(64)90254-4}{\emph{Annals Phys.}
  {\bfseries 26} (1964) 322--335}.

\bibitem{Hepp:1964}
K.~Hepp, \emph{{Lorentz invariant analytic S-matrix amplitudes}},
  {\emph{{Helvetica Physica Acta }} {\bfseries {37}} (1964) {55--73}}.

\bibitem{Williams:1963zz}
D.~N. Williams, \emph{{Construction Of Invariant Scalar Amplitudes Without
  Kinematical Singularities For Arbitrary Spin Nonzero Mass Two-Body Scattering
  Processes}}, .

\bibitem{PhysRev.160.1251}
H.~P. Stapp, \emph{{Analyticity Properties of Helicity Amplitudes}},
  \href{http://dx.doi.org/10.1103/PhysRev.160.1251}{\emph{Phys. Rev.}
  {\bfseries 160} (Aug, 1967) 1251--1256}.

\bibitem{Barut:1963zzb}
A.~O. Barut, I.~Muzinich and D.~N. Williams, \emph{{Construction of Invariant
  Scattering Amplitudes for Arbitrary Spins and Analytic Continuation in Total
  Angular Momentum}},
  \href{http://dx.doi.org/10.1103/PhysRev.130.442}{\emph{Phys. Rev.} {\bfseries
  130} (1963) 442--457}.

\bibitem{Scadron:1969rw}
M.~D. Scadron and H.~F. Jones, \emph{{Covariant m functions for higher spin}},
  \href{http://dx.doi.org/10.1103/PhysRev.173.1734}{\emph{Phys. Rev.}
  {\bfseries 173} (1968) 1734--1744}.

\bibitem{PhysRev.130.436}
A.~O. Barut, \emph{Crossing symmetry in $s$-matrix theory},
  \href{http://dx.doi.org/10.1103/PhysRev.130.436}{\emph{Phys. Rev.} {\bfseries
  130} (Apr, 1963) 436--439}.

\bibitem{barut1967theory}
A.~Barut, \emph{The Theory of the Scattering Matrix}.
\newblock Macmillan, 1967.

\bibitem{Mahoux:1969um}
G.~Mahoux and A.~Martin, \emph{{Extension of axiomatic analyticity properties
  for particles with spin, and proof of superconvergence relations}},
  \href{http://dx.doi.org/10.1103/PhysRev.174.2140}{\emph{Phys. Rev.}
  {\bfseries 174} (1968) 2140--2150}.

\bibitem{Kotanski:1965zz}
A.~Kotanski, \emph{Diagonalization of helicity-crossing matrices}, {\emph{Acta
  Physica Polonica} {\bfseries 29} (1966) }.

\bibitem{wigner_algebra_1931}
E.~Wigner, \emph{Algebra der {Darstellungstheorie}},  in \emph{Gruppentheorie
  und ihre {Anwendung} auf die {Quantenmechanik} der {Atomspektren}},
  pp.~120--133.
\newblock Vieweg+Teubner Verlag, 1931.

\bibitem{Richman:1984gh}
J.~D. Richman, \emph{{An Experimenter's Guide to the Helicity Formalism}}, .

\bibitem{Hara:1964zza}
Y.~Hara, \emph{{Analyticity Properties of Helicity Amplitudes and Construction
  of Kinematical Singularity-Free Amplitudes for Any Spin}},
  \href{http://dx.doi.org/10.1103/PhysRev.136.B507}{\emph{Phys. Rev.}
  {\bfseries 136} (1964) B507--B514}.

\bibitem{Wang:1966zza}
L.-L.~C. Wang, \emph{{General Method of Constructing Helicity Amplitudes Free
  from Kinematic Singularities and Zeros}},
  \href{http://dx.doi.org/10.1103/PhysRev.142.1187}{\emph{Phys. Rev.}
  {\bfseries 142} (1966) 1187--1194}.

\bibitem{cohen-tannoudji_kinematical_1968}
G.~Cohen-Tannoudji, A.~Morel and H.~Navelet, \emph{Kinematical singularities,
  crossing matrix and kinematical constraints for two-body helicity
  amplitudes},
  \href{http://dx.doi.org/http://dx.doi.org/10.1016/0003-4916(68)90243-1}{\emph{Annals
  of Physics} {\bfseries 46} (1968) 239 -- 316}.

\bibitem{kotanski_transversity_1970}
A.~Kotański, \emph{Transversity amplitudes and their application to the study
  of particles with spin}, {\emph{Acta Phys Pol B1 45} (1970) }.

\bibitem{Hara:1970gc}
Y.~Hara, \emph{{On crossing relations for helicity amplitudes}},
  \href{http://dx.doi.org/10.1063/1.1665056}{\emph{J. Math. Phys.} {\bfseries
  11} (1970) 253--257}.

\bibitem{Hara:1971kj}
Y.~Hara, \emph{{Crossing relations for helicity amplitudes}},
  \href{http://dx.doi.org/10.1143/PTP.45.584}{\emph{Prog. Theor. Phys.}
  {\bfseries 45} (1971) 584--595}.

\bibitem{martin_rigorous_1971}
A.~Martin, \emph{Rigorous {Results} from {Field} {Theory} and {Unitarity} {A}2
  - {Zichichi}, {A}.},  in \emph{Elementary {Processes} {At} {High} {Energy}},
  pp.~22--61.
\newblock Academic Press, 1971.

\bibitem{Peters:2004qw}
K.~J. Peters, \emph{{A Primer on partial wave analysis}},
  \href{http://dx.doi.org/10.1142/S0217751X06034811}{\emph{Int. J. Mod. Phys.}
  {\bfseries A21} (2006) 5618--5624},
  [\href{https://arxiv.org/abs/hep-ph/0412069}{{\ttfamily hep-ph/0412069}}].

\bibitem{Martin:1965jj}
A.~Martin, \emph{{Extension of the axiomatic analyticity domain of scattering
  amplitudes by unitarity. 1.}},
  \href{http://dx.doi.org/10.1007/BF02720568}{\emph{Nuovo Cim.} {\bfseries A42}
  (1965) 930--953}.

\bibitem{Jin:1964zza}
Y.~S. Jin and A.~Martin, \emph{{Number of Subtractions in Fixed-Transfer
  Dispersion Relations}},
  \href{http://dx.doi.org/10.1103/PhysRev.135.B1375}{\emph{Phys. Rev.}
  {\bfseries 135} (1964) B1375--B1377}.

\bibitem{Bargmann:1948ck}
V.~Bargmann and E.~P. Wigner, \emph{{Group Theoretical Discussion of
  Relativistic Wave Equations}},
  \href{http://dx.doi.org/10.1073/pnas.34.5.211}{\emph{Proc. Nat. Acad. Sci.}
  {\bfseries 34} (1948) 211}.

\bibitem{Schwinger:1966zz}
J.~Schwinger, \emph{{Particles and Sources}},
  \href{http://dx.doi.org/10.1103/PhysRev.152.1219}{\emph{Phys. Rev.}
  {\bfseries 152} (1966) 1219--1226}.

\bibitem{PhysRev.161.1316}
S.-J. Chang, \emph{Quantization of multispinor fields},
  \href{http://dx.doi.org/10.1103/PhysRev.161.1316}{\emph{Phys. Rev.}
  {\bfseries 161} (Sep, 1967) 1316--1326}.

\bibitem{Schwinger:1978ra}
J.~S. Schwinger, \emph{{Multispinor Basis of Fermi-bose Transformations}},
  \href{http://dx.doi.org/10.1016/0003-4916(79)90255-0}{\emph{Annals Phys.}
  {\bfseries 119} (1979) 192}.

\bibitem{dRMTZ}
C.~de~Rham, S.~Melville, A.~J. Tolley and S.-Y. Zhou, \emph{{To appear}}, .

\bibitem{bogoliubov1959introduction}
N.~N. Bogoliubov, D.~V. Shirkov and S.~Chomet, \emph{Introduction to the theory
  of quantized fields}, vol.~59.
\newblock Interscience New York, 1959.

\bibitem{Hepp_1964}
K.~Hepp, \emph{On the analyticity properties of the scattering amplitude in
  relativistic quantum field theory}, {\emph{Helvetica Physica Acta
  (Switzerland)} {\bfseries Vol: 37} (Jan, 1964) }.

\bibitem{Bremermann:1958zz}
H.~J. Bremermann, R.~Oehme and J.~G. Taylor, \emph{{Proof of Dispersion
  Relations in Quantized Field Theories}},
  \href{http://dx.doi.org/10.1103/PhysRev.109.2178}{\emph{Phys. Rev.}
  {\bfseries 109} (1958) 2178--2190}.

\bibitem{Scadron:1968zz}
M.~D. Scadron, \emph{{Covariant Propagators and Vertex Functions for Any
  Spin}}, \href{http://dx.doi.org/10.1103/PhysRev.165.1640}{\emph{Phys. Rev.}
  {\bfseries 165} (1968) 1640--1647}.

\bibitem{Doughty:1986ya}
N.~A. Doughty and G.~P. Collins, \emph{{Multispinor Symmetries for Massless
  Arbitrary Spin Fierz-pauli and Rarita-schwinger Wave Equations}},
  \href{http://dx.doi.org/10.1063/1.527079}{\emph{J. Math. Phys.} {\bfseries
  27} (1986) 1639--1645}.

\bibitem{Sakurai:1167961}
J.~J. Sakurai, \emph{{Modern quantum mechanics; rev. ed.}}
\newblock Addison-Wesley, Reading, MA, 1994.

\bibitem{Feng:2015mqa}
X.~M. Feng, P.~Wang, W.~Yang and G.~R. Jin, \emph{{High-precision evaluation of
  Wigner's d-matrix by exact diagonalization}},
  \href{http://dx.doi.org/10.1103/PhysRevE.92.043307}{\emph{Phys. Rev.}
  {\bfseries E92} (2015) 043307},
  [\href{https://arxiv.org/abs/1507.04535}{{\ttfamily 1507.04535}}].

\bibitem{Tajima:2015owa}
N.~Tajima, \emph{{Analytical formula for numerical evaluations of the Wigner
  rotation matrices at high spins}},
  \href{http://dx.doi.org/10.1103/PhysRevC.91.014320}{\emph{Phys. Rev.}
  {\bfseries C91} (2015) 014320},
  [\href{https://arxiv.org/abs/1501.06347}{{\ttfamily 1501.06347}}].

\end{thebibliography}\endgroup

\end{document}